\documentclass[aps,pra,twocolumn,showpacs,nofootinbib,longbibliography,notitlepage]{revtex4-2}
\usepackage[english,french]{babel}
\makeatletter
\@ifundefined{l@en}{\let\l@en\l@english}{}
\makeatother
\babeltags{en=english, fr=french}
\usepackage{etex}
\usepackage[normalem]{ulem}
\usepackage{pgfplots}
\pgfplotsset{compat=1.18}
\usepackage{xcolor}
\usepackage{amsmath,amssymb,amsthm}
\usepackage[colorlinks=true,citecolor=blue,urlcolor=blue]{hyperref}
\usepackage{cleveref}
\usepackage{times,txfonts}
\usepackage{braket}
\usepackage{color}
\usepackage{physics}
\usepackage{natbib}
\usepackage{amsmath,blkarray}
\usepackage{mathtools}
\usepackage{latexsym}
\usepackage{tabularx, booktabs}
\usepackage{graphics,epstopdf}
\usepackage{graphicx}
\usepackage{graphicx}
\usepackage{amsfonts}
\usepackage{cleveref}
\usepackage{comment}
\usepackage{color,soul}
\newcommand{\X}{\sigma_x}
\newcommand{\Z}{\sigma_z}

\newcommand{\be}{\begin{equation}}
\newcommand{\ee}{\end{equation}}
\newcommand{\ba}{\begin{eqnarray}}
\newcommand{\ea}{\end{eqnarray}}
\newcommand{\eac}[2]{\expval{\acomm{#1}{#2}}}

\newcommand{\ps}{\ket{\psi}}
\usepackage{mathrsfs}
\newcommand{\na}{\nonumber}

\newcommand{\tcr}{\textcolor{red}}

\begin{document}
	\title{Robust self-testing and certified randomness based on arbitrary-input  Bell inequality} \author{Rajdeep Paul}\email{rajdeeppaul22@gmail.com}
		\author{Sneha Munshi}
        \email{snehamunshi5@gmail.com}
\author{ A. K. Pan }
	\email{akp@phy.iith.ac.in}
	\affiliation{Department of Physics, Indian Institute of Technology Hyderabad, Kandi, Sangareddy, Telengana-502284, India.}
	
	\begin{abstract}
Self-testing is the strongest certification procedure that uniquely characterizes the physical system based on the observed statistics, without any knowledge of the inner workings of the devices. The optimal quantum violation of a Bell inequality enables such a device-independent (DI) self-testing of the source and the measurement devices. In this work, we demonstrate the DI self-testing based on the arbitrary-input chained Bell inequality. We devise a systematic and elegant sum-of-squares (SOS) technique enabling dimension-independent optimization of the quantum violation. Our approach enables the derivation of the state along with the relationship between the local observables directly from the optimization condition.   One significant aspect is the robustness of such self-testing in real experimental situations involving noise and imperfection, leading to deviation from the optimal quantum violation. We provide an analytical technique for robust self-testing in the presence of noise.  As an application of our scheme, we demonstrate the generation of two bit DI randomness and analyze the robustness of such randomness. Our optimization method is both simple and elegant, making it suitable for deriving the optimal quantum violation of various arbitrary-input Bell inequalities.
	\end{abstract}
	\pacs{} 
	\maketitle
\section{Introduction}

The entangled states and incompatible measurements in quantum theory generate non-trivial correlations, which do not admit a local-realist description. Such a feature, widely known as nonlocality, is usually revealed through the quantum violation of suitable Bell inequalities \cite{Bell1964, Clauser1969, Brunner2014, Braunstein1990}. It is now established that the optimal quantum violation of a Bell inequality enables the certification of untrusted sources and uncharacterized devices, commonly referred as DI self-testing. Such a certification is solely based on the input-output statistics, without knowing the inner working of the devices and dimension of the quantum system. The self-testing is the purest form of certification protocol and has profound implications in the development of quantum technologies, such as in DI randomness generation \cite{Acin2012, Colbeck2011, Wooltorton2024}, and DI quantum key distribution \cite{Wooltorton2024, Pironio2009, Acin2021, Acin2007}.

Since the introduction of the self-testing strategy by Mayer and Yao \cite{mayers1998,mayers2004s}, a wide range of protocols have been developed, such as the DI self-testing of pure non-maximally entangled two-qubit states \cite{Acin2012,  Coladangelo2017, Rai2021, Wooltron2022, Rai2022, Bamps2015}, parallel self-testing of multiple maximally entangled two-qubit states \cite{McKague2016, Wu2016}, and so on. Extending this approach to multipartite systems,  self-testing of tripartite W states \cite{Wu2014}, $N$-partite GHZ states \cite{Panwar2023, Singh2025}, maximally entangled two-qudit states \cite{Sarkar2021}, and multipartite graph states \cite{McKague2014} has also been demonstrated. The self-testing of Pauli observables \cite{Bowles2018, Pan2021, Bowles2018/2}, quantum instruments \cite{Wagner2020,mohan2019}, and quantum memory \cite{SekatskiPrl2018, SekatskiPrl2023}, and unsharp measurements \cite{Roy2023, Paul2024,Gomez2016}  has been demonstrated. In the network nonlocality scenario, self-testing of all entangled states \cite{Supic2023}, commuting observables \cite{Munshi2022,Munshi2023PRA}, and complex quantum theory \cite{Renou2021} have been reported.   Relaxing the assumption of dimension independence in DI protocols, significant progress has been made in semi-DI scenarios such as in prepare-measure frameworks \cite{Tavakoli2018, Miklin2020, supic2020input, Abhyoudai2023,Singh2025prepare,  Paulunitary}. Experimental self-testing of states and measurements has also been reported \cite{Smania2020, Gomez2018,zhang2018, Hu2018, Hu2023}. For a comprehensive review, we refer \cite{Supic2020rev} to the reader.

In a real experimental setup, the inevitable noise in the quantum system hinders achieving the optimal quantum violation of a Bell inequality. This necessitates that self-testing techniques be resilient to a noisy state and imperfect devices. The robust self-testing of bipartite entangled states and local measurements with an emphasis on noise tolerance has been studied \cite{McKague2012,Kaniewski2016, McKague2014, Bamps2015, Supic2016, Wu2014, Bowles2018/2}. This was extended to three-qubit W states \cite{Wu2014} and multipartite states \cite{Singh2025, Panwar2023, Sarkar2022, Zhang2019}. Recently, the self-testing protocols is explored to quantum networks \cite{Agresti2021}.

In this paper, we propose a DI self-testing protocol based on the arbitrary-input chained Bell inequality \cite{Braunstein1990}, featuring an arbitrary $n$ number of measurement settings for both parties. Note that the DI self-testing corresponding to a Bell test featuring two inputs commonly uses the Jordan Lemma \selectlanguage{french}\cite{Jordanlemma}, \selectlanguage{english} which proves that there exists a basis so that two dichotomic observables can be block diagonalized, in blocks of dimension $2\times 2$. However, the Jordan lemma does not extend to cases involving more than two measurements or outcomes, which makes DI self-testing quite challenging. We circumvent this shortcoming by introducing a systematic and elegant SOS approach that enables a dimension-independent derivation of the optimal quantum violation. Crucially, the quantum state and observables are directly self-tested from the optimization conditions of the SOS method.

Further, to authenticate the DI certification, we introduce the swap-circuit method to prove the existence of the local isometry necessary for self-testing both the quantum state and the measurement settings. Furthermore, we develop a robust self-testing scheme to account for noise, ensuring the effectiveness of the generated randomness in practical experimental implementation. Crucially, we demonstrate that with higher value of $n$, the DI self-testing becomes more robust so that the state can be extracted with improved fidelity. As an application of our scheme, we demonstrate the generation of two-bit DI randomness for a higher value of odd $n$ and analyzed the robustness of the  generated randomness.

The paper is organized as follows. In Section~\ref{SECI}, we start by discussing the notion of the chained Bell inequality and provide the contrast between the existing results and our work. In Section~\ref{SecII}, we demonstrate the construction of the SOS approach to describe the dimension-independent analytical derivation of the optimal quantum violation of the chained Bell inequality with an arbitrary $n$ number of measurement settings for each party. Moreover, we derive the unique state and measurement settings from the optimization conditions arising from the SOS method. Furthermore,  we provide explicit examples of the derivation of optimal quantum violation of the chained Bell inequality along with the state and observables for odd values of $n$ in  Appendix.~\ref {SOS 3,5,7,11},  and for even values of $n$ in Appendix.~\ref{SOS 4,6}. Then, in Section~\ref{secVI}, we provide the explicit construction of the swap circuit and the illustration of the self-testing of the maximally entangled bipartite state and the corresponding observables using the local isometry. The robustness analysis for self-testing based on the swap circuit is provided in Section~\ref{SecVII}. Then, in Section~\ref{sectionseven}, we provide randomness certification based on the optimal quantum violation of the chained Bell inequality. Finally, we summarize our results and discuss the future direction in Section~\ref{SecIX}.

\section{The arbitrary-input chained Bell inequality}\label{SECI}
In a bipartite Bell scenario, the arbitrary-input chained Bell inequality was proposed in \cite{Braunstein1990}, where two parties, Alice and Bob, both perform an arbitrary $n$ number of dichotomic measurements $A_i$ and $B_j$ with $i,j\in[n]$. The inequality is of the form \cite{Braunstein1990}
  \ba
  \label{BmAB}
  \mathscr{C}_{n}=\sum\limits_{i=1}^{n}(A_i+A_{i+1})B_i\leq {2n-2}
  \ea
   such that 
  
  where $A_{n+1}=-A_1$. The Trielsen bound of the chained Bell inequality was derived in \cite{ Wehner2006}, which is $2n\cos\frac{\pi}{2n}$. The set of measurements and state was provided for a two-qubit system are given by 
 \ba 
 A_i=\sin\theta_i\X+\cos\theta_i\Z, \quad B_i=\sin\phi_i\X+\cos\phi_i\Z, \ea 
 with $\theta_i = \frac{\qty(i-1)\pi}{n}$, $\phi_i=\frac{\qty(2i-1)\pi}{2n}$ and for a two-qubit  maximally entangled state $|\phi^+\rangle=\frac{1}{\sqrt{2}}\qty(\ket{00}+\ket{11})$.

In \cite{Supic2016}, the optimal quantum violation of the chained Bell inequality was derived using an SOS approach,  however, by assuming a two-qubit system. Moreover, the maximally entangled two-qubit state was assumed rather than derived from the optimization conditions. More recently, in \cite{Xiao2023}, the optimal quantum value was obtained using the NPA hierarchy, but the state and the observables cannot be derived.
 
Here, we introduce an analytical method to derive the optimal quantum violation of the chained Bell inequality by developing an elegant SOS approach without requiring the dimension of the quantum system. The nature of the observables and the required entangled state are derived solely from the optimization conditions, thereby showcasing  This showcases a significant advancement from the previous work. Also, in \cite{Supic2016},  the derivation of the local isometry in terms of the observables was imprecise, leading to the self-testing of the plane spanned by only the measurement settings of the local qubit of two parties. We also provide a more straightforward and general definition of robustness \cite{Supic2016}.

\section{An elegant SOS approach for deriving the optimal quantum violation}\label{SecII}
As mentioned above, we formulate an elegant SOS method to derive the optimal quantum violation of the chained Bell inequality in Eq. (\ref{BmAB}) without assuming the dimension of the system. For this, we introduce a suitable positive semi-definite operator $\Gamma_{n}$ such that the Bell functional $\mathscr{C}_n$ can be written as $\expval{\mathscr{C}_n}=\beta_n-\expval{\Gamma_n}$. Since $\expval{\Gamma_n}\geq 0$, we find that the optimal value of $\mathscr{C}_n$ is obtained when $\expval{\Gamma_n}=0$ which, in turn, provides $\expval{\mathscr{C}_n}^{opt}_Q=\beta_n$. We define the operator $\Gamma_{n}$ as 
\begin{eqnarray}\label{gcn}
    \Gamma_{n} = \sum_{i=1}^n\frac{\nu_{n,i}}{2}\mathcal{L}_{n,i}^\dagger\mathcal{L}_{n,i}
\end{eqnarray}  
where 
\ba \label{lni}\mathcal{L}_{n,i}=\left(\frac{A_i+A_{i+1}}{\nu_{n,i}} \otimes \openone_d -\openone_d \otimes B_i\right), \forall i\in[n]\ea
Note that the choice of $\mathcal{L}_{n,i}$ is not unique. But here we specifically choose this form by looking at the structure of the Bell functional.
The coefficients $\nu_{n,i}$s are defined to be suitable  norms given by $\nu_{n,i}=||{A}_{i}+{A}_{i+1}||_{\rho_{AB}}=\sqrt{2+\langle\{A_{i},A_{i+1}\}\rangle}$ and $\langle\{A_{i},A_{i+1}\}\rangle=\Tr[\{A_{i},A_{i+1}\}\rho_{AB}]$. Substituting $\mathcal{L}_{n,i}$ from Eq. (\ref{lni}) into Eq. (\ref{gcn}), we get $\expval{\Gamma_n}=\sum\limits_{i=1}^n\nu_{n,i} -\expval{\mathscr{C}_n}$. Therefore, the optimal quantum value is achieved when  $\expval{\Gamma_n}=0$,  i.e., \ba \label{cn}\sum_i\Tr[\mathcal{L}_{n,i}^\dagger\mathcal{L}_{n,i} \ \rho_{AB}]=0, \quad  \forall i\in[n]\ea Hence\ba\expval{\mathscr{C}_n}^{opt}_Q=\max\qty[\sum\limits_{i=1}^n\nu_{n,i}]=\max\qty[\sum\limits_{i=1}^n\sqrt{2+\langle\{A_{i},A_{i+1}\}\rangle}]\ea 
As $\mathcal{L}_{n,i}^\dagger\mathcal{L}_{n,i}$s are positive hermitian operators, Eq. (\ref{cn}) leads us to the condition  $\Tr[\mathcal{L}_{n,i} \ \rho_{AB}]=0,  \forall i\in[n]$, which further implies
\begin{eqnarray}\label{selfn}
\Tr[(\mathcal{A}_i\otimes B_i) \ \rho_{AB}]&=&1, \quad \forall i\in[n]\nonumber\\
    \mathcal{A}_i\otimes B_i \ps_{AB}&=&\ps_{AB}
\end{eqnarray}
 where $\mathcal{A}_i=\frac{{A}_{i}+{A}_{i+1}}{\nu_{n,i}}$. 
We use  the following inequality   \ba \label{cnv} \sum\limits_{i=1}^{n}f_i\leq  \sqrt{n\sum\limits_{i=1}^{n}f_i^2}, \quad\forall f_i\geq 0, \ea 
where the equality holds if  $f_i=f_j, \forall i\neq j\in[n]$ . Hence, we can write
 \begin{eqnarray}
 \mathscr{
C}_{n}&\leq&\sqrt{n\big[\sum\limits_{i=1}^n (\nu_{n,i})^2\big]}\\
&=&\begin{cases}
     \Bigg[n\bigg(2n +\eac{A_2}{(A_1+A_3)}+\langle\{A_4,(A_3+A_5)\}\rangle\nonumber\\+\eac{A_6}{(A_5+A_7)}
  +...+\eac{A_i}{(A_{i-1}+A_{i+1})}+...\nonumber\\
  \eac{A_n}{(A_{n-1}-A_1)}\bigg)\Bigg]^{\frac{1}{2}}, \quad\text{for even $n$}\label{evenn}\\\\
  \Bigg[n\bigg(2n +\eac{A_2}{(A_1+A_3)}+\langle\{A_4,(A_3+A_5)\}\rangle\nonumber\\+\eac{A_6}{(A_5+A_7)}
  +...+\eac{A_i}{(A_{i-1}+A_{i+1})}+...\nonumber\\
  -\eac{A_1}{A_n}\bigg)\Bigg]^{\frac{1}{2}}, \quad \text{for odd $n$} \label{oddn}\end{cases}
 \end{eqnarray}
 
The optimization conditions provide that  Alice's observables satisfy the following relations.
\begin{eqnarray}\label{ai}\na
    A_i&=&\frac{A_{i-1}+A_{i+1}}{\sqrt{2+\langle\{A_{i-1}, A_{i+1}\}\rangle}},   \ A_{n+1}=-A_1, \  A_0=-A_n\forall \  i\in [n]\\
\end{eqnarray}
\begin{eqnarray}\label{aiix}
    \big{\langle}\{A_i,A_{i+x}\}\big{\rangle}&=&2\cos\frac{\pi \ x}{n}, \forall i\in[n], x\in[n-i]
\end{eqnarray}
where $ \big{\langle}\{A_i,A_{i+x}\}\big{\rangle}=\Tr[\{A_i,A_{i+x}\}\rho_{AB}]$, which means that the quantity $\big{\langle}\{A_i,A_{i+x}\}\big{\rangle}$ is defined only on the support of the state $\rho_{AB}$. We provide the form of the state shortly. 
Using Eq. (\ref{aiix}), we obtain $\nu_{n,i}=||{A}_{i}+{A}_{i+1}||_{\rho_{AB}}=\sqrt{2+\langle\{A_{i},A_{i+1}\}\rangle}=2\cos\frac{\pi}{2n}, \forall i\in[n]$.
Consequently, we obtain the optimal quantum value
\ba\label{cnopt}(\mathscr{C}_{n})^{opt}_{Q}=2n\cos\frac{\pi}{2n}\ea
 Note that the derivation of $ (\mathscr{C}_n)^{opt}_Q$ for even $n$ and odd $n$ requires a slightly different technique. In Appendix.~\ref{SOS 3,5,7,11}, and \ref{SOS 4,6}, the detailed derivations for the optimal quantum values for odd $n$ and even $n$ are provided.

From the optimization conditions, we also get that  for all $i\in[n]$
\begin{eqnarray}\label{rand0m}
 \bra{\psi}_{AB}A_i\otimes B_{i+x}\ket{\psi}_{AB}
    &=&\frac{\cos{\frac{\pi x}{n}}+\cos{\frac{\pi(x+1)}{n}}}{2\cos{\frac{\pi}{2n}}}
\end{eqnarray}
where $i\in[n], x\in \mathbb{Z}$ and $x\in[-n, n],  \forall n\in\mathbb{N},\forall|i+x|\leq n$. The detailed proof of Eq. (\ref{rand0m}) is provided in Appendix.~\ref{Bobo}. 

As shown in the Appendix.~\ref{Bobo}, the optimization condition provides that Bob's observables satisfy the following relations.
\begin{eqnarray}\label{bobi}
   &&B_i=\frac{B_{i-1}+B_{i+1}}{\sqrt{2+\langle\{B_{i-1}, B_{i+1}\}\rangle}},   \  B_0=-B_n \ \forall i\in [n]
\end{eqnarray}
and
\begin{eqnarray}
   &&\big{\langle}\{B_i,B_{i+x}\}\big{\rangle}=2\cos\frac{\pi \ x}{n}, \forall i\in[n], x\in[n-i]\label{biix}
\end{eqnarray}
which leads to
\begin{eqnarray}
    &&A_i\otimes \mathcal{B}_i\ket{\psi}_{AB}=\ket{\psi}_{AB}, \  \forall i\in[n]\label{Bobsc1}
\end{eqnarray}
where $\mathcal{B}_i=\frac{B_i+B_{i-1}}{\nu'_{n,i}}$ and $\nu'_{n,i}=||{B}_{i}+{B}_{i-1}||_{\rho_{AB}}$.
From the construction of the $\mathcal{B}_i$ we show that \ba \label{Bobscmain}\langle\{\mathcal{B}_i,\mathcal{B}_{i+x}\}\rangle=2\cos\frac{\pi \ x}{n}, \forall i\in[n], x\in[n-i]\ea  Thus, we show that, at the optimal quantum value, the sets of observables of Alice and Bob satisfy a similar set of relations. Interestingly, under the optimal condition, both the Eq.~(\ref{selfn}) and (\ref{Bobsc1}) hold simultaneously. 

From the optimization condition of Eq.~(\ref{selfn}), we see that   $\rho_{AB}$ must be a common eigen state of the operators $\qty(\mathcal{A}_i\otimes B_i),  \forall i\in[n]$. As these operators yield the maximum eigenvalue with the normalized state $\rho_{AB}$, which implies that  $\rho_{AB}=|\psi_{AB}\rangle\langle\psi_{AB}|$ needs to be a maximally entangled pure state. We can then write the state $\rho_{AB}\in \mathbb{C}^d\times \mathbb{C}^d$ in this form \cite{Paul2024}
\begin{eqnarray}\label{rho odd}
\rho_{AB} = \frac{1}{d^2} \qty[\openone_d\otimes\openone_d + \sum_{\bar{i}=1}^{d^2-1} C_{\bar{i}} \otimes C_{\bar{i}}]
\end{eqnarray}
Since $\rho_{AB}$ is a pure state, for each $\bar{i}\neq \bar{j}  \in[d^2-1]$, the terms $C_{\bar{i}}\otimes C_{\bar{i}}$ and $C_{\bar{j}}\otimes C_{\bar{j}}$ in Eq.~(\ref{rho odd}) has to be  mutually commuting and $\Tr[C_{\bar{i}}]=0$. To explicitly derive
$C_{\bar{i}} \otimes C_{\bar{i}}$, we use  the following conditions i) $\Tr[(\mathcal{A}_i\otimes B_i) \ \rho_{AB}]=1$, ii) $\Tr[(C_{\bar{i}} \otimes C_{\bar{i}})\rho_{AB}]=1$, and 
    iii) $\langle[C_{\bar{i}}\otimes C_{\bar{i}}, C_{\bar{j}}\otimes C_{\bar{j}}]_{\bar{i}\neq \bar{j}}\rangle_{\rho_{AB}}=0$.
Using these conditions, we obtain explicit expressions for any two \(C_{\bar{i}} \otimes C_{\bar{i}}\) say,  \(C_1 \otimes C_1\) and \(C_2 \otimes C_2\), in such a way that only these two terms will contribute for \(\Tr[(\mathcal{A}_i \otimes B_i) \rho_{AB}] = 1\), $\forall i \in [n]$. Detailed derivations are provided in Appendix~\ref{cseodd} and \ref{stateeven}.

For odd $n$, the observables in Eq. (\ref{rho odd}) take the form as follows. 
\ba\label{C'so}
C_1 \otimes C_1 &=& \mathcal{A}_{\frac{n+1}{2}} \otimes B_{\frac{n+1}{2}}\\\na
C_2 \otimes C_2&=&\frac{1}{\lfloor \frac{n}{2} \rfloor}\sum_{i=1}^{{\lfloor \frac{n}{2}\rfloor}}\frac{(\mathcal{A}_{i}-\mathcal{A}_{n+1-i}) \otimes (B_{i}-B_{n+1-i})}{(2-\langle\{B_{i},B_{n+1-i}\}\rangle)}\ea
and 
\ba C_3 \otimes C_3 &=&\frac{1}{\lfloor \frac{n}{2} \rfloor}\sum_{i=1}^{{\lfloor \frac{n}{2}\rfloor}}\frac{\mathcal{A}_{\frac{n+1}{2}}(\mathcal{A}_{i}-\mathcal{A}_{n+1-i}) \otimes B_{\frac{n+1}{2}}(B_{i}-B_{n+1-i})}{(2-\langle\{B_{i},B_{n+1-i}\}\rangle)}\nonumber\\
&=&(C_1 \otimes C_1)\cdot (C_2 \otimes C_2)\na
\ea
where $\mathcal{A}_i = \frac{A_i+A_{i+1}}{2\cos\frac{\pi }{2 n}}$. The detailed derivation is placed in  Appendix.~\ref{cseodd},  along with an explicit derivation for $n=7$.\\

Similarly, for even $n$,
\ba\label{C'se}
C_1 \otimes C_1 &=& {A}_1 \otimes \mathcal{B}_1\\\na
C_2 \otimes C_2&=&{A}_{\frac{n}{2}+1} \otimes \mathcal{B}_{\frac{n}{2}+1}\ea 
and \ba 
C_3 \otimes C_3 &=& \frac{1}{4(\frac{n}{2}-1)}\sum\limits_{i=2}^{\frac{n}{2}}\big[A_i, A_{i+\frac{n}{2}}\big]\otimes \big[\mathcal{B}_i,\mathcal{B}_{i+\frac{n}{2}}\big]\nonumber\\
&=&(C_1 \otimes C_1)\cdot (C_2 \otimes C_2)
\ea
Detailed derivation is referred in the Appendix.~\ref{stateeven}, together with an explicit derivation for $n=4$.
%%%%%%%%%%%%%%%%%%%%%%%%%%%%%%%%%%%%%%%%%%%%%%%%%%%%%%%%%%%%%%%%%%%%%%%%%%%%%%%%%%%%%%%%%%%%%%%%%%%%%%%%%%%%%%%%% 
\section{Self-testing of state and observables using Swap-circuit} \label{secVI}

\begin{figure}
    \centering
    \includegraphics[width=8cm, height=4cm]{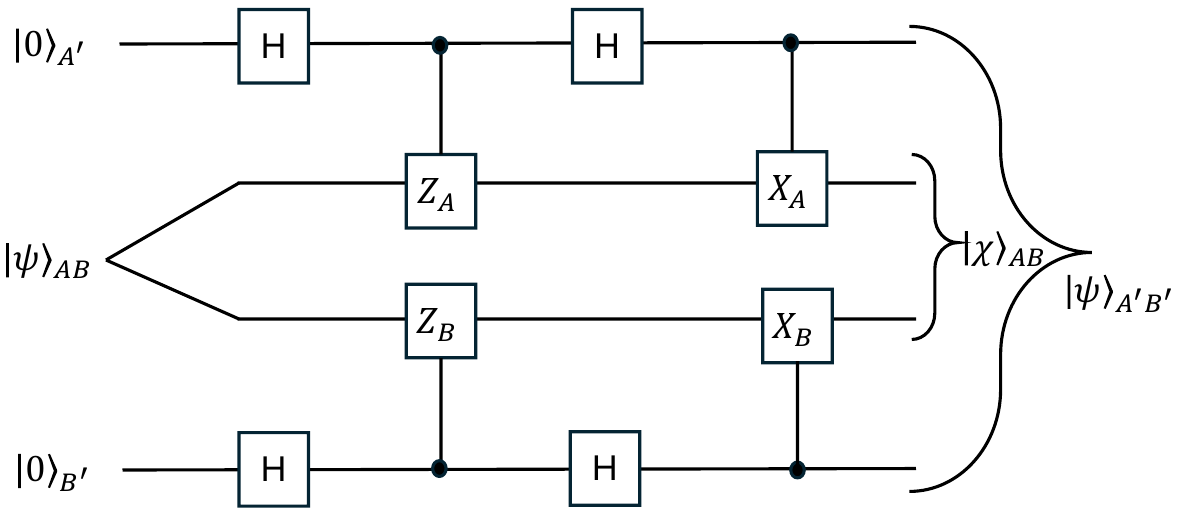}
    \caption{Swap circuit for state and observables for arbitrary $n$}\label{Oddc}
\end{figure}

 We now demonstrate that the state and measurements can be uniquely characterized when the optimal quantum violation of the chained Bell inequality is obtained. The swap circuit in the self-testing protocol is used to mimic a physical experiment in a black-box setting, presuming that the system functions with a minimum-dimensional quantum state. In such cases, the action of a local unitary transfers the properties of the physical state to the ancillary system, thus enabling the certification of both the state and the measurements. The optimal quantum violation of the chained Bell inequality requires the condition in Eq.~(\ref{selfn}) to be satisfied. 
 
 Consider that the required bipartite state is $\ket{\psi}_{AB}$, and the corresponding local observables of Alice(Bob) are $X_m$ and $Z_m$, with $m\in\{A,B\}$ Swap-circuit aims to replicate the same measurement statistics as the physical setup while creating an isometry that acts jointly on the physical system and the reference system. This is  mathematically expressed as 
\begin{eqnarray}
\label{stest}
\Phi(\mathcal{O}_t\ket{\psi}_{AB}\otimes\ket{ancilla}_{A'B'} 
 )&=&\ket{\chi}_{AB}\otimes U_t \ket{ent}_{A'B'}
\end{eqnarray}
where $\mathcal{O}_1=\openone_d, \mathcal{O}_2=X_m$ and $\mathcal{O}_3=Z_m, \forall m\in[A, B]$. Also,  $\ket{\psi}_{AB}$ is the physical state, $\ket{ancilla}_{A'B'}$ is ancillary's system, $\ket{\chi}_{AB}$ is the so-called junk' state, $\ket{ent}_{A'B'}$ is the entangled state of a Ancillary systems and $U_t$ is the unitary operation acting on the ancillary system. For $t=1$, $U_1=\openone$. We construct the observables $X_m$ and $Z_m$  in such a way that they satisfy the normalization condition as well as the self-testing conditions. The construction of observables is different for different values of input $n$. For odd $n$,
\begin{eqnarray}
\label{stest1}\na
    Z_A&=&\mathcal{A}_\frac{n+1}{2}, 
     X_A =\frac{1}{\lfloor\frac{n}{2}\rfloor}\sum_{i=1}^{{\lfloor \frac{n}{2}\rfloor}}\frac{\mathcal{A}_{i}-\mathcal{A}_{n+1-i}}{\sqrt{2-\langle\{\mathcal{A}_{i},\mathcal{A}_{n+1-i}\}\rangle}}\\
    Z_B&=&B_\frac{n+1}{2},
    X_B = \frac{1}{\lfloor\frac{n}{2}\rfloor}\sum_{i=1}^{{\lfloor \frac{n}{2}\rfloor}}\frac{B_{i}-B_{n+1-i}}{\sqrt{2-\langle\{B_{i},B_{n+1-i}\}\rangle}}\label{n4}
\end{eqnarray}
and for even $n$,
\begin{eqnarray}
    Z_A &=&\frac{2}{n}\sum_{i=1}^{ \frac{n}{2}}\hspace{-0.2cm}\frac{\mathcal{A}_{i}+\mathcal{A}_{n+1-i}}{\sqrt{2+\langle\{\mathcal{A}_{i},\mathcal{A}_{n+1-i}\}\rangle}},
    X_A =\frac{2}{n}\sum_{i=1}^{ \frac{n}{2}}\hspace{-0.2cm}\frac{\mathcal{A}_{i}-\mathcal{A}_{n+1-i}}{\sqrt{2-\langle\{\mathcal{A}_{i},\mathcal{A}_{n+1-i}\}\rangle}}\nonumber\\ \na 
    Z_B &=&\frac{2}{n}\sum_{i=1}^{ \frac{n}{2}}\hspace{-0.2cm}\frac{B_{i}+B_{n+1-i}}{\sqrt{2-\langle\{B_{i},B_{n+1+i}\}\rangle}},
    X_B =\frac{2}{n}\sum_{i=1}^{ \frac{n}{2}}\hspace{-0.2cm}\frac{B_{i}-B_{n+1-i}}{\sqrt{2-\langle\{B_{i},B_{n+1-i}\}\rangle}}\\   
\end{eqnarray}
    It is straightforward to check that
\begin{eqnarray}\label{circuitself}
    Z_A\ket{\psi}_{AB}=Z_B\ket{\psi}_{AB}&,& X_A\ket{\psi}_{AB}=X_B\ket{\psi}_{AB}\nonumber\\
    \{Z_A, X_A\}\ket{\psi}_{AB} &=& \{Z_B, X_A\}\ket{\psi}_{AB}=0.
\end{eqnarray}
 A detailed derivation of Eq.(\ref{stest}) and the construction of the isometries for even and odd values of  $n$ is provided in Appendix.~\ref{c5} and \ref{c4}. 

For experimental authentication, it will be more helpful to find the isometry by considering known dimensions. If $\ps_{AB}\in \mathcal{H}_A\otimes \mathcal{H}_B$ is the physical state, which satisfies Eq.~(\ref{circuitself}) and  $A_i\in \mathcal{H}_A$, $B_j\in \mathcal{H}_B$ are the observables which gives the optimal value of $\mathscr{C}_n$. Hence, from the swap-circuit in Fig.~\ref{Oddc}, we prove that there exists a local isometry $\Phi$ and a local ancilla $\ket{00}_{A'B'}$ so that
\begin{eqnarray}
    \Phi(\mathcal{O}_t\ket{\psi}_{AB}\otimes\ket{00}_{A'B'} 
 )&=&\ket{\chi}_{AB}\otimes U_t \ \ket{\phi^+}_{A'B'}
\end{eqnarray}
where $\ket{\chi}_{AB}=\frac{1+Z_A}{\sqrt{2}}\ket{\psi}_{AB}$ is the junk state. $\mathcal{O}_1=\openone_2, \mathcal{O}_2=X_m$ and $\mathcal{O}_3=Z_m,  \forall m\in\{A, B\}$, and $U_t$s are Pauli observables. Now we can express $A_i$ and $B_i$ in terms of $Z_A, Z_B, X_A, X_B$ hence, the same relations will hold for $A_i$ and $B_j$ i.e.,
\begin{eqnarray}
\Phi(A_i\ket{\psi}_{AB}\otimes\ket{00}_{A'B'} 
 )&=&\ket{\chi}_{AB}\otimes (A'_i\otimes \openone_2)\ket{\phi^+}_{A'B'}\\
\Phi(B_j\ket{\psi}_{AB}\otimes\ket{00}_{A'B'} 
 )&=&\ket{\chi}_{AB}\otimes (\openone_2\otimes B'_j)\ket{\phi^+}_{A'B'}\\
 \Phi(A_i\otimes B_j\ket{\psi}_{AB}\otimes\ket{00}_{A'B'} 
 )&=&\ket{\chi}_{AB}\otimes (A'_i\otimes B'_j)\ket{\phi^+}_{A'B'}
\end{eqnarray}
The detailed calculation is provided in Appendix.~\ref{c5} and \ref{c4}. We thus demonstrate that the self-testing protocol based on the optimal quantum value  $(\mathscr{C}_n)_Q^{opt}$ given in Eq.~(\ref{cnopt}) provides the equivalence between the reference and the physical experiments. 
%%%%%%%%%%%%%%%%%%%%%%%%%%%%%%%%%%%%%%%%%%%%%%%%%%%%%%%%%%%%%%%%%%%%%%%%%%%%%%%%%%%%%%%%%%%%%%%%%%%%%%%%%%

\section{Robust self-testing using swap circuit} \label{SecVII}
We note that in a real experiment, noise and imperfection in implementation do not allow to achieve the optimal quantum value. Then it is crucial to analyze the robustness of a self-testing protocol to the noise. Here, we provide a rigorous analysis of the robustness of our protocol in the presence of noise that leads to sub-optimal quantum violation. This analysis does not require the exact root cause of the error; instead, we may assume that it arises solely from the imperfect implementation of observables \cite{Bamps2015, Acin2020}.

Assume that the imperfect version of the observables ($\Tilde{X}_m, \Tilde{Z}_m$) differ from the ideal observables ($X_m, Z_m$) with $m\in\{A,B\}$, so that

\begin{eqnarray}
    ||(\Tilde{X}_m-X_m)\ket{\psi}_{AB}||\leq \alpha_m, \ \ \ ||(\Tilde{Z}_m-Z_m)\ket{\psi}_{AB}||\leq \beta_m
\end{eqnarray}
where $(\alpha_m, \beta_m)\in \mathbb{R}^+$. In an ideal condition, $ \alpha_m=\beta_m=0$. However, due to the presence of noise, the observables are not necessarily unitary. Therefore, the revised self-testing conditions are $0\leq||\Tilde{\mathcal{L}}_{n,i}\ket{\psi}_{AB}||\leq \xi_{n,i}$ instead of $||\mathcal{L}_{n,i}\ket{\psi}_{AB}||=0$. Consequently, 
\begin{eqnarray}
    \Tr[\Tilde{\Gamma}_n\  \rho_{AB}]&=& \frac{1}{2}\sum_{i=1}^{n} \nu_{n,i} \bra{\psi}_{AB}\Tilde{\mathcal{L}}_{n,i}^{\dag} \Tilde{\mathcal{L}}_{n,i}\ket{\psi}_{AB}\nonumber\\
    &=& \frac{1}{2}\sum_{i=1}^{n} \nu_{n,i}  \ \xi_{n,i}^2
\end{eqnarray}
This leads to a sub-optimal quantum value of the Bell functional
\begin{eqnarray}\label{Cn noise}
    \Tilde{(\mathscr{C}_{n})}_{Q}=&2n\cos {\frac{\pi}{2n}}-\xi
\end{eqnarray}
where $\xi=\frac{1}{2}\sum_{i=1}^{n} \nu_{n,i}  \ \xi_{n,i}^2\geq 0$. For robust self-testing of state and observables, we show that the trace distance between the output of the perfect and imperfect implementations is
\begin{eqnarray}
    ||\Tilde{\Phi}(U_t\ket{\psi}_{AB}\otimes\ket{00}_{A'B'})&-&\Phi(\ket{\psi}_{AB}\otimes\ket{00}_{A'B'})||\nonumber\\
 &&\hspace{8mm} \leq F_t(\alpha_A,\alpha_B,\beta_A,\beta_B)
\end{eqnarray}
where $U_{t=1}=\openone, $ $U_{t=2}=\Tilde{X}_m$ , $U_{t=3}=\Tilde{Z}_m$ and \ba \na F_{t=1}(\alpha_A,\alpha_B,\beta_A,\beta_B)&=&F_S(\alpha_A,\alpha_B,\beta_A,\beta_B) \\
F_{t=2}(\alpha_A,\alpha_B,\beta_A,\beta_B)&=&F_{O_x}(\alpha_A,\alpha_B,\beta_A,\beta_B)\\\na 
F_{t=3}(\alpha_A,\alpha_B,\beta_A,\beta_B)&=&F_{O_Z}(\alpha_A,\alpha_B,\beta_A,\beta_B)\ea with 
\begin{eqnarray}
    &&\lim_{\{\alpha_A,\alpha_B,\beta_A,\beta_B\}\to 0} F_t(\alpha_A,\alpha_B,\beta_A,\beta_B)=0.
\end{eqnarray}
We have given the detailed derivations in Appendix.~\ref{robustselftesting}.

Note that errors can arise from imperfections in state preparation, observable implementation, or both. However, this is extremely difficult to demonstrate, and therefore, we restrict our analysis by considering the imperfection that arises from one party's implementation of observables.  
If Bob implements imperfect observables and the error is the same for both implementations, then $\alpha_B=\beta_B=\epsilon\geq 0$. If the error of each observable of Bob is $\delta$, then \ba ||(\Tilde{B}_i-B_i)\ket{\psi}_{AB}||\leq \delta\implies \Tilde{B}_i\approx B_i+\delta \ \openone_d, \forall i\in[n],\ \delta \geq 0\na \\
\ea  which implies that $||(\Tilde{Z}_B-Z_B)\ket{\psi}_{AB}||=||(\Tilde{B}_{\frac{n+1}{2}}-B_{\frac{n+1}{2}})\ket{\psi}_{AB}||\leq \delta$,  and thus we get  $  \delta\approx\epsilon$. Then the output isometry differs from the ideal one as
\begin{eqnarray}
    &&||\Tilde{\Phi}(\ket{\psi}_{AB}\otimes\ket{00}_{A'B'})-\Phi(\ket{\psi}_{AB}\otimes\ket{00}_{A'B'})||\leq 4\epsilon +\epsilon^2\\
    &&||\Tilde{\Phi}(\Tilde{O}\ket{\psi}_{AB}\otimes\ket{00}_{A'B'})-\Phi(O\ket{\psi}_{AB}\otimes\ket{00}_{A'B'})||\nonumber\\
    && \quad \hspace{25mm}\leq \epsilon^3+5\epsilon^2+8\epsilon, \quad \forall O\in\{X_B,Z_B\}
\end{eqnarray}
We derive the relation between $\epsilon$ and  the deviation $\xi$ from the optimal value (in Eq.~(\ref{Cn noise})) as $\epsilon =\sqrt{\frac{\xi }{n \cos \left(\frac{\pi }{2 n}\right)}}$. Detailed derivation is provided in the Appendix~\ref{spodd}. To determine the robust self-testing bounds for both state and observables, we define relative observed violations $(r)$ as a function of $n$ and $\xi$, where
\begin{eqnarray}
    &&r=\frac{\Tilde{(\mathscr{C}_{n})}_{Q}-(\mathscr{C}_n)_C}{(\mathscr{C}_{n})^{opt}_{Q}-(\mathscr{C}_n)_C}= 1-\frac{\xi}{2 n \cos{\frac{\pi}{2n}}-2n+2}
\end{eqnarray}
Alternatively, $\xi$ can be written in terms of $r$ and $n$ as
\begin{eqnarray}
        &&\xi =(1-r) \left(2 n \cos \left(\frac{\pi }{2 n}\right)-2n+2)\right)
\end{eqnarray}
where $\Tilde{(\mathscr{C}_{n})}_{Q}$ and $ (\mathscr{C}_{n})^{opt}_{Q}$ are defined in Eqs.~(\ref{Cn noise}) and (\ref{cnopt}) respectively. Here $(\mathscr{C}_n)_c =2n-2$ is the classical bound of the chained Bell inequality.

Hence, the trace distance between the observed and ideal state in terms of the relative observed value $r$ is
\begin{eqnarray}
    &&||\Tilde{\Phi}(\ket{\psi}_{AB}\otimes\ket{00}_{A'B'})-\Phi(\ket{\psi}_{AB}\otimes\ket{00}_{A'B'})||\leq  f_s(r)\quad
\end{eqnarray}
 where  
 \begin{eqnarray}
 f_s(r)&=&\frac{2 (n-1) (r-1) \sec \left(\frac{\pi }{2 n}\right)}{n}\\
 \nonumber
 &&+4 \sqrt{2} \sqrt{\frac{(r-1) \left((n-1) \sec \left(\frac{\pi }{2 n}\right)-n\right)}{n}}-2 r+2
 \end{eqnarray} 
 The detailed derivation is provided in Appendix.~\ref{spodd}. Now, using  Fuchs-Van de Graaf inequality \cite{Wilde2013}, the approximate relation between the trace distance $f_s(r)$ and robust fidelity $F_s(r)$ is 
 \begin{eqnarray}
     2\qty(1-\sqrt{F_s(r)})\leq f_s(r)\leq 2\sqrt{1-F_s(r)}
 \end{eqnarray}
 This gives the lower bound of fidelity in terms of trace distance as
 \begin{eqnarray}
     F_k(r)\geq \qty(1-\frac{1}{2}f_k(r))^2\quad \forall k\in\{s,o\}
 \end{eqnarray}
where, for the observables, we define the fidelity $F_o(r)$ in terms of trace distance $f_o(r)$ where  
\begin{eqnarray}
\nonumber
 f_o(r)&=&2 \sqrt{2} \left(\frac{(r-1) \left((n-1) \sec \left(\frac{\pi }{2 n}\right)-n\right)}{n}\right)^{3/2}\\
 &&+\frac{10 (r-1) \left((n-1) \sec \left(\frac{\pi }{2 n}\right)-n\right)}{n}\\
 \nonumber
 &&+8 \sqrt{2} \sqrt{\frac{(r-1) \left((n-1) \sec \left(\frac{\pi }{2 n}\right)-n\right)}{n}}   
\end{eqnarray}
 The explicit derivation is provided in Appendix.~\ref{spodd}.

 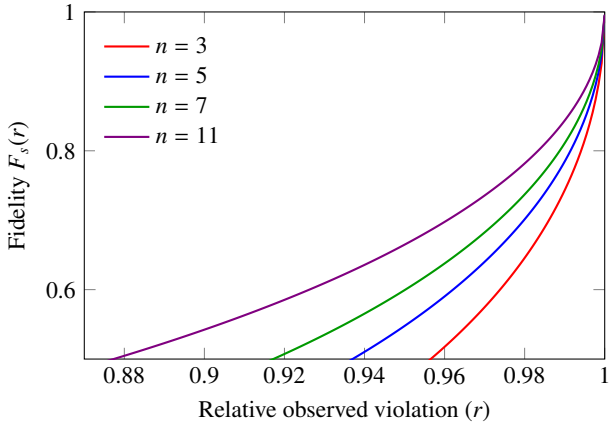
\begin{figure}[ht]\label{osn1}
\centering
\begin{tikzpicture}
\begin{axis}[
    xlabel={Relative observed violation $(r)$},
    ylabel={Fidelity $F_s(r)$},
    xmin=0.87, xmax=1,
    ymin=0.5, ymax=1,
    legend style={at={(0.02,0.58)}, anchor=south west, draw=none},
    width=8.5cm,
    height=6.2cm,
    domain=0.85:1,
    samples=200,
    axis lines=box,
    tick label style={font=\small},
    label style={font=\small},
    legend cell align={left}
]

% n = 3
\addplot[red, thick]
    {(1 + 0.5*(-2 + 2*x - (2*(3 - 1)*(x - 1)/3)*sec(deg(pi/(2*3))) - 4*sqrt(2)*sqrt(((x - 1)*(-3 + (3 - 1)*sec(deg(pi/(2*3))))/3))))^2};
\addlegendentry{$n = 3$}

% n = 5
\addplot[blue, thick]
    {(1 + 0.5*(-2 + 2*x - (2*(5 - 1)*(x - 1)/5)*sec(deg(pi/(2*5))) - 4*sqrt(2)*sqrt(((x - 1)*(-5 + (5 - 1)*sec(deg(pi/(2*5))))/5))))^2};
\addlegendentry{$n = 5$}

% n = 7
\addplot[green!60!black, thick]
    {(1 + 0.5*(-2 + 2*x - (2*(7 - 1)*(x - 1)/7)*sec(deg(pi/(2*7))) - 4*sqrt(2)*sqrt(((x - 1)*(-7 + (7 - 1)*sec(deg(pi/(2*7))))/7))))^2};
\addlegendentry{$n = 7$}

% n = 11
\addplot[violet, thick]
    {(1 + 0.5*(-2 + 2*x - (2*(11 - 1)*(x - 1)/11)*sec(deg(pi/(2*11))) - 4*sqrt(2)*sqrt(((x - 1)*(-11 + (11 - 1)*sec(deg(pi/(2*11))))/11))))^2};
\addlegendentry{$n = 11$}

\end{axis}
\end{tikzpicture}
\caption{Trade-off between relative observed violation ($r$) and fidelity  $F_s(r)$ for $n=3,5,7,11$. Here, we are taking only that region where fidelity converges to one.}
\end{figure}

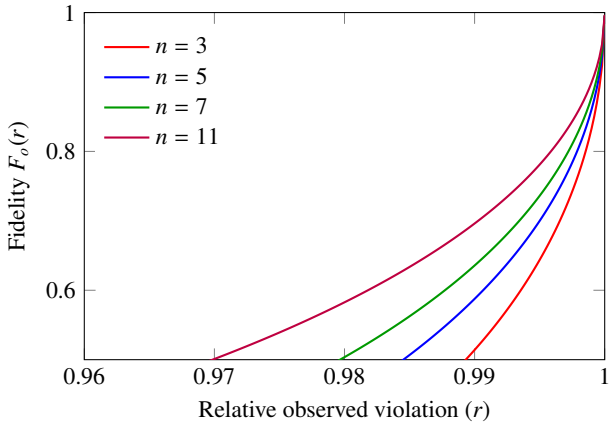
\begin{figure}[ht]
\centering
\begin{tikzpicture}
\begin{axis}[
    xlabel={Relative observed violation $(r)$},
    ylabel={Fidelity $F_o(r)$},
    xmin=0.96, xmax=1,
    ymin=0.5, ymax=1,
    legend style={at={(0.02,0.58)}, anchor=south west, draw=none},
    width=8.5cm,
    height=6.2cm,
    domain=0.96:1,
    samples=200,
    axis lines=box,
    tick label style={font=\small},
    label style={font=\small},
    legend cell align={left}
]

% Define the common expression for the square root part
% f = ((-1 + r) (-n + (-1 + n) Sec[π/(2n)]))/n

% n = 3
\addplot[red, thick]
    {(1 + 0.5*(-(10*(x - 1)*(-3 + (3 - 1)*sec(deg(pi/(2*3))))/3)
               - 8*sqrt(2)*sqrt(((x - 1)*(-3 + (3 - 1)*sec(deg(pi/(2*3))))/3))
               - 2*sqrt(2)*((x - 1)*(-3 + (3 - 1)*sec(deg(pi/(2*3))))/3)^(3/2)))^2};
\addlegendentry{$n = 3$}

% n = 5
\addplot[blue, thick]
    {(1 + 0.5*(-(10*(x - 1)*(-5 + (5 - 1)*sec(deg(pi/(2*5))))/5)
               - 8*sqrt(2)*sqrt(((x - 1)*(-5 + (5 - 1)*sec(deg(pi/(2*5))))/5))
               - 2*sqrt(2)*((x - 1)*(-5 + (5 - 1)*sec(deg(pi/(2*5))))/5)^(3/2)))^2};
\addlegendentry{$n = 5$}

% n = 7
\addplot[green!60!black, thick]
    {(1 + 0.5*(-(10*(x - 1)*(-7 + (7 - 1)*sec(deg(pi/(2*7))))/7)
               - 8*sqrt(2)*sqrt(((x - 1)*(-7 + (7 - 1)*sec(deg(pi/(2*7))))/7))
               - 2*sqrt(2)*((x - 1)*(-7 + (7 - 1)*sec(deg(pi/(2*7))))/7)^(3/2)))^2};
\addlegendentry{$n = 7$}

% n = 11
\addplot[purple, thick]
    {(1 + 0.5*(-(10*(x - 1)*(-11 + (11 - 1)*sec(deg(pi/(2*11))))/11)
               - 8*sqrt(2)*sqrt(((x - 1)*(-11 + (11 - 1)*sec(deg(pi/(2*11))))/11))
               - 2*sqrt(2)*((x - 1)*(-11 + (11 - 1)*sec(deg(pi/(2*11))))/11)^(3/2)))^2};
\addlegendentry{$n = 11$}

\end{axis}
\end{tikzpicture}
 \caption{Trade-off between relative observed violation ($r$) and fidelity  $F_o(r) \ \forall O\in\{X_B,Z_B\}$ for $n=3,5,7,11$. Here, we are taking only that region where fidelity converges to one.}\label{osn2}

\end{figure}

 Figures~\ref{osn1} and \ref{osn2} illustrate the trade-off between the lower bound of the fidelity function and the relative observed violation. In particular, the extractability of both quantum states and observables remains robust when employing the swap circuit, provided that the fidelity exceeds $\frac{1}{2}$. This threshold is reached at $r = 0.8774 \ (0\leq\epsilon\leq 0.1414)$ for the state and $r = 0.97 \ (0\leq\epsilon\leq 0.0701)$ for the observables, where $n=11$. The maximum tolerance $\xi$ is constrained to $0.052$ when $n = 3$; it increases to $0.22$ for $n = 11$. This indicates that increasing the number of measurement settings $n$ enables successful extraction even with smaller observed violations, suggesting that comparatively weak violations of the Bell inequality can still be sufficient to achieve the desired fidelity threshold. In Appendix~\ref{spodd}, Fig.\ref{highn}, we have shown how the relative observed violation $r$ changes with sufficiently large measurement settings.

%%%%%%%%%%%%%%%%%%%%%%%%%%%%%%%%%%%%%%%%%%%%%%%%%%%%%%%%%%%%%%%%%%%%%%%%%%%%%%%%%%%%%%%%%%%%%%%%%%%%%%%%%%%%%%%%%%%%%%%%%%%%%%%%%%%%%%%%%%%%
\section{DI Randomness generation}\label{sectionseven}
It is well known that DI randomness can only be generated from the Bell test \cite{colbeck2011arxiv, Colbeck2011, Pironio2010}. In a bipartite scenario with two outcome measurements for each subsystem,   up to two bits of randomness can be generated, in principle. However, the optimal quantum violation of the Clauser-Horne-Shimony-Holt (CHSH) inequality can generate $1.23$ bits of randomness only \cite{Acin2012}.  Later, it was demonstrated that two bits of randomness are achievable, with an arbitrarily small level of nonlocality, while implementing the tilted Bell inequality \cite{Acin2012, Wooltron2022}; however, the robustness analysis remained unaddressed.  Wooltorton \textit{et al.,}\cite{Wooltorton2024} came up with a protocol demonstrating DI self-testing of two bits of randomness using a tilted CHSH inequality. However, this requires unbounded precision in practical implementation due to the small difference between classical and quantum values, which makes robust self-testing extremely challenging.

In a bipartite Bell experiment with two outcomes per party, one can certify at most 2 bits of global randomness. This follows from the fact that for two local dichotomic observables per party, there are four joint outcome probabilities, and maximum randomness corresponds to each of them being $1/4$. The certification of 2 bits of randomness was derived  \cite{Supic2016} through the chained Bell inequality. But they needed a modification of the original Bell functional. More recently, employing the NPA hierarchy, 2 bit randomness was demonstrated \cite{Xiao2023}. However, they did not demonstrate self-testing of the correlations. In this work, without modifying the inequality and only considering self-testing conditions, we demonstrate the certification of 2 bits of randomness.
 
 The certified randomness \(\mathcal{R}_{ij}\) corresponding to Alice's and Bob's measurement choices  \(A_i\) and  \(B_j\) is observed for any behaviors \(\mathcal{P}\equiv\{P(a,b|A_i,B_j)\}\), is determined as \cite{Acin2012}

\begin{equation}\label{random}
\mathcal{R}_{ij} = - \log_{2}\left[\max_{\mathcal{P},a,b}P\left(a,b,|A_i,B_j\right)\right]
\end{equation}
Note that $\mathcal{R}_{ij}$ explicitly depends on the measurement choices of Alice and Bob, and therefore the certified randomness $\mathcal{R}_{ij}$ is derived by considering all possible correlation choices $\langle A_i\otimes B_j\rangle$ for $i,j\in[n]$.  For a given  measurement settings $A_i$ and $B_j$, the probability $P\left(a,b|A_i,B_j\right)$ is defined as follows.

\begin{eqnarray}\label{pbc}
P\left(a,b|A_i,B_j\right)&=&\Tr[ \bigg(\frac{\openone_d+a A_i}{2}\otimes \frac{\openone_d+bB_j}{2}\bigg)\ \rho_{AB}]
\end{eqnarray}
where $ a,b\in \pm 1$. We know, for a maximally entangled state, $\langle{A_i}\rangle_{\rho_{AB}}=0, \langle{B_j}\rangle_{\rho_{AB}}=0, \forall i,j\in[n]$.  Hence, we get 
\begin{eqnarray}\label{probaibj}
    P\left(a,b|A_i,B_j\right)=\frac{1}{4}\Bigg[1+ab \ \langle{A_{i}B_{j}}\rangle_{\rho_{AB}}\Bigg]
\end{eqnarray}
Using  Eq. (\ref{selfn}),  we have  
\begin{eqnarray}\label{r52}
      \big(A_i\otimes B_i+A_{i+1}\otimes B_i\big)\ket{\psi}_{AB}&=&2\cos{\frac{\pi}{2n}}\ket{\psi}_{AB}
\end{eqnarray}
Note that, for $i=n$, $A_{n+1}=-A_1$. In Appendix.~\ref{randoma},  it is proven that for any value of $n$, the following relation holds  \ba\label{trcos} A_i\otimes B_i\ket{\psi}_{AB}=A_{i+1}\otimes B_i\ket{\psi}_{AB}\ea 
 Hence we can write \ba\Tr[\left(A_i \otimes B_j \right)\rho_{AB}]= \pm\cos{\frac{\pi}{2n}}, \ \forall j=i,i+1\ea 
Using Eq.~(\ref{rand0m}), we can write that  for each $i\in[n]$, 
\begin{eqnarray}
    \Tr[\left(A_i \otimes B_{i+x} \right)\rho_{AB}]
    &=&\frac{\cos{\frac{\pi x}{n}}+\cos{\frac{\pi (x+1)}{n}}}{2\cos{\frac{\pi}{2n}}}
\end{eqnarray}
where $i\in[n], x\in \mathbb{Z}$ and $x\in[-n,n]\setminus \{0,1\}, \forall n\in\mathbb{N},\forall|i+x|\leq n$. Since, we know  \ba \cos{\frac{\pi}{2n}}\geq\frac{\cos{\frac{\pi x}{n}}+\cos{\frac{\pi (x+1)}{n}}}{2\cos{\frac{\pi}{2n}}},\quad  \forall n\geq 3, 
\ea 
the maximum value of the probability is  \ba P_{max}\left(a,b|A_i,B_j\right)=\frac{1}{4}\Bigg[1+\cos{\frac{\pi}{2n}}\Bigg]\ea 
which implies that  the minimum certified  achievable randomness is given  by 
\begin{eqnarray}
    \mathcal{R}_{\text{min}} = \log_2 \left(\frac{4}{1 + \cos \frac{\pi}{2n}}\right)
\end{eqnarray}
Note that, for the increasing values of $n$, randomness $\mathcal{R}_{\text{min}}$ decreases, and for $n\to \infty$,  it  approaches $1$ bit of randomness.

We note again that Alice and Bob randomly receive $n$ input. However, not all such correlations $\langle A_i\otimes B_j\rangle$ are included in the chained Bell inequality, for example $
\langle A_i\otimes B_{i+\frac{n-1}{2}}\rangle$. It is a consequence of our quantum optimal value that confirms that $\langle A_i\otimes B_{i+\frac{n-1}{2}}\rangle_{\rho_{AB}}=0$. In Appendix.~\ref{randoma}, we have provided the derivation. Hence from Eq.~(\ref{probaibj}) the probability is 
\begin{eqnarray}
    P\left(a,b|A_i,B_{i+\frac{n-1}{2}}\right)=\frac{1}{4}, \quad \forall i\in [n]
\end{eqnarray}
where $n$ is odd, and hence the maximum randomness $\mathcal{R}_{\text{max}} =-\log_2\qty(\frac{1}{4})=2$. We note here that all the above works \cite{Acin2012,Wooltron2022,Wooltorton2024,Dhara2013,Supic2016,Xiao2023} generally derive $\mathcal{R}_{\text{max}}$ not the $\mathcal{R}_{\text{min}}$.

We analyze the robustness of generated randomness. In the presence of noise, the expression of randomness becomes

\begin{eqnarray}
    &&\Tilde{\mathcal{R}}_{\text{min}} = \log_2 \left(\frac{4\sqrt{1+\epsilon^2}}{\sqrt{1+\epsilon^2}+\epsilon+\cos{\frac{\pi}{2n}}}\right)\\
    &&\Tilde{\mathcal{R}}_{\text{max}} = \log_2 \left(\frac{4\sqrt{1+\epsilon^2}}{\sqrt{1+\epsilon^2}+\epsilon}\right)
\end{eqnarray}

where the noise parameter $\epsilon$ is  defined as $||(\Tilde{B}_i-B_i)\ket{\psi}_{AB}||\leq \epsilon, \forall i\in[n]$ and $\epsilon =\sqrt{\frac{\xi }{n \cos \left(\frac{\pi }{2 n}\right)}}$. The detailed derivations are provided in the Appendix.~\ref{Rob RC}.
%%%%%%%%%%%%%%%%%%%%%%%%%%%%%%%%%%%%%%%%%%%%%%%%%%%%%%%%%%%%%%
\section{Summary and Discussion}  
\label{SecIX}
In sum, we demonstrated the self-testing of state and measurements based on the optimal quantum violation of the chained Bell inequality \cite{Braunstein1990}, which features an arbitrary $n$ number of dichotomic measurements for each party.  We developed a systematic analytically elegant SOS approach to derive the optimal quantum violation, which are quite straightforward compared to earlier approach \cite{Supic2016}. We explicitly derived the functional relation between local observables  solely from the optimization conditions of the SOS approach. Finding the dimension-independent state required for optimal quantum violation was a complicated task which we have also efficiently derived. Our approach is so systematic that it can be used for deriving optimal value of many other arbitrary-input Bell functionals. 

Further, we introduced a self-testing approach based on a swap circuit, which certifies a black-box quantum system by constructing a local isometry to swap its properties with a known reference system, assuming only minimal dimension. To bridge the gap between theory and practice, we have performed a rigorous robustness analysis of our swap-circuit self-testing scheme. Our robustness analysis focuses on measurement implementation errors, a dominant source of experimental noise, by quantifying the deviation of the certified state from its ideal target. Interestingly, if only one party (say, Bob) implements imperfect observables, the self-testing becomes more robust with the increasing number of settings per party.  

As an application, we demonstrated the generation of DI randomness. We have shown that the least achievable randomness in this context is $\mathcal{R}_{\text{min}}$ goes to one for higher values of odd $n$. We also demonstrated that for suitable choices of the measurement settings of Alice and Bob, provide the maximum randomness $\mathcal{R}_{\text{max}}=2$ for odd values of $n$. To make it more practically relevant, we have provided the robustness of our generated randomness in the presence of noise.

This  DI self-testing protocol featuring an arbitrary number of measurement settings indeed certifies a large set of observables, leading towards a wide range of applications in quantum computation processes.       Another interesting aspect for future research is the formation of similar Bell inequalities featuring the measurement settings with more than two outcomes, for bipartite or multipartite scenarios,  and investigate whether such a protocol can provide useful DI robust self-testing and certification of genuine randomness. This calls for further studies.    Note that the  DI self-testing protocols provide a significant advancement in generating a secure key rate. Hence,  the security analysis of such complex protocols with an increasing number of measurement settings leads to an exciting avenue for future research. 

\section*{Acknowledgments}
RP acknowledges the financial support from the Council of Scientific and Industrial Research (CSIR, 09/1001(12429)/2021-EMR-I), Government of India. SM acknowledges the support from the research grant  I-HUB/PDF/2022-23/06, Government of India.  AKP acknowledges the support
from Research Grant No. SERB/CRG/2021/004258, Government of India.
\vspace{3cm}
\appendix
\begin{widetext}
\section{Relation between the Bob's observables}\label{Bobo}
From the main text, Eq.~(\ref{aiix}) depicts the relation between Alice's observables. In order to derive the relations between Bob's observables, we use the following approach. The optimization conditions (as in Eq.~(\ref{selfn}) in the main text) are given by 
\begin{eqnarray}\label{rum1}
    \mathcal{A}_{j}\otimes B_{j}\ket{\psi}_{AB}=\ket{\psi}_{AB}\quad \forall j\in[n]
\end{eqnarray}
where we have   $\mathcal{A}_{j}=\frac{A_{j}+A_{j+1}}{2\cos{\frac{\pi}{2n}}}$.
Hence,  we can rewrite Eq.~(\ref{rum1}) as follows. 
\begin{eqnarray}
&&\frac{A_{j}+A_{j+1}}{2\cos{\frac{\pi}{2n}}}\otimes B_j\ket{\psi}_{AB}=\ket{\psi}_{AB}\nonumber\\&&\openone_d\otimes B_{j}\ket{\psi}_{AB}=\frac{A_{j}+A_{j+1}}{2\cos{\frac{\pi}{2n}}}\otimes\openone_d\ket{\psi}_{AB}\label{rum2}\\
    &&\bra{\psi}_{AB}\openone_d\otimes B_{j}=\bra{\psi}_{AB}\frac{A_{j}+A_{j+1}}{2\cos{\frac{\pi}{2n}}}\otimes\openone_d\label{rum2t}
\end{eqnarray}
Since, the each of the observables $A_i$ and $B_j$ are hermitian, $\bra{\psi}_{AB} A_i\otimes B_j\ket{\psi}_{AB}$ is  real, for each $i,j\in[n]$. Hence, using Eq.~(\ref{rum2}) and (\ref{rum2t}), we get 
\begin{eqnarray}
    \bra{\psi}_{AB}A_i\otimes B_{j}\ket{\psi}_{AB}&=& \begin{cases}
        \bra{\psi}_{AB}A_i\frac{A_{j}+A_{j+1}}{2\cos{\frac{\pi}{2n}}}\otimes \openone_d\ket{\psi}_{AB}\nonumber\\
\\
\bra{\psi}_{AB}\frac{A_{j}+A_{j+1}}{2\cos{\frac{\pi}{2n}}} A_i\otimes \openone_d\ket{\psi}_{AB}
    \end{cases}\nonumber\\
    &=&\frac{1}{2}\bra{\psi}_{AB}\qty(\frac{\{A_i,A_{j}\}+\{A_i,A_{j+1}\}}{2\cos{\frac{\pi}{2n}}})\otimes\openone_d\ket{\psi}_{AB}\label{rand0'}\\
    &=&\frac{2\cos{\frac{\pi x}{n}}+2\cos{\frac{\pi(x+1)}{n}}}{4\cos{\frac{\pi}{2n}}}\quad \qty(\text{For $j=i+x$, $\langle\{A_i,A_{i+x}\}\rangle=2\cos\frac{\pi \ x}{n}$})\nonumber\\
    &=& \frac{\cos{\frac{\pi x}{n}}+\cos{\frac{\pi(x+1)}{n}}}{2\cos{\frac{\pi}{2n}}}\label{rand0}
\end{eqnarray}
Now, substituting $j=i-1$ and $j=i$ in the Eq.~(\ref{rand0}) respectively such that  $x=-1$ and $0$, we get 
\begin{eqnarray}
    \bra{\psi}_{AB}A_i\otimes B_{i-1}\ket{\psi}_{AB}    &=&\bra{\psi}_{AB}A_i\otimes B_{i}\ket{\psi}_{AB}  =\frac{1+\cos{\frac{\pi}{n}}}{2\cos{\frac{\pi}{2n}}}
\end{eqnarray}
This implies that 
\begin{eqnarray}
    A_i\otimes \frac{B_i+B_{i-1}}{2\cos{\frac{\pi}{2n}}}\ket{\psi}_{AB}=\ket{\psi}_{AB}\label{aibibi-1}
\end{eqnarray}

From Eq.~(\ref{aibibi-1}), it is straightforward to see that  $\frac{B_i+B_{i-1}}{2\cos{\frac{\pi}{2n}}}$ is normalized, and   hence, the suitable norms $\nu'_{n,i}$s are defined as $ \nu'_{n,i}=||{B}_{i}+{B}_{i-1}||_{\rho_{AB}}=\sqrt{2+\langle\{B_{i},B_{i-1}\}\rangle}=2\cos{\frac{\pi}{2n}}$ which  implies that \ba \langle\{B_{i},B_{i-1}\}\rangle=2\cos{\frac{\pi}{n}}\label{bi1}\ea 
Substituting  $i=i-1$ in Eq.~(\ref{bi1}), we get, 
\begin{eqnarray}
    &&\langle\{B_{i},B_{i+1}\}\rangle=2\cos{\frac{\pi}{n}}\label{bi2} 
\end{eqnarray}
Adding  Eq.~(\ref{bi1}) and (\ref{bi2}) 
and simplifying, we get
\begin{eqnarray}\label{biii1}
    &&\Bigg\langle\qty{B_{i},\frac{B_{i-1}+B_{i+1}}{2\cos{\frac{\pi}{n}}}}\Bigg\rangle=2\label{bi3} 
\end{eqnarray}
As each observable $B_i$ is dichotomic and normalized, Eq. (\ref{biii1})  holds only  if  \ba \label{bi} B_i=\frac{B_{i-1}+B_{i+1}}{2\cos{\frac{\pi}{n}}}\label{bi4}\ea
which  implies that  $\langle\{B_{i-1},B_{i+1}\}\rangle=2\cos{\frac{2\pi}{n}}$. Now substituting $i=i+1$ in Eq. (\ref{biii1}), we  get
$\langle\{B_{i},B_{i+2}\}\rangle=2\cos{\frac{2\pi}{n}}$. Again for  $i=i+2$ in Eq.~(\ref{bi4}), we get $B_{i+2}=\frac{B_{i+1}+B_{i+3}}{2\cos{\frac{\pi}{n}}}$. Hence, we get \ba \langle\{B_i,B_{i+3}\}\rangle=2 \cos{\frac{\pi}{n}}\langle\{B_i,B_{i+2}\}\rangle-\langle\{B_i,B_{i+1}\}\rangle=2 \cos{\frac{3\pi}{n}}\ea  Following the similar way, we can construct the general relation between Bob's observables as follows.
\begin{eqnarray}\label{bibix}
    \langle\{B_i,B_{i+x}\}\rangle&=&2\cos\frac{\pi \ x}{n}\quad \text{$\forall i\in[n],\  x\in[n-i]$}
\end{eqnarray}
Again, Eq.~(\ref{aibibi-1}) can be written as
    \begin{eqnarray}\label{Bobsc}
        A_i\otimes \mathcal{B}_i\ket{\psi}_{AB}&=&\ket{\psi}_{AB}, \  \forall i\in[n].
    \end{eqnarray}
where $\mathcal{B}_i=\frac{B_i+B_{i-1}}{\nu'_{n,i}}$ and $\nu'_{n,i}=||{B}_{i}+{B}_{i-1}||_{\rho_{AB}}$. Hence, using Eq. (\ref{bibix}), it is straightforward to show that  \ba \langle\{\mathcal{B}_i,\mathcal{B}_{i+x}\}\rangle=2\cos\frac{\pi \ x}{n}, \ \forall i\in[n],\  x\in[n-i].\ea 
\section{Explicit derivation of the optimal quantum violation for \texorpdfstring{$n=3,5,7$ and $11$}{n=3 and n=5}}\label{SOS 3,5,7,11}
 Here, we provide an explicit derivation of the optimal quantum violation $ (\mathscr{C}_n)^{opt}_Q$ for $n=3,5,7$ and $11$ along with the state and observables, using our dimension-independent SOS approach.
%%%%%%%%%%%%%%%%%%%%%%%%%%%%%%%%%%%%%%%%%%%%%%%%%%%%%%%%%%%%%%%%%%%%%%%%%%%%%%%%%%%%%%%%%%%%%%%%%%%%%%%%%%%%%%%%%%%%%%%%%%%%%%%%%%%%%%%%%%
\subsection{Detailed derivation for \texorpdfstring{$n=3$}{n=3}}\label{n=3}
Substituting  $n=3$ in Eq. (\ref{BmAB}) of the main text, we get the  chained Bell inequality
\ba 
\mathscr{C}_3=(A_1+A_2)B_1+(A_2+A_3)B_2+(A_3-A_1)B_3\leq 4\ea
Following the  SOS approach as outlined in Sec.~\ref{SecII}, we get that  the optimal value of $(\mathscr{ C}_{3})_Q$ is obtained if   $\Tr[ \Gamma_3\ \rho_{AB}]=0$ and  thus we get  
\begin{eqnarray}
   (\mathscr{ C}_{3})^{opt}_Q=\max \limits_{\{\{A_i\}, \rho_{AB}\}}[\nu_{3,1}+\nu_{3,2}+\nu_{3,3}] 
\end{eqnarray}
where $\nu_{3,i}=||A_i+A_{i+1}||_{\rho_{AB}}, \  \forall i\in[3]$. Using the inequality in Eq. (\ref{cnv}), we get 
 \ba 
 (\mathscr{ C}_{3})_Q&\leq&\sqrt{3\big[(\nu_{3,1})^2+(\nu_{3,2})^2+(\nu_{3,3})^2\big]}
 \ea
The equality  holds only if  $\nu_{3,1}=\nu_{3,2}=\nu_{3,3}$, which implies $\langle \{A_i,A_{i+1}\}\rangle=\langle \{A_j,A_{j+1}\}\rangle , \forall i\neq j\in[3]$ and $A_{4}=-A_1$. We can write
 \begin{eqnarray}
     (\mathscr{ C}_{3})_Q&\leq&\sqrt{3\big[6+\langle \{A_2,(A_1+A_3)\}\rangle-\langle\{A_1,A_3\}\rangle\big]}
 \end{eqnarray}
Considering $A_2=(A_1+A_3)/\nu^1_3$  where $v^1_3=\sqrt{2+\langle\{A_1,A_3\}\rangle}$, we can write \ba
\label{opt3} (\mathscr{
C}_{3})_Q&\leq&\sqrt{3\big[6+2\sqrt{2+\langle\{A_1,A_3\}\rangle}-\langle\{A_1,A_3)\}\rangle\big]}
\ea
The right-hand side of Eq. (\ref{opt3}) is  maximized when  $\langle\{A_1,A_3\}\rangle=-1$ which implies that $\nu^1_3=\sqrt{2+\langle\{A_1,A_3\}\rangle}=1$ and $A_2=A_1+A_3$. Hence, we get  $\langle\{A_1,A_2\}\rangle=\langle\{A_2,A_3\}\rangle=1$,  and consequently  $\nu_{3,1}=\nu_{3,2}=\nu_{3,3}=\sqrt{3}$.
 This, in turn, provides  
 \begin{eqnarray}
    (\mathscr{ C}_{3})^{opt}_Q&=&3\sqrt{3}=6\cos\frac{\pi }{3}
\end{eqnarray}
Clearly, Alice's  observables follow the relation 
     \begin{eqnarray} 
     \langle\{A_i,A_{i+x}\}\rangle&=&2\cos\frac{\pi x}{3},\quad \forall i\in[2], \ x\in[3-i]
     \end{eqnarray}
     Following  Eq. (\ref{rho odd}) (more details are given in Appendix \ref{cseodd}) the main text, we can write the required state $\rho_{AB}$ with  
\begin{eqnarray}
 C_1 \otimes C_1 &=& \mathcal{A}_2 \otimes B_2,\quad
C_2 \otimes C_2=\frac{(\mathcal{A}_1-\mathcal{A}_3) \otimes (B_1-B_3)}{(2-\langle\{B_1,B_3\}\rangle)}\end{eqnarray} and 
$C_3 \otimes C_3 =(C_1 \otimes C_1)\cdot (C_2 \otimes C_2)$
where    $\mathcal{A}_1 = \frac{A_1+A_2}{\sqrt{3}}, \mathcal{A}_2 = \frac{A_2+A_3}{\sqrt{3}}, \mathcal{A}_3 = \frac{A_3-A_1}{\sqrt{3}}$.\\

In a two-qubit system, a  set of choices of observables is given by  \ba A_1=\sigma_z,\  A_2=\frac{\sqrt{3}\sigma_x+\sigma_z}{2},\  A_3=\frac{\sqrt{3}\sigma_x-\sigma_z}{2},\ 
B_1=\frac{\sigma_x+\sqrt{3}\sigma_z}{2},\   B_2=\X, \ B_3=\frac{\sigma_x-\sqrt{3}\sigma_z}{2}\ea 
And consequently, we get 
\begin{eqnarray}\label{cop3}
    C_1 \otimes C_1 = \mathcal{A}_2 \otimes B_2=\sigma_x\otimes \sigma_x,\quad C_2 \otimes C_2=\frac{(\mathcal{A}_1-\mathcal{A}_3) \otimes (B_1-B_3)}{(2-\langle\{B_1,B_3\}\rangle)}=\sigma_z\otimes \sigma_z,\  C_3 \otimes C_3 =(C_1 \otimes C_1)\cdot (C_2 \otimes C_2)=-\sigma_y\otimes \sigma_y\nonumber\\
\end{eqnarray}
Hence, using Eq.~(\ref{cop3}),  from  Eq.~(\ref{rho odd}) of the main text, we get the two-qubit maximally entangled state \ba \label{rhophi+}\rho_{\phi^+}=\frac{1}{4}\qty(\openone_2\otimes\openone_2+\sigma_x\otimes \sigma_x+\sigma_z\otimes \sigma_z-\sigma_y\otimes \sigma_y)\ea that produces the optimal quantum violation $(\mathscr{ C}_{3})^{opt}_Q=6\cos\frac{\pi }{6}$. 
    
\subsection{Detailed derivation for \texorpdfstring{$n=5$}{n=5}}
Substituting $n=5$ in  Eq. (\ref{BmAB}) of the main text, we get the  chained Bell inequality
\begin{eqnarray}\label{Chain5}
\mathscr{C}_5&=&(A_1+A_2)B_1+(A_2+A_3)B_2+(A_3+A_4)B_3+(A_4+A_5)B_4+(A_5-A_1)B_5\leq 8
\end{eqnarray}
Following the similar SOS approach, we get that  the optimal value of $(\mathscr{C}_5)_Q$ is obtained when  $\Tr[ \Gamma_5\ \rho_{AB}]=0$, implying that 
\begin{eqnarray}(\mathscr{C}_{5})^{opt}_{Q}&=& \max \limits_{\{\{A_i\}, \rho_{AB}\}}\left[\nu_{5,1}+\nu_{5,2}+\nu_{5,3}+\nu_{5,4}+\nu_{5,5}\right]
\end{eqnarray}
where $\nu_{5,i}=||A_i+A_{i+1}||_{\rho_{AB}} \forall i\in[5]$. Using the inequality in Eq. (\ref{cnv}) of the main text, we get
\ba
(\mathscr{ C}_{5})_Q&\leq&\sqrt{5\sum\limits_{i=1}^5\left[\nu_{5,i}\right]^2}
 \ea
 The equality  holds only  when $\nu_{5,i}=\nu_{5,j}, \forall i\neq  j\in[5]$, which implies $\langle \{A_i,A_{i+1}\}\rangle=\langle \{A_j,A_{j+1}\}\rangle,  \forall i\neq j\in[5]$ and $A_{6}=-A_1$. Further, we can write
 \ba
(\mathscr{ C}_{5})_Q
 \leq\sqrt{5\big(10+ \delta_5 \big)}
 \ea
  where \ba\delta_5 =\langle\{A_2,(A_1+A_3)\}\rangle+\langle \{A_4,(A_3+A_5)\}\rangle-\langle\{A_1,A_5\}\rangle \ea
By considering $A_2=(A_1+A_3)/\nu^1_5, A_4=(A_3+A_5)/\nu^2_5$ , where $v^1_5=\sqrt{2+\langle\{A_1,A_3\}\rangle}$ and $v^2_5=\sqrt{2+\langle\{A_3,A_5\}\rangle}$,  we get
 \ba\na \delta_5&=&2\sqrt{2+\langle\{A_1,A_3\}\rangle}+2\sqrt{2+\langle\{A_3,A_5\}\rangle}-\langle\{A_1,A_5\}\rangle
\\&\leq&2\sqrt{2\bigg(4+\langle\{A_1,A_3\}\rangle+\langle\{A_3,A_5\}\rangle\bigg)}-\langle\{A_1,A_5\}\rangle\ea
Again the equality holds only when $v^1_5=v^2_5$ which needs $\langle\{A_1,A_3\}\rangle=\langle\{A_3, A_5\}\rangle$.
Considering  $A_3=(A_1+A_5)/\nu^3_5$ with $\nu^3_5=\sqrt{2+\langle\{A_1,A_5\}\rangle}$, we get   \ba\delta_5&\leq& 2\sqrt{2\bigg(4+2\sqrt{2+\langle\{A_1,A_5\}\rangle}\bigg)}-\langle\{A_1,A_5\}\rangle \ea
 We find that $\delta_5$ will be maximized when $\langle\{A_1,A_5\}\rangle=-(\sqrt{5}+1)/2=2\cos\frac{4\pi}{5}$. This  implies \ba \nu^3_5=\sqrt{2+\langle\{A_1,A_5\}\rangle}=\sqrt{(3-\sqrt{5})/2}\ea 
 Hence we get $A_3=\frac{A_1+A_5}{\sqrt{(3-\sqrt{5})/2}}$, which implies that   that $\langle\{A_1,A_3\}\rangle=\langle\{A_3,A_5\}\rangle=2\cos\frac{2\pi}{5}$.
 Following a similar way,  we derive $\nu^1_5=\nu^2_5=\sqrt{(7-\sqrt{5})/2}$, which leads to  \ba A_2=\frac{(A_1+A_3)}{\sqrt{(7-\sqrt{5})/2}}, \ A_4=\frac{(A_3+A_5)}{\sqrt{(7-\sqrt{5})/2}}\ea 
 It is straight forward to check that $\langle\{A_1,A_2\}\rangle=\langle\{A_2,A_3\}\rangle=\langle\{A_3,A_4\}\rangle=\langle\{A_4,A_5\}\rangle= 2\cos\frac{\pi}{5}$. Similarly,  using $ A_4=\frac{(A_3+A_5)}{\sqrt{(7-\sqrt{5})/2}}$ and $A_3=\frac{A_1+A_5}{\sqrt{(3-\sqrt{5})/2}}$ we get $\langle\{A_1,A_4\}\rangle= 2\cos\frac{3\pi}{5}, \langle\{A_2,A_4\}\rangle= 2\cos\frac{2\pi}{5}$ and $\langle\{A_2,A_3\}\rangle= 2\cos\frac{\pi}{5}$. Hence, Alice's observables follow the relation
 %$\{A_1,A_3\}=\{A_3,A_5\}=\{A_2,A_4\}=-\{A_1,A_4\}=-\{A_2,A_5\}=(\sqrt{5}-1)/2, \quad \{A_1,A_2\}=\{A_2,A_3\}=\{A_3,A_4\}=\{A_4,A_5\}=(\sqrt{5}+1)/2, \{A_1,A_4\}=\{A_2,A_5\}=-(\sqrt{5}-1)/2,$  %
 \begin{eqnarray}
     \langle\{A_i,A_{i+x}\}\rangle&=&2\cos\frac{\pi \ x}{5}, \quad \text{$\forall i\in[4], x\in[5-i]$}
 \end{eqnarray}
 Consequently,  we get \ba \nu_{5,i}=\sqrt{ (5 + \sqrt{5})/2}=2\cos \frac{\pi}{10},\ \forall i\in[5]\ea  
 Finally, we get\ba(\mathscr{C}_{5})^{opt}_{Q}=5\sqrt{ (5 + \sqrt{5})/2}=10\cos\frac{\pi}{10}\ea
 Following Eq. (\ref{rho odd})  (more details are given in Appendix \ref{cseodd})  of the main text, we can write the required state $\rho_{AB}$ by substituting  $n=5$ in Eq.~(\ref{C'so}). Thus we get  
\ba C_1 \otimes C_1 &=& \mathcal{A}_3 \otimes B_3,
C_2 \otimes C_2=\dfrac{1}{2}\Bigg[\frac{(\mathcal{A}_1-\mathcal{A}_5) \otimes (B_1-B_5)}{(2-\langle\{B_1,B_5\}\rangle)}+\frac{(\mathcal{A}_2-\mathcal{A}_4) \otimes (B_2-B_4)}{(2-\langle\{B_2,B_4\}\rangle)}\Bigg],
C_3 \otimes C_3 =(C_1 \otimes C_1)\cdot (C_2 \otimes C_2)\quad
\ea
where, $\mathcal{A}_1=\frac{A_1+A_2}{\sqrt{ (5 + \sqrt{5})/2}},  \mathcal{A}_2=\frac{A_2+A_3}{\sqrt{ (5 + \sqrt{5})/2}}, \mathcal{A}_3=\frac{A_3+A_2}{\sqrt{ (5 + \sqrt{5})/2}},\mathcal{A}_4=\frac{A_4+A_5}{\sqrt{ (5 + \sqrt{5})/2}}, \mathcal{A}_5=\frac{A_5-A_1}{\sqrt{ (5 + \sqrt{5})/2}} $.\\
As an example, in a two-qubit system, we provide a set of observables are the following. 
\ba \na 
 A_1&=&\Z, \ A_2=\frac{\sqrt{2(5-\sqrt{5})}\X+(\sqrt{5}+1)\Z}{4}, \  A_3=\frac{\sqrt{2(5+\sqrt{5})}\X+(\sqrt{5}-1)\Z}{4},\  A_4=\frac{\sqrt{2(5+\sqrt{5})}\X+(1-\sqrt{5})\Z}{4},\\ \na 
 A_5&=&\frac{\sqrt{2(5-\sqrt{5})}\X-(\sqrt{5}+1)\Z}{4}, \  B_1=\frac{(\sqrt{5}-1)\X+\sqrt{2(5+\sqrt{5})}\Z}{4}, \ B_2=\frac{(\sqrt{5}+1)\X+\sqrt{2(5-\sqrt{5})}\Z}{4},\ B_3=\X,\\ B_4&=&\frac{(\sqrt{5}+1)\X-\sqrt{2(5-\sqrt{5})}\Z}{4},\  B_5=\frac{(\sqrt{5}-1)\X-\sqrt{2(5+\sqrt{5})}\Z}{4}, \ea  which again provides
\begin{eqnarray}
    &&C_1 \otimes C_1 = \mathcal{A}_3 \otimes B_3=\sigma_x\otimes \sigma_x,\quad C_2 \otimes C_2=\dfrac{1}{2}\Bigg[\frac{(\mathcal{A}_1-\mathcal{A}_5) \otimes (B_1-B_5)}{(2-\langle\{B_1,B_5\}\rangle)}+\frac{(\mathcal{A}_2-\mathcal{A}_4) \otimes (B_2-B_4)}{(2-\langle\{B_2,B_4\}\rangle)}\Bigg]=\sigma_z\otimes \sigma_z\\
    &&C_3 \otimes C_3 =(C_1 \otimes C_1)\cdot (C_2 \otimes C_2)=-\sigma_y\otimes \sigma_y\nonumber
\end{eqnarray}
This again leads to the entangled state  $\rho_{\phi^+}$ as in Eq. (\ref{rhophi+}), providing $(\mathscr{ C}_{5})^{opt}_Q=10\cos\frac{\pi }{10}$.
 \subsection{Detailed derivation for  \texorpdfstring{$n=7$}{n=7}}\label{SOS7}
  Considering  the  chained Bell inequality  in Eq. (\ref{BmAB}) of the main text for $n=7$, we get  
\ba
\mathscr{
C}_{7}&=&\sum\limits_{i=1}^7(A_i+A_{i+1})B_i\leq 12\hspace{0.6cm}
 \ea
where $A_8=-A_1$. Following the SOS approach as presented in sec.~\ref{SecII}, we get 	\begin{eqnarray}
    (\mathscr{C}_{7})_Q&=& \max\sum\limits_{i=1}\nu_{7,i}\hspace{1mm}
	\end{eqnarray}
where $\nu_{7,i}=||(A_i+A_{i+1})||_{\rho_{AB}}=\sqrt{2+\langle\{A_{i},A_{i+1}\}\rangle},  \ \forall i\in [7]$.
 Using Eq. (\ref{cnv}) of  the main text, we get 
\ba
(\mathscr{C}_{7})_Q&\leq&\sqrt{7\sum\limits_{i=1}^7\left[\nu_{7,i}\right]^2}
 \ea
The  equality holds only when $\nu_{7,i}=\nu_{7,j}, \forall i\neq j\in[7]$, which implies $\langle \{A_i,A_{i+1}\}\rangle=\langle \{A_j,A_{j+1}\}\rangle, \forall i\neq j\in[7]$ and $A_{8}=-A_1$. Hence
 \ba
(\mathscr{C}_{7})_Q
 =\sqrt{7\bigg(14+ \delta_7 \bigg)}
 \ea
where 
  \begin{eqnarray}
      \delta_7 &=&\langle\{A_2,(A_1+A_3)\}\rangle+\langle \{A_4,(A_3+A_5)\}\rangle+\langle \{A_7,(A_6-A_1)\}\rangle+\langle  \{A_5,A_6\}\rangle
  \end{eqnarray}
 Considering $A_2=(A_1+A_3)/\nu^1_7,\  A_4=(A_3+A_5)/\nu^2_7$, and $A_7=(A_6-A_1)/\nu^3_7,$ we get
\begin{eqnarray}
 \delta_7&=&2\Big(\sqrt{2+\langle\{A_1,A_3\}\rangle}+\sqrt{2+\langle\{A_3,A_5\}\rangle}+\sqrt{2-\langle\{A_1,A_6\}\rangle}\Big)\nonumber+ \langle \{A_5,A_6\}\rangle\nonumber\\
&\leq&2\sqrt{3\bigg(6+\langle\{A_1,A_3\}\rangle+\langle\{A_3,A_5\}\rangle-\langle\{A_1,A_6\}\rangle}\bigg)+ \langle \{A_5,A_6\}\rangle
 \end{eqnarray}
Using Eq.~(\ref{cnv}) of the main text, we can depict that the equality sign holds only when $\nu^1_7=\nu^2_7=\nu^3_7$, implies $\langle\{A_1, A_3\}\rangle=\langle\{A_3, A_5\}\rangle=-\langle\{A_1, A_6\}\rangle$. Hence, we can derive the following
 \begin{eqnarray}
\delta_7
&\leq&2\sqrt{3\bigg(6+\langle\{(A_1+A_5),A_3\}\rangle-\langle\{A_1,A_6\}\rangle\bigg)}+ \langle \{A_5,A_6\}\rangle
 \end{eqnarray}
Considering   $A_3=(A_1+A_5)/\nu^4_7$, we further set  \ba\delta_7&\leq& 2\sqrt{3\bigg(6+2\sqrt{2+\langle\{A_1,A_5\}\rangle}-\langle\{A_1,A_6\}\rangle\bigg)}+ \langle \{A_5,A_6\}\rangle\ea
Without loss of generality, let us consider $A_1=\frac{a\ A_5-A_6}{\sqrt{1+a^2-a\eac{A_5}{A_6}}}$, where $a\in[0,1]$ which implies that
\begin{eqnarray}
    \delta_7&\leq& 2\Bigg[3\bigg(6+2\sqrt{2+\frac{2a-\eac{A_5}{A_6}}{\sqrt{1+a^2-a\eac{A_5}{A_6}}}}-\frac{a\eac{A_5}{A_6}-2}{\sqrt{1+a^2-a\eac{A_5}{A_6}}}\bigg)\Bigg]^{\frac{1}{2}}+ \langle \{A_5,A_6\}\rangle
\end{eqnarray}
 Following a few  simple steps, we get that $\delta_7$  is  maximized for  $\langle\{A_5,A_6\}\rangle=2\cos{\frac{\pi}{7}}$ and $a=\cos{\frac{10\pi}{49}}$ which leads to $(\delta_7)_{\max}=14\cos{\frac{\pi}{7}}$. Hence, we can write the following

 \begin{eqnarray}
     (\mathscr{C}_{7})^{opt}_Q=\sqrt{7(14+\delta_7)}=14\cos{\frac{\pi}{14}}
 \end{eqnarray}
This also implies that $A_1=\frac{\cos{\frac{10\pi}{49}}\ A_5-A_6}{\sqrt{1+\cos^2{\frac{10\pi}{49}}-2\cos{\frac{10\pi}{49}}\cos{\frac{\pi}{7}}}}$ which further leads to  $\langle\{A_1,A_5\}\rangle=2\cos{\frac{4\pi}{7}}$ and $\nu^4_7=\sqrt{2+\langle\{A_1,A_5\}\rangle}=2\cos{\frac{2\pi}{7}}$. Following a similar  way, we find that  Alice's observables follow the relation
\begin{eqnarray}
    \expval{\{A_i,A_{i+x}\}}&=&2\cos\frac{\pi \ x}{7}, \quad \text{$\forall i\in[6], \ x\in[7-i]$}
\end{eqnarray}
Hence $\nu^3_7=\sqrt{2-\langle\{A_1,A_6\}\rangle}=\sqrt{2-2\cos{\frac{5\pi}{7}}}$, $\nu^2_7=\sqrt{2+\langle\{A_3,A_5\}\rangle}=\sqrt{2+2\cos{\frac{2\pi}{7}}}$, $\nu^1_7=\sqrt{2+\langle\{A_1,A_3\}\rangle}=\sqrt{2+2\cos{\frac{2\pi}{7}}}$ and $\nu_{7,i}=\sqrt{2+\langle\{A_i,A_{i+1}\}\rangle}=\sqrt{2+2\cos{\frac{\pi}{7}}}=2\cos{\frac{\pi}{14}}$.
We can show that Bob's observables also hold similar relations due to the symmetric property of the inequality.
\begin{eqnarray}\label{b7} \langle\{B_i,B_{i+x}\}\rangle&=&2\cos\frac{\pi \ x}{7}\quad \text{$\forall i\in[6], x\in[7-i]$}\end{eqnarray}
Following  Eq. (\ref{rho odd}) of the main text, we can write the required state $\rho_{AB}$ by substituting  $n=7$ in Eq.~(\ref{C'so}). Thus we get
\ba C_1 \otimes C_1 &=& \mathcal{A}_4 \otimes B_4, \quad C_2 \otimes C_2=\frac{1}{3}\Bigg[\frac{(\mathcal{A}_1-\mathcal{A}_7) \otimes (B_1-B_7)}{2-\expval{\{B_1,B_7\}}}+\frac{(\mathcal{A}_3 -\mathcal{A}_5)\otimes (B_3-B_5)}{2-\expval{\{B_3,B_5\}}}+\frac{(\mathcal{A}_2 -\mathcal{A}_6)\otimes (B_2-B_6)}{2-\expval{\{B_2,B_6\}}}\Bigg]\ea and $C_3 \otimes C_3 =(C_1 \otimes C_1)\cdot (C_2 \otimes C_2)$
where $\mathcal{A}_i=\frac{A_i+A_{i+1}}{\nu_{7,i}}$,  and $\nu_{7,i}=2\cos{\frac{\pi}{14}}$. 

For instance, one can consider a two-qubit system where a  set of observable choices are given by  \ba A_i=\sin\theta_i\X+\cos\theta_i\Z, \  B_i=\sin\phi_i\X+\cos\phi_i\Z , \ \theta_i = \frac{\qty(i-1)\pi}{7}, \ \phi_i=\frac{\qty(2i-1)\pi}{14}\ea  which further provides 
\begin{eqnarray}
    &&C_1 \otimes C_1 = \mathcal{A}_4 \otimes B_4=\sigma_x\otimes \sigma_x\nonumber\\
    &&C_2 \otimes C_2=\frac{1}{3}\Bigg[\frac{(\mathcal{A}_1-\mathcal{A}_7) \otimes (B_1-B_7)}{2-\expval{\{B_1,B_7\}}}+\frac{(\mathcal{A}_3 -\mathcal{A}_5)\otimes (B_3-B_5)}{2-\expval{\{B_3,B_5\}}}+\frac{(\mathcal{A}_2 -\mathcal{A}_6)\otimes (B_2-B_6)}{2-\expval{\{B_2,B_6\}}}\Bigg]=\sigma_z\otimes \sigma_z\\
    &&C_3 \otimes C_3 =(C_1 \otimes C_1)\cdot (C_2 \otimes C_2)=-\sigma_y\otimes \sigma_y\nonumber
\end{eqnarray}This, in turn, provides the  entangled state  $\rho_{\phi^+}$ as in Eq. (\ref{rhophi+}) that produces $(\mathscr{C}_{7})^{opt}_Q=14\cos{\frac{\pi}{14}}$. 
%%%%%%%%%%%%%%%%%%%%%%%%%%%%%%%%%%%%%%%%%%%%%%%%%%%%%%%%%%%%%%%%%%%%%%%%%%%%%%%%%%%%%%%%%%%%%%%%%%
\subsection{Detailed derivation for  \texorpdfstring{$n=11$}{n=11}}
  Substituting $n=11$ in  Eq. (\ref{BmAB}) of the main text, we get the  chained Bell inequality  
\ba
\mathscr{
C}_{11}&=&\sum\limits_{i=1}^{11}(A_i+A_{i+1})B_i\leq 20\hspace{0.6cm}
 \ea
where $A_{12}=-A_1$. Following a similar SOS approach, we get 	
\begin{eqnarray}
    (\mathscr{C}_{11})_Q&=& \max\sum\limits_{i=1}\nu_{11,i}\hspace{1mm}
	\end{eqnarray}
where $||\nu_{11,i}||=(A_i+A_{i+1})||_{\rho_{AB}}=\sqrt{2+\langle\{A_{i},A_{i+1}\}\rangle},  \ \forall i\in[11]$.
 Using Eq. (\ref{cnv}) of  the main text, we get 
 \ba
(\mathscr{C}_{11})_Q&\leq&\sqrt{11\sum\limits_{i=1}^{11}\left[\nu_{11,i}\right]^2}
 \ea
The equality holds only if  $\nu_{7,i}=\nu_{7,j}, \forall i\neq j\in[7]$. Hence we get 
   \ba
(\mathscr{C}_{11})_Q
 \leq\sqrt{11\bigg[22+ \delta_{11} \bigg]}
 \ea
  where 
  \begin{eqnarray}
      \delta_{11} &=&\langle\{A_2,(A_1+A_3)\}\rangle+\langle \{A_4,(A_3+A_5)\}\rangle+\langle \{A_6,(A_5+A_7)\}\rangle+\langle \{A_8,(A_7+A_9)\}\rangle+\langle \{A_{11},(A_{10}-A_1)\}\rangle+\langle \{A_9,A_{10}\}\rangle
  \end{eqnarray}
\\
 Considering $A_2=(A_1+A_3)/\nu^1_{11}, A_4=(A_3+A_5)/\nu^2_{11}$, $A_6=(A_5+A_7)/\nu^3_{11}, A_8=(A_7+A_9)/\nu^4_{11}$ and $A_{11}=(A_{10}-A_1)/\nu^5_{11}$, we get
 
\begin{eqnarray}
 \delta_{11}&=&2\Big(\sqrt{2+\langle\{A_1,A_3\}\rangle}+\sqrt{2+\langle\{A_3,A_5\}\rangle}+\sqrt{2+\langle\{A_5,A_7\}\rangle}+\sqrt{2+\langle\{A_7,A_9\}\rangle}+\sqrt{2-\langle\{A_1,A_{10}\}\rangle}\Big)+\langle \{A_9,A_{10}\}\rangle\nonumber\\
 &\leq&2\Bigg[5\bigg(10+\langle\{A_1,A_3\}\rangle+\langle\{A_3,A_5\}\rangle+\langle\{A_5,A_7\}\rangle+\langle\{A_7,A_9\}\rangle-\langle\{A_1,A_{10}\}\rangle\bigg)\Bigg]^{\frac{1}{2}}+ \langle \{A_9,A_{10}\}\rangle
\end{eqnarray}
Again using Eq.~(\ref{cnv}) of the main text, we can say that the equality sign holds only if  $\nu^1_{11}=\nu^2_{11}=\nu^3_{11}=\nu^4_{11}=\nu^5_{11}$, implying that  $\langle\{A_1, A_3\}\rangle=\langle\{A_3, A_5\}\rangle=\langle\{A_5, A_7\}\rangle=\langle\{A_7, A_9\}\rangle=-\langle\{A_1, A_{10}\}\rangle$. Hence, we  derive the following.
\begin{eqnarray}
\delta_{11}
&\leq&2\Bigg[5\bigg(10+\langle\{A_1+A_5,A_3\}\rangle+\langle\{A_5+A_9,A_7\}\rangle-\langle\{A_1,A_{10}\}\rangle\bigg)\Bigg]^{\frac{1}{2}}+ \langle \{A_9,A_{10}\}\rangle
 \end{eqnarray}
 Now, let us consider $A_3=(A_1+A_5)/\nu^6_{11}$ and $A_7=(A_5+A_9)/\nu^7_{11}$ which implies
\begin{eqnarray}
    \delta_{11}&\leq&2\Bigg[5\bigg(10+2\sqrt{2+\eac{A_1}{A_5}}+2\sqrt{2+\eac{A_5}{A_9}}-\langle\{A_1,A_{10}\}\rangle\bigg)\Bigg]^{\frac{1}{2}}+ \langle \{A_9,A_{10}\}\rangle\nonumber\\
    &\leq&2\Bigg[5\bigg(10+2\sqrt{2(4+\eac{A_1+A_9}{A_5})}-\langle\{A_1,A_{10}\}\rangle\bigg)\Bigg]^{\frac{1}{2}} + \langle \{A_9,A_{10}\}\rangle
\end{eqnarray}
Again from Eq.~(\ref{cnv}) of the  main text, equality holds only when $\nu^6_{11}=\nu^7_{11}$, implying that  $\eac{A_1}{A_5}=\eac{A_5}{A_9}$. Hence we get 
\begin{eqnarray}
    \delta_{11}
    &\leq &2\Bigg[5\bigg(10+2\sqrt{2(4+\eac{A_1+A_9}{A_5})}-\langle\{A_1,A_{10}\}\rangle\bigg)\Bigg]^{\frac{1}{2}} + \langle \{A_9,A_{10}\}\rangle
\end{eqnarray}
Considering $A_5=(A_1+A_9)/\nu^8_{11}$, we get
\begin{eqnarray}
    \delta_{11}&\leq&2\Bigg[5\bigg(10+2\sqrt{2(4+2\sqrt{2+\eac{A_1}{A_9}})}-\langle\{A_1,A_{10}\}\rangle\bigg)\Bigg]^{\frac{1}{2}}+\langle \{A_9,A_{10}\}\rangle
\end{eqnarray}
Without loss of generality, let us  consider $A_1=\frac{a\ A_9-A_{10}}{\sqrt{1+a^2-a\eac{A_9}{A_{10}}}}$, where $a\in[0,1]$ which implies that
\begin{eqnarray}
    \delta_{11}&\leq&2\Bigg[5\bigg(10+2\sqrt{2\bigg(4+2\sqrt{2+\frac{2a-\eac{A_9}{A_{10}}}{\sqrt{1+a^2-a\eac{A_9}{A_{10}}}}}\bigg)}-\frac{a\eac{A_9}{A_{10}}-2}{\sqrt{1+a^2-a\eac{A_9}{A_{10}}}}\bigg)\Bigg]^{\frac{1}{2}}+\langle \{A_9,A_{10}\}\rangle
\end{eqnarray}
  
Following a few  simple steps, we get $\delta_{11}$  is  maximized for  $\langle\{A_9,A_{10}\}\rangle=2\cos{\frac{\pi}{11}}$ and $a=\cos{\frac{\pi}{4}}$,  which gives $\delta_{11}=22\cos{\frac{\pi}{11}}$. Hence, we can write the following
 
 \begin{eqnarray}
     (\mathscr{C}_{11})^{opt}_Q=\sqrt{11(22+\delta_{11})}=22\cos{\frac{\pi}{22}}
 \end{eqnarray}
 We can further conclude that $A_1=\frac{\cos{\frac{\pi}{4}}\ A_9-A_{10}}{\sqrt{1+\cos^2{\frac{\pi}{4}}-2\cos{\frac{\pi}{4}}\cos{\frac{\pi}{11}}}}$ which implies that  $\langle\{A_1,A_9\}\rangle=2\cos{\frac{8\pi}{11}}$ and $\nu^8_{11}=\sqrt{2+\langle\{A_1,A_9\}\rangle}=\sqrt{2+2\cos{\frac{8\pi}{11}}}=2\cos{\frac{4\pi}{11}}$. Following the similar way, we get that 
 Alice's observables follow the following relation:

\begin{eqnarray}
    \langle\{A_i,A_{i+x}\}\rangle&=&2\cos\frac{\pi  x}{11}\quad \text{$\forall i\in[10], x\in [11-i]$}
\end{eqnarray}
Hence  we get 
\ba \nu^1_{11}=\nu^2_{11}=\nu^3_{11}=\nu^4_{11}=\nu^5_{11}=\sqrt{2+2\cos{\frac{2\pi}{11}}}=2\cos{\frac{\pi}{11}} ,\  \nu^6_{11}=\nu^7_{11}=2\cos{\frac{2\pi}{11}}\ea 
Thus we get, $ \nu_{11,i}=\sqrt{2+\langle\{A_i,A_{i+1}\}\rangle}=2\cos{\frac{\pi}{22}}$. Following  Eq. (\ref{rho odd}) of the main text, we can write the required state $\rho_{AB}$ by substituting  $n=11$ in Eq.~(\ref{C'so}). Hence, we get 
\ba C_1 \otimes C_1 &=& \mathcal{A}_6 \otimes B_6\\ \na 
C_2 \otimes C_2&=&\frac{1}{5}\Bigg[\frac{(\mathcal{A}_1-\mathcal{A}_{11}) \otimes (B_1-B_{11})}{2-\langle\{B_1,B_{11}\}\rangle}+\frac{(\mathcal{A}_2 -\mathcal{A}_{10})\otimes (B_2-B_{10})}{2-\langle\{B_2,B_{10}\}\rangle}+\frac{(\mathcal{A}_3-\mathcal{A}_9) \otimes (B_3-B_9)}{2-\langle\{B_3,B_{9}\}\rangle}\\\na &&
\hspace{4mm}+\frac{(\mathcal{A}_4- \mathcal{A}_8) \otimes (B_4-B_8)}{2-\langle\{B_4,B_8\}\rangle} +\frac{(\mathcal{A}_5- \mathcal{A}_7) \otimes (B_5-B_7)}{2-\langle\{B_5,B_7\}\rangle}\Bigg]\ea 

and $C_3 \otimes C_3 =(C_1 \otimes C_1)\cdot (C_2 \otimes C_2)$
where, $\mathcal{A}_i=\frac{A_i+A_{i+1}}{\nu_{11,i}}$ and $\nu_{11,i}=\sqrt{2+2\cos{\frac{\pi}{11}}}$. 

For a two-qubit system, a suitable set of observables is given by  \ba A_i=\sin\theta_i\X+\cos\theta_i\Z, \ B_i=\sin\phi_i\X+\cos\phi_i\Z,\  \theta_i = \frac{\qty(i-1)\pi}{11},  \ \phi_i=\frac{\qty(2i-1)\pi}{22}\ea which further provides 
\begin{eqnarray}
    &&C_1 \otimes C_1 =\mathcal{A}_6 \otimes B_6=\sigma_x\otimes \sigma_x\nonumber\\
    &&C_2 \otimes C_2=\frac{1}{5}\Bigg[\frac{(\mathcal{A}_1-\mathcal{A}_{11}) \otimes (B_1-B_{11})}{2-\langle\{B_1,B_{11}\}\rangle}+\frac{(\mathcal{A}_2 -\mathcal{A}_{10})\otimes (B_2-B_{10})}{2-\langle\{B_2,B_{10}\}\rangle}+\frac{(\mathcal{A}_3-\mathcal{A}_9) \otimes (B_3-B_9)}{2-\langle\{B_3,B_{9}\}\rangle}\\\na &&
\hspace{1.5cm}+\frac{(\mathcal{A}_4- \mathcal{A}_8) \otimes (B_4-B_8)}{2-\langle\{B_4,B_8\}\rangle} +\frac{(\mathcal{A}_5- \mathcal{A}_7) \otimes (B_5-B_7)}{2-\langle\{B_5,B_7\}\rangle}\Bigg]=\sigma_z\otimes \sigma_z\\
    &&C_3 \otimes C_3 =(C_1 \otimes C_1)\cdot (C_2 \otimes C_2)=-\sigma_y\otimes \sigma_y\nonumber
\end{eqnarray}
Hence we get the  entangled state  $\rho_{\phi^+}$ as in Eq. (\ref{rhophi+}) that produces $(\mathscr{C}_{11})^{opt}_Q=22\cos{\frac{\pi}{22}}$.

%%%%%%%%%%%%%%%%%%%%%%%%%%%%%%%%%%%%%%%%%%%%%%%%%%%%%%%%%%%%%%%%%%%%%%%%%%%%%%%%
\section{Explicit derivation of the optimal quantum violation for \texorpdfstring{$n=4$ and $n=6$}{n=4 and n=6}}\label{SOS 4,6}
Here, we provide a detailed derivation of the optimal quantum value $ (\mathscr{C}_n)^{opt}_Q$ along with the state and observables, using the dimension-independent SOS approach, for  $n=4$ and $n=6$.
\subsection{Detailed derivation for \texorpdfstring{$n=4$}{n=4}}Substituting $n=4$ in Eq. (\ref{BmAB}) of  the main text, we get the  chained Bell inequality
\begin{eqnarray}\label{Chain4}
    \mathscr{C}_{4}&=&(A_1+A_2)B_1+(A_2+A_3)B_2+(A_3+A_4)B_3+(A_4-A_1)B_4\leq 6
\end{eqnarray}
Following the similar SOS approach, we get that  the optimal value of $(\mathscr{C}_4)_Q$ is obtained if   $\Tr[ \Gamma_4\ \rho_{AB}]=0$, implying that 
\begin{eqnarray}(\mathscr{
C}_{4})^{opt}_{Q}&=& \max \limits_{\{\{A_i\}, \rho_{AB}\}}\left[\nu_{4,1}+\nu_{4,2}+\nu_{4,3}+\nu_{4,4}\right]
\end{eqnarray}
where $\nu_{4,i}=||A_i+A_{i+1}||_{\rho_{AB}}=\sqrt{2+\langle \{A_i,A_{i+1}\}\rangle},  \forall i\in[4], A_{5}=-A_1$. Using the inequality in Eq. (\ref{cnv}) of the main text, we get 
 \ba
 (\mathscr{
C}_{4})_Q&\leq&\sqrt{4\big[(\nu_{4,1})^2+(\nu_{4,2})^2+(\nu_{4,3})^2+(\nu_{4,4})^2\big]} \ea
The equality holds only when $\nu_{4,i}=\nu_{4,j}, \forall i\neq j\in[4]$, which implies $\langle \{A_i,A_{i+1}\}\rangle=\langle \{A_j,A_{j+1}\}\rangle \forall i\neq j\in[4]$ and $A_{5}=-A_1$. Hence
 \ba
 (\mathscr{
C}_{4})_Q&\leq&\sqrt{4\big[8+\langle\{A_2,(A_1+A_3)\}\rangle+\langle\{A_4,(A_3-A_1)\}\rangle\big]}
\ea
 Considering $A_2=(A_1+A_3)/v^1_4$ and $A_4=(A_3-A_1)/v^2_4$, where $v^1_4=\sqrt{2+\langle\{A_1,A_3\}\rangle}$ and $v^2_4=\sqrt{2-\langle\{A_1,A_3\}\rangle}$, we get 
 \ba 
 (\mathscr{
C}_{4})_Q&\leq&\sqrt{4\big[8+2\sqrt{2+\langle\{A_1,A_3)\}\rangle}+2\sqrt{2-\langle\{A_1,A_3)\}\rangle}\big]}
\ea 
 Clearly, $(\mathscr{
C}_{4})_Q$ is optimized when $\langle\{A_1,A_3\}\rangle=0$,  which implies $\nu^1_4=\nu^2_4=\sqrt{2}$. Hence $A_2=(A_1+A_3)/\sqrt{2}$ and $A_4=(A_3-A_1)/\sqrt{2}$, which further implies  $\langle\{A_1,A_2\}\rangle= \langle\{A_2,A_3\}\rangle=\sqrt{2}=2\cos{\frac{\pi}{4}}$, $\langle\{A_1,A_4\}\rangle=-\langle\{A_3,A_4\}\rangle=\sqrt{2}=2\cos{\frac{\pi}{4}}$ and $\langle\{A_2,A_4\}\rangle=2\cos{\frac{\pi}{2}}$.  Hence, we can write 
 \begin{eqnarray}\label{a4}
    \langle\{A_i,A_{i+x}\}\rangle&=&2\cos\frac{\pi \ x}{4}\quad \text{$\forall i\in[3], x\in[4-i]$}
\end{eqnarray}
We can then explicitly find that \ba \nu_{4,i}=\sqrt{2+\sqrt{2}}=2\cos\frac{\pi}{8} ,\  \forall i\in[4]\ea   Hence, the optimal quantum value is given by 
 \ba(\mathscr{
C}_{4})^{opt}_{Q}=4\sqrt{2+\sqrt{2}}=8\cos\frac{\pi}{8}\ea   
 Following  Eq. (\ref{rho odd}) (more details in Appendix.~\ref{stateeven}) of the main text, we can write the required state $\rho_{AB}$ by putting $n=4$ in Eq.~(\ref{C'se}). Thus we get  
\ba C_1 \otimes C_1 &=& {A}_1 \otimes \mathcal{B}_1, \quad 
C_2 \otimes C_2={A}_3 \otimes \mathcal{B}_3,\quad
C_3 \otimes C_3 = \frac{1}{4}\big[A_2, A_4\big]\otimes \big[\mathcal{B}_2, \mathcal{B}_4\big]=(C_1 \otimes C_1)\cdot (C_2 \otimes C_2)\ea 
and 
$\mathcal{B}_1 = \frac{B_1-B_4}{\sqrt{2+\sqrt{2}}}, \ \mathcal{B}_2 = \frac{B_1+B_2}{\sqrt{2+\sqrt{2}}},
\mathcal{B}_3 = \frac{B_2+B_3}{\sqrt{2+\sqrt{2}}}, \ \mathcal{B}_4 = \frac{B_3+B_4}{\sqrt{2+\sqrt{2}}}
$.

For example, let us consider the two-qubit set of observables \ba A_i=\sin\theta_i\X+\cos\theta_i\Z, \quad B_i=\sin\phi_i\X+\cos\phi_i\Z,\ \theta_i = \frac{\qty(i-1)\pi}{4}, \ \phi_i=\frac{\qty(2i-1)\pi}{8}\ea which in turn  provides
\begin{eqnarray}
    &&C_1 \otimes C_1 = {A}_1 \otimes \mathcal{B}_1=\sigma_x\otimes \sigma_x,\quad C_2 \otimes C_2={A}_3 \otimes \mathcal{B}_3=\sigma_z\otimes \sigma_z,\quad C_3 \otimes C_3 =(C_1 \otimes C_1)\cdot (C_2 \otimes C_2)=-\sigma_y\otimes \sigma_y
\end{eqnarray}
This, in turn, provides the  entangled state  $\rho_{\phi^+}$ as in Eq. (\ref{rhophi+}) that produces $(\mathscr{C}_{4})^{opt}_Q=8\cos{\frac{\pi}{8}}$.

%%%%%%%%%%%%%%%%%%%%%%%%%%%%%%%%%%%%%%%%%%%%%%%%%%%%%%%%%%%%%%%%%%%%%%%%%%%%%%%%%%%%%%
\subsection{Detailed derivation for \texorpdfstring{$n=6$}{n=6}}
Similarly, substituting $n=6$ in Eq. (\ref{BmAB}) of  the main text,  we get the  chained Bell inequality
\ba
\mathscr{
C}_{6}&=&(A_1+A_2)B_1+(A_2+A_3)B_2+(A_3+A_4)B_3+(A_4+A_5)B_4+(A_5+A_6)B_5+(A_6-A_1)\leq 10
 \ea
Following the similar SOS approach as presented, we get that  the optimal value of $(\mathscr{C}_6)_Q$ is obtained when  $\Tr[ \Gamma_6\ \rho_{AB}]=0$, implying that 
\begin{eqnarray}(\mathscr{
C}_{6})^{opt}_{Q}&=& \max \limits_{A_i, \rho_{AB}}\left[\nu_{6,1}+\nu_{6,2}+\nu_{6,3}+\nu_{6,4}+\nu_{6,5}+\nu_{6,6}\right]
\end{eqnarray} 
where $\nu_{6,i}=||A_i+A_{i+1}||_{\rho_{AB}}, \forall i\in[6], A_{7}=-A_1$. Using the inequality in Eq.~(\ref{cnv}) of the main text, we get 
 \ba
(\mathscr{C}_{6})_Q&\leq&\sqrt{6\sum\limits_{i=1}^6\left[\nu_{6,i}\right]^2}
 \ea
The equality holds only if  $\nu_{6,i}=\nu_{6,j}, \forall i\neq j\in[6]$, which implies $\langle \{A_i,A_{i+1}\}\rangle=\langle \{A_j,A_{j+1}\}\rangle \forall i\neq j\in[6]$ and $A_{7}=-A_1$. Hence 
 \ba
(\mathscr{C}_{6})_Q&\leq&\sqrt{6\big(12+ \delta_6 \big)}
 \ea
  where \ba\na\delta_6 &=&\langle\{A_2,(A_1+A_3)\}\rangle+\langle \{A_4,(A_3+A_5)\}\rangle+\langle \{A_6,(A_5-A_1)\}\rangle\\\ea
 Let us consider  $A_2=(A_1+A_3)/\nu^1_6, A_4=(A_3+A_5)/\nu^2_6$, and $A_6=(A_5-A_1)/\nu^3_6,$ where $v^1_6=\sqrt{2+\langle\{A_1,A_3\}\rangle}$, $v^2_6=\sqrt{2+\langle\{A_3,A_5\}\rangle}$  and $v^3_6=\sqrt{2-\langle\{A_1,A_5\}\rangle}$.  This  implies that 
 \ba\na\delta_6&=&2\Big(\sqrt{2+\langle\{A_1,A_3\}\rangle}+\sqrt{2+\langle\{A_3,A_5\}\rangle}+\sqrt{2-\langle\{A_1,A_5\}\rangle}\Big)\nonumber\\
&\leq&2\sqrt{3\bigg(6+\langle\{A_1,A_3\}\rangle+\langle\{A_3,A_5\}\rangle-\langle\{A_1,A_5\}\rangle}\bigg)\hspace{1mm}\ea
Again the equality holds only when $\nu^1_6=\nu^2_6=\nu^3_6$, which implies $\langle\{A_1,A_3\}\rangle=\langle\{A_3,A_5\}\rangle=-\langle\{A_1,A_5\}\rangle$. Hence, further, we can write
\ba
\delta_6&\leq&2\sqrt{3\bigg(6+\langle\{(A_1+A_5),A_3\}\rangle-\langle\{A_1,A_5\}\rangle\bigg)}
 \ea
Considering   $A_3=(A_1+A_5)/\nu^4_6$ where $v^4_6=\sqrt{2+\langle\{A_1,A_5\}\rangle}$, we get  \ba\label{opt6}\delta_6&\leq& 2\sqrt{3\bigg(6+2\sqrt{2+\langle\{A_1,A_5\}\rangle}-\langle\{A_1,A_5\}\rangle\bigg)}\ea
 
 Clearly, $\delta_6$ of Eq. (\ref{opt6}) will be maximized when $\langle\{A_1,A_5\}\rangle=-1$ i.e.,  $\nu^4_6=1$ which leads to the condition $A_3=A_1+A_5$. Hence $\langle\{A_1,A_3\}\rangle=\langle\{A_3,A_5\}\rangle=1=2\cos{\frac{2\pi}{6}}$. Following  the same way,  we can find that $\nu^1_6=\nu^2_6=\nu^3_6=\sqrt{3}$, which further gives $\langle\{A_1,A_2\}\rangle=\langle\{A_2,A_3\}\rangle=\langle\{A_3,A_4\}\rangle=\langle\{A_4,A_5\}\rangle=\langle\{A_5,A_6\}\rangle=-\langle\{A_1,A_6\}\rangle= \sqrt{3}=2\cos{\frac{\pi}{6}}$. Further, we also get  $A_4=(A_3+A_5)/\nu^2_6$, and $A_6=(A_5-A_1)/\nu^3_6$, which  further shows that $\langle\{A_1,A_4\}\rangle=\langle\{A_2,A_5\}\rangle=\langle\{A_3,A_6\}\rangle=0, \langle\{A_1,A_5\}\rangle=\langle\{A_2,A_6\}\rangle=2\cos{\frac{2\pi}{3}}$ and $\langle\{A_2,A_4\}\rangle=2\cos{\frac{\pi}{3}}$. Hence, we can write the following 
 \begin{eqnarray}
    \langle\{A_i,A_{i+x}\}\rangle&=&2\cos\frac{\pi \ x}{6}, \quad \text{$\forall i\in[5], x\in[6-i]$}
\end{eqnarray}
Consequently we get   $\nu_{6,i}=\sqrt{2+\sqrt{3}}= 2\cos\dfrac{\pi}{12}, \forall i\in[6]$. Finally,  we get  \ba (\mathscr{C}_{6})_Q^{opt}=6\sqrt{2+\sqrt{3}}=12\cos\frac{\pi}{12}\ea 
Following  Eq. (\ref{rho odd}) (more details in Appendix.~\ref{stateeven})of the main text, we can write the required state $\rho_{AB}$ by putting $n=6$ in Eq.~(\ref{C'se}). Thus we get,
\ba C_1 \otimes C_1 &=& {A}_1 \otimes \mathcal{B}_1, \ C_2 \otimes C_2={A}_4 \otimes \mathcal{B}_4,\  
C_3 \otimes C_3 = \frac{1}{8}\qty(\big[A_2,A_5]\otimes \big[\mathcal{B}_2 , \mathcal{B}_5\big] +\big[A_3, A_6]\otimes \big[\mathcal{B}_3, \mathcal{B}_6\big])=(C_1 \otimes C_1)\cdot (C_2 \otimes C_2)\nonumber\ea
with $\mathcal{B}_i = \frac{B_i+B_{i-1}}{w},  w=\sqrt{2+\sqrt{3}}$ and $B_{0}=-B_n$. 

For a two-qubit system,  a suitable  set of observables is given by \ba  
 A_i=\sin\theta_i\X+\cos\theta_i\Z, \quad B_i=\sin\phi_i\X+\cos\phi_i\Z, \ \theta_i = \frac{\qty(i-1)\pi}{6},\  \phi_i=\frac{\qty(2i-1)\pi}{12}\ea which   provides
\begin{eqnarray}
    &&C_1 \otimes C_1 = {A}_1 \otimes \mathcal{B}_1=\sigma_x\otimes \sigma_x,\quad
    C_2 \otimes C_2={A}_4 \otimes \mathcal{B}_4=\sigma_z\otimes \sigma_z,\quad C_3 \otimes C_3 =(C_1 \otimes C_1)\cdot (C_2 \otimes C_2)=-\sigma_y\otimes \sigma_y
\end{eqnarray}
This, in turn, provides the  entangled state  $\rho_{\phi^+}$ as in Eq. (\ref{rhophi+}) that produces  $(\mathscr{ C}_{6})^{opt}_Q=12\cos\frac{\pi }{12}$. For $n=8,9,10,12, \cdots$, the derivations are very similar and straightforward and less cumbersome compared to odd $n$.
%%%%%%%%%%%%%%%%%%%%%%%%%%%%%%%%%%%%%%%%%%%%%%%%%%%%%%%%%%%%%%%%%%%%%%%%%%%%%%%%%%%%%%%%%%%%%%%%%%%%%%%%%%%
\section{Derivation of the  required  state for optimal quantum violation}\label{cse}
\subsection{Derivation of the  required  state for odd \texorpdfstring{$n$}{n}}\label{cseodd}
The optimal quantum vale   $(\mathscr{C}_{n})^{opt}_Q$  in the main text  provides the condition 
    \begin{eqnarray}\label{L1}
        \mathcal{A}_i\otimes B_i\ket{\psi}_{AB}&=&\ket{\psi}_{AB}, \ \forall i\in[n].
    \end{eqnarray}
    
    For $i=1$ and $i=n$, we have the relations 
     \begin{eqnarray}\label{i1}
        \mathcal{A}_1\otimes B_1\ket{\psi}_{AB}&=&\ket{\psi}_{AB}\\\label{in}
        \mathcal{A}_n\otimes B_n\ket{\psi}_{AB}&=&\ket{\psi}_{AB}
    \end{eqnarray}
First we consider $C_1\otimes C_1=\mathcal{A}_{\frac{n+1}{2}}\otimes B_{\frac{n+1}{2}}$, thus implying $\Tr[C_1\otimes C_1 \ \rho_{AB}]=1$. Pre-multiplying $\openone_d\otimes B_n B_1$ and $\openone_d\otimes B_1 B_n$ in  the Eq.~(\ref{i1}) and Eq.~(\ref{in}) respectively, we get,
    \begin{eqnarray}\label{A1B5}
        \mathcal{A}_1\otimes B_n\ket{\psi}_{AB}&=&\openone_d\otimes B_n B_1\ket{\psi}_{AB}
    \\\label{A5B1}
        \mathcal{A}_n\otimes B_1\ket{\psi}_{AB}&=&\openone_d\otimes B_1 B_n\ket{\psi}_{AB}
    \end{eqnarray}
    Adding Eq.~(\ref{i1}),(\ref{in}) and subtracting Eq.~(\ref{A1B5}),(\ref{A5B1}) we get,
    \begin{eqnarray}
        &&\mathcal{A}_1\otimes B_1+\mathcal{A}_n\otimes B_n-\mathcal{A}_1\otimes B_n-\mathcal{A}_n\otimes B_1\ket{\psi}_{AB}= (2\openone_d\otimes\openone_d-\openone_d\otimes\{B_1,B_n\})\ket{\psi}_{AB}\nonumber\\
        &&\bra{\psi}_{AB}\mathcal{A}_1\otimes B_1+\mathcal{A}_n\otimes B_n-\mathcal{A}_1\otimes B_n-\mathcal{A}_n\otimes B_1\ket{\psi}_{AB}= (2-\langle\{B_1,B_n\}\rangle)\nonumber\\
        &&\bra{\psi}_{AB}\frac{\mathcal{A}_1\otimes B_1+\mathcal{A}_n\otimes B_n-\mathcal{A}_1\otimes B_n-\mathcal{A}_n\otimes B_1}{(2-\langle\{B_1,B_n\}\rangle)}\ket{\psi}_{AB}=1\nonumber\\
    &&\frac{\mathcal{A}_1\otimes B_1+\mathcal{A}_n\otimes B_n-\mathcal{A}_1\otimes B_n-\mathcal{A}_n\otimes B_1}{(2-\langle\{B_1,B_n\}\rangle)}\ket{\psi}_{AB}= \ket{\psi}_{AB}\quad\bigg[\text{$(2-\langle\{B_1,B_n\}\rangle)$ is normalization constant}\bigg]\label{A1AnB1Bn}
    \end{eqnarray}
Following the similar steps, we also get for any pair $\mathcal{A}_i,\mathcal{A}_{n+1-i}, \  i\in[n]$ we get
\ba\label{AiBi}
\frac{(\mathcal{A}_{i} \otimes B_{i} + \mathcal{A}_{n+1-i} \otimes B_{n+1-i} - \mathcal{A}_{i} \otimes B_{n+1-i} - \mathcal{A}_{n+1-i} \otimes B_{i})}{(2-\langle\{B_{i},B_{n+1-i}\}\rangle)}\ket{\psi}_{AB}&=& \ket{\psi}_{AB}
\ea 
Adding all these for $i\in\bigg[\lfloor\frac{n}{2}\rfloor\bigg]$ and   simplifying ,  we get
\begin{eqnarray}
    \frac{1}{\lfloor \frac{n}{2} \rfloor}\sum_{i=1}^{{\lfloor \frac{n}{2}\rfloor}}\frac{(\mathcal{A}_{i}-\mathcal{A}_{n+1-i}) \otimes (B_{i}-B_{n+1-i})}{(2-\langle\{B_{i},B_{n+1-i}\}\rangle)}\ket{\psi}_{AB}&=& \ket{\psi}_{AB}\label{c2c2}
\end{eqnarray}
Now, let us consider  
\ba C_2\otimes C_2= \frac{1}{\lfloor \frac{n}{2} \rfloor}\sum_{i=1}^{{\lfloor \frac{n}{2}\rfloor}}\frac{(\mathcal{A}_{i}-\mathcal{A}_{n+1-i}) \otimes (B_{i}-B_{n+1-i})}{(2-\langle\{B_{i},B_{n+1-i}\}\rangle)}
\ea 
From Eq.~(\ref{c2c2}), clearly we get $\Tr[C_2\otimes C_2 \ \rho_{AB}]=1$. Further premultiplying 
$\mathcal{A}_{\frac{n+1}{2}}\otimes B_{\frac{n+1}{2}}$ in  Eq.~(\ref{c2c2}) we get
\begin{eqnarray}
     \frac{1}{\lfloor \frac{n}{2} \rfloor}\sum_{i=1}^{{\lfloor \frac{n}{2}\rfloor}}\frac{\mathcal{A}_{\frac{n+1}{2}}(\mathcal{A}_{i}-\mathcal{A}_{n+1-i}) \otimes B_{\frac{n+1}{2}}(B_{i}-B_{n+1-i})}{(2-\langle\{B_{i},B_{n+1-i}\}\rangle)}\ket{\psi}_{AB}&=& \ket{\psi}_{AB}\nonumber
\end{eqnarray}
We consider  
\ba C_3\otimes C_3= \frac{1}{\lfloor \frac{n}{2} \rfloor}\sum_{i=1}^{{\lfloor \frac{n}{2}\rfloor}}\frac{\mathcal{A}_{\frac{n+1}{2}}(\mathcal{A}_{i}-\mathcal{A}_{n+1-i}) \otimes B_{\frac{n+1}{2}}(B_{i}-B_{n+1-i})}{(2-\langle\{B_{i},B_{n+1-i}\}\rangle)}\ea 
Consequently, $\Tr[C_3\otimes C_3 \ \rho_{AB}]=1$. It is straightforward to show that $C_3\otimes C_3 = (C_1\otimes C_1)(C_2\otimes C_2)$ and $\langle[C_{\bar{i}}\otimes C_{\bar{i}}, C_{\bar{j}}\otimes C_{\bar{j}}]_{\bar{i}\neq {\bar{j}}}\rangle_{\rho_{AB}}=0,  \ \forall {\bar{i}}\neq {\bar{j}}\in [d^2-1]$ implying that  $\Tr[C_{\bar{i}}\otimes C_{\bar{i}} \ \rho_{AB}]=1, {\bar{i}}\in[d^2-1]$. Now, any arbitrary $d$- dimensional bipartite state can be written as a function of $C_{\bar{i}}\otimes C_{\bar{i}}$ as 
\begin{eqnarray}\label{state 7}
    \rho_{AB} = \frac{1}{d^2} \qty[\openone_d\otimes\openone_d + C_1 \otimes C_1+C_2 \otimes C_2+C_3 \otimes C_3 + \sum_{\bar{i}=4}^{d^2-1} C_{\bar{i}} \otimes C_{\bar{i}}]
\end{eqnarray}
Again, altogether we get
\begin{eqnarray}
    C_1\otimes C_1\ket{\psi}_{AB}=C_2\otimes C_2\ket{\psi}_{AB}=C_3\otimes C_3\ket{\psi}_{AB}=\ket{\psi}_{AB}
\end{eqnarray} 
Hence, we can directly show that $\langle [C_{\bar{i}}\otimes C_{\bar{i}},C_{\bar{j}}\otimes C_{\bar{j}}]\rangle_{\rho_{AB}}=0,\forall \bar{i}\neq \bar{j}\in\{1,2,3\}$.

The state  $\rho_{AB}$ provides the optimal quantum violation if $\Tr[(\mathcal{A}_i\otimes B_i) \ \rho_{AB}]=1,  \ \forall i\in[n]$. Clearly, for higher values of $n$, the derivation is much more cumbersome and lengthy. For better understanding,  we show the derivation for  $n=7$. Following the above argument and using the optimization conditions from  Eq. (\ref{b7}), we get 
\ba C_1 \otimes C_1 &=& \mathcal{A}_4 \otimes B_4\\\na 
C_2 \otimes C_2&=&\frac{1}{3}\Bigg[\frac{(\mathcal{A}_1-\mathcal{A}_7) \otimes (B_1-B_7)}{2+2\cos{\frac{\pi}{7}}}+\frac{(\mathcal{A}_2 -\mathcal{A}_6)\otimes (B_2-B_6)}{2+2\cos{\frac{3\pi}{7}}}+\frac{(\mathcal{A}_3 -\mathcal{A}_5)\otimes (B_3-B_5)}{2+2\cos{\frac{5\pi}{7}}}\Bigg]\\\na 
C_3 \otimes C_3 &=&\frac{1}{3}\Bigg[\frac{\mathcal{A}_4(\mathcal{A}_1-\mathcal{A}_7) \otimes B_4 (B_1 -B_7)}{2+2\cos{\frac{\pi}{7}}}+\frac{\mathcal{A}_4(\mathcal{A}_2-\mathcal{A}_6) \otimes B_4 (B_2-B_6)}{2+2\cos{\frac{3\pi}{7}}}+\frac{\mathcal{A}_4(\mathcal{A}_3-\mathcal{A}_5) \otimes B_4 (B_3-B_5)}{2+2\cos{\frac{5\pi}{7}}}\Bigg]\ea 

From the derivation of the optimal quantum bound, we have $\langle\{\mathcal{A}_i,\mathcal{A}_{i+x}\}\rangle=\langle\{B_i, B_{i+x}\}\rangle=2 \cos{\frac{\pi x}{7}}$ which can be further generalized to $\{\mathcal{A}_i,\mathcal{A}_{i+x}\}=\{B_i, B_{i+x}\}=2\openone_d \cos{\frac{\pi x}{7}} \text{ only for $\rho_{AB}$}$,  \textcolor{red}{}which directly implies 
\ba \label{ix} \Tr[\mathcal{A}_i \mathcal{A}_{i+x}]=\Tr[B_i B_{i+x}]=d\cos{\frac{\pi x}{7}}\ea
For proving $\Tr[\mathcal{A}_1\otimes B_1 \ \rho_{AB}]=1$, we use Eq. (\ref{ix}) and calculate the following.
\begin{eqnarray}\label{A1C1B1C1}
    \Tr[(\mathcal{A}_1\otimes B_1) (C_1 \otimes  C_1)]&=& \Tr[\mathcal{A}_1 \mathcal{A}_4 \otimes B_1 B_4]= d^2 \cos^2{\frac{3\pi}{7}}\\
\label{A1C2B1C2}
    \Tr[(\mathcal{A}_1\otimes B_1)( C_2 \otimes C_2)]&=& \frac{1}{3}\Bigg[\Tr\bigg(\frac{\openone_d \otimes \openone_d-\openone_d \otimes B_1 B_7 - \mathcal{A}_1 \mathcal{A}_7\otimes\openone_d+\mathcal{A}_1 \mathcal{A}_7\otimes B_1 B_7}{2(1+\cos{\frac{\pi}{7}})}\bigg)\nonumber\\
    &&\phantom{\frac{1}{3}\Bigg[}+\Tr\bigg(\frac{\mathcal{A}_1 \mathcal{A}_3\otimes B_1 B_3-\mathcal{A}_1 \mathcal{A}_3\otimes B_1 B_5 - \mathcal{A}_1 \mathcal{A}_5\otimes B_1 B_3+\mathcal{A}_1 \mathcal{A}_5\otimes B_1 B_5}{2(1+\cos{\frac{5\pi}{7}})}\bigg)\nonumber\\
    &&\phantom{\frac{1}{3}\Bigg[}+\Tr\bigg(\frac{\mathcal{A}_1 \mathcal{A}_2\otimes B_1 B_2-\mathcal{A}_1 \mathcal{A}_2\otimes B_1 B_6 - \mathcal{A}_1 \mathcal{A}_6\otimes B_1 B_2+\mathcal{A}_1 \mathcal{A}_6\otimes B_1 B_6}{2(1+\cos{\frac{3\pi}{7}})}\bigg) \ \Bigg]\nonumber\\
      &=& \frac{d^2}{6}\Bigg[\frac{1+\cos^2 \left(\frac{\pi }{7}\right)+2 \cos\left(\frac{\pi }{7}\right)}{\left(1+\cos \left(\frac{\pi }{7}\right)\right)}+\frac{\cos^2 \left(\frac{2 \pi }{7}\right)+\cos^2 \left(\frac{4 \pi }{7}\right)-2\cos \left(\frac{2 \pi }{7}\right)\cos \left(\frac{4 \pi }{7}\right)}{1+\cos \left(\frac{5 \pi }{7}\right)}\nonumber\\
      &&\phantom{\frac{d^2}{6}\Bigg[}+\frac{\cos^2 \left(\frac{\pi }{7}\right)+\cos^2 \left(\frac{5 \pi }{7}\right)-2\cos \left(\frac{\pi }{7}\right) \cos \left(\frac{5 \pi }{7}\right)}{1+\cos \left(\frac{3 \pi }{7}\right)}\Bigg]\nonumber\\
    &=& \frac{d^2}{2}\Bigg[1+\cos \left(\frac{\pi }{7}\right)\Bigg]\\ \label{A1C3B1C3}
     \Tr[(\mathcal{A}_1\otimes B_1)( C_3 \otimes C_3)]&=&\frac{1}{3}\Bigg[\Tr\bigg(\frac{(\mathcal{A}_1\mathcal{A}_4\mathcal{A}_1 \otimes B_1 B_4 B_1 + \mathcal{A}_1 \mathcal{A}_4\mathcal{A}_7 \otimes B_1 B_4 B_7 - \mathcal{A}_1 \mathcal{A}_4\mathcal{A}_1 \otimes B_1 B_4 B_7 -\mathcal{A}_1 \mathcal{A}_4\mathcal{A}_7 \otimes B_1 B_4 B_1)}{2+2\cos{\frac{\pi}{7}}}\bigg)\nonumber\\
    &&\phantom{\frac{1}{3}\Bigg[}+\Tr\bigg(\frac{(\mathcal{A}_1\mathcal{A}_4\mathcal{A}_3 \otimes B_1 B_4 B_3 + \mathcal{A}_1 \mathcal{A}_4\mathcal{A}_5 \otimes B_1 B_4 B_5 - \mathcal{A}_1 \mathcal{A}_4\mathcal{A}_3 \otimes B_1 B_4 B_5 -\mathcal{A}_1 \mathcal{A}_4 \mathcal{A}_5 \otimes B_1 B_4 B_3)}{2+2\cos{\frac{5\pi}{7}}}\bigg)\nonumber\\\na
    &&\phantom{\frac{1}{3}\Bigg[}+\Tr\bigg(\frac{(\mathcal{A}_1 \mathcal{A}_4\mathcal{A}_2 \otimes B_1 B_4 B_2 + \mathcal{A}_1 \mathcal{A}_4\mathcal{A}_6 \otimes B_1 B_4 B_6 - \mathcal{A}_1 \mathcal{A}_4\mathcal{A}_2 \otimes B_1 B_4 B_6 - \mathcal{A}_1 \mathcal{A}_4\mathcal{A}_6 \otimes B_1 B_4 B_2)}{2+2\cos{\frac{3\pi}{7}}}\bigg)\Bigg]\\
\end{eqnarray}
From the optimization in Appendix.~\ref{SOS7} and \ref{Bobo}, we have found that Bob's observables have the following relations
\ba \label{b72}B_1=\frac{a\ B_5-B_6}{a'}, \ B_2=(B_1+B_3)/\nu^1_7, \  B_4=(B_3+B_5)/\nu^2_7, B_7=(B_6-B_1)/\nu^3_7, B_3=(B_1+B_5)/\nu^4_7 \ea 
  where $a'=\sqrt{1+a^2-a\eac{B_5}{B_6}}, \langle\{B_5,B_6\}\rangle=2\cos{\frac{\pi}{7}}, a=\cos{\frac{10\pi}{49}}$ and $\nu^1_7=\nu^2_7=$, $\nu^3_7=\nu^4_7=2\cos{\frac{\pi}{7}}$. Since each observable is dichotomic, we have $(B_i)^2= \openone_d$ and $\Tr[B_i]=0,\forall i\in[7]$. 
Now, using the cyclic property of the trace operator, we get \ba \label{tr1}\Tr [B_iB_jB_i]=\Tr[B_j]=0, \ \forall i\neq j\in[7]\ea  Hence using Eq. (\ref{b7}) and (\ref{tr1}),  we calculate the following.

    \ba\na  \Tr[B_1 B_4 B_7]&=&\Tr[B_1 B_4 \left(\frac{B_6-B_1}{\nu^3_7}\right)]=\frac{1}{\nu^3_7} \Tr[B_1 B_4 B_6]=\frac{1}{\nu^3_7}\Tr[\left(\frac{a B_5-B_6}{a'}\right) B_4 B_6]=\frac{a}{a'\nu^3_7}\Tr[B_5 B_4 B_6] = \frac{a}{a'\nu^3_7}\Tr[B_5\left(\frac{B_3+B_5}{\nu^2_7}\right) B_6]\\\na &&=\frac{a}{a'\nu^2_7\nu^3_7}\Tr[B_5 B_3 B_6]=\frac{a}{a'\nu^2_7\nu^3_7}\Tr[B_5\left(\frac{B_1+B_5}{\nu^4_7}\right) B_6]= \frac{a}{a'\nu^2_7\nu^3_7\nu^4_7}\Tr[B_5 B_1 B_6]
    =  \frac{a}{a'\nu^2_7\nu^3_7\nu^4_7}\Tr[B_5 \left(\frac{a B_5-B_6}{a'}\right) B_6 ]=0\\\ea 
    We can also calculate 
\ba \Tr[B_1 B_4 B_3]&=&\Tr[B_1 B_4 B_5]=\frac{1}{\nu^2_7\nu^4_7} \Tr[B_1+B_5]=0\\\label{tr4}
    \Tr[B_1 B_4 B_5]&=&\Tr[B_1 B_4 B_2]=\frac{1}{\nu^1_7}\Tr[B_1 B_4 B_3]=0
\end{eqnarray}
Following the similar steps, we also get 
\ba\label{tr6}
    \Tr[B_1 B_4 B_6]=\frac{1}{\nu_7^4}\Tr[\left(\frac{aB_6-B_5}{a'}\right)B_5B_6]=0
\ea
Using  Eq.~(\ref{tr1})-(\ref{tr6}) in Eq.~(\ref{A1C3B1C3}),  we get
\begin{eqnarray}
    \Tr[(\mathcal{A}_1\otimes B_1)( C_3 \otimes C_3)]=0
\end{eqnarray}
Hence, we see that  \(\frac{1}{d^2}\Tr[(\mathcal{A}_1 \otimes B_1)(C_1 \otimes C_1 + C_2 \otimes C_2)] = 1\), that gives $\Tr[(\mathcal{A}_1 \otimes B_1)\rho_{AB}]=1$, and hence, the other terms $(C_{\bar{i}} \otimes C_{\bar{i}}), \forall {\bar{i}}\in[d^2-1]\setminus [2]$ has no contribution. The same conclusion applies for the other correlation terms \((\mathcal{A}_i \otimes B_i)\). Thus, for all \(i \in [n]\), the only contributing terms for \(\Tr[(\mathcal{A}_i \otimes B_i) \rho_{AB}] = 1\) are \(C_1 \otimes C_1\) and \(C_2 \otimes C_2\). For all \(i \in [7]\), this results in the condition \(\Tr\left[\sum_{{\bar{i}}=3}^{d^2-1} (\mathcal{A}_j \otimes B_j)(C_{\bar{i}} \otimes C_{\bar{i}})\right] = 0\).
Since we have  $\Tr[(\mathcal{A}_1 \otimes B_1)\openone_d]=0$, we get 
\begin{eqnarray}
    \Tr[\mathcal{A}_1\otimes B_1 \ \rho_{AB}]&=&\frac{1}{d^2}\qty[ d^2 \cos^2{\frac{3\pi}{7}}+\frac{d^2}{2}\Bigg(1+\cos \left(\frac{\pi }{7}\right)\Bigg)]=1
\end{eqnarray}
Following the similar procedure,  we can show that $\Tr[\mathcal{A}_i\otimes B_i \ \rho_{AB}]=1, \forall i\in[7]$.  This shows that the state defined in Eq. (\ref{state 7}) is indeed required to obtain the optimal quantum violation $(\mathscr{C}_7)^{opt}_Q$.
%%%%%%%%%%%%%%%%%%%%%%%%%%%%%%%%%%%%%%%%%%%%%%%%%%%%%%%%%%%%%%%%%%%%%%%%%%%%%%%%%%%%%%
\subsection{Derivation of the  required  state for even  \texorpdfstring{$n$}{n}}\label{stateeven}
The optimization condition  for deriving y  $(\mathscr{C}_n)^{opt}_Q$ is given by 
   \begin{eqnarray}
        A_i\otimes \mathcal{B}_i\ket{\psi}_{AB}&=&\ket{\psi}_{AB}, \forall i\in[n].
    \end{eqnarray}
as derived in  Eq.~(\ref{Bobsc}),  where $\mathcal{B}_i=\frac{B_i+B_{i-1}}{\nu'_{n,i}}$ where $B_0=-B_n$. For $i=1$ and $i=\frac{n}{2}+1$, we have the relations 
 \ba A_{1}\otimes \mathcal{B}_1 \ket{\psi}_{AB} &=& \ket{\psi}_{AB}, \ 
A_{\frac{n}{2}+1}\otimes  \mathcal{B}_{\frac{n}{2}+1} \ket{\psi}_{AB} = \ket{\psi}_{AB}
 \ea 
We consider $C_1\otimes C_1=A_1\otimes \mathcal{B}_{1}$ and $C_2\otimes C_2=A_{\frac{n}{2}+1}\otimes \mathcal{B}_{\frac{n}{2}+1}$ which  implies  that \ba \Tr[C_1\otimes C_1 \ \rho_{AB}]=\Tr[C_2\otimes C_2 \ \rho_{AB}]=1\ea 
For $i=2$ and $i=\frac{n}{2}+2$, we have the relations 
 \ba \label{even2}A_{2}\otimes \mathcal{B}_2 \ket{\psi}_{AB} &=& \ket{\psi}_{AB}, \\ \label{evenn/2}
A_{\frac{n}{2}+2}\otimes  \mathcal{B}_{\frac{n}{2}+2} \ket{\psi}_{AB} &=& \ket{\psi}_{AB}
 \ea 
From the optimization, we have $\langle\{A_i,A_{i+x}\}\rangle=2\cos\frac{\pi \ x}{n},  \forall i\in[n], x\in[n-i]$ implying that \ba \label{antin/2}\langle\{A_i,A_{i+\frac{n}{2}}\}\rangle=0\ea 
Pre-multiplying $A_{\frac{n}{2}+2}\otimes  \mathcal{B}_{\frac{n}{2}+2}$ and $A_{2}\otimes \mathcal{B}_2$  in  Eq.~(\ref{even2}) and Eq.~(\ref{evenn/2}) respectively, and using Eq. (\ref{antin/2}),  \textcolor{red}{}we have
 \begin{eqnarray}
    A_{\frac{n}{2}+2} A_2\otimes \mathcal{B}_{\frac{n}{2}+2}\mathcal{B}_2 \ket{\psi}_{AB}&=&\ket{\psi}_{AB}\label{es9}\\
   A_2 A_{\frac{n}{2}+2} \otimes \mathcal{B}_{\frac{n}{2}+2}\mathcal{B}_2 \ket{\psi}_{AB}&=&-\ket{\psi}_{AB}\label{es10}\\
   A_2 A_{\frac{n}{2}+2}\otimes \mathcal{B}_2 \mathcal{B}_{\frac{n}{2}+2} \ket{\psi}_{AB}&=&\ket{\psi}_{AB}\label{es11}\\
   A_{\frac{n}{2}+2} A_2 \otimes \mathcal{B}_2 \mathcal{B}_{\frac{n}{2}+2} \ket{\psi}_{AB}&=&-\ket{\psi}_{AB}\label{es12}
\end{eqnarray}
Now adding Eq.~(\ref{es9}), Eq.~(\ref{es11}) and subtracting Eq.~(\ref{es10}), Eq.~(\ref{es12}) we get
\begin{eqnarray}
    \frac{1}{4}\bigg(A_{\frac{n}{2}+2} A_2\otimes \mathcal{B}_{\frac{n}{2}+2}\mathcal{B}_2+A_2 A_{\frac{n}{2}+2}\otimes \mathcal{B}_2 \mathcal{B}_{\frac{n}{2}+2}-A_2 A_{\frac{n}{2}+2} \otimes \mathcal{B}_{\frac{n}{2}+2}\mathcal{B}_2-A_{\frac{n}{2}+2} A_2 \otimes \mathcal{B}_2 \mathcal{B}_{\frac{n}{2}+2}\bigg)\ket{\psi}_{AB}&=&\ket{\psi}_{AB}
\end{eqnarray}

Similarly, for each pair $\mathcal{B}_{i}, \mathcal{B}_{\frac{n}{2}+i}$, we  find the relations
\ba 
 \frac{1}{4}\bigg(A_{\frac{n}{2}+i} A_i\otimes \mathcal{B}_{\frac{n}{2}+i}\mathcal{B}_i+A_i A_{\frac{n}{2}+i}\otimes \mathcal{B}_i \mathcal{B}_{\frac{n}{2}+i}-A_i A_{\frac{n}{2}+i} \otimes \mathcal{B}_{\frac{n}{2}+i}\mathcal{B}_i-A_{\frac{n}{2}+i} A_i \otimes \mathcal{B}_i \mathcal{B}_{\frac{n}{2}+i}\bigg)\ket{\psi}_{AB}&=&\ket{\psi}_{AB}\label{es13}
\ea

Adding all these for \tcr{}$i\in[\frac{n}{2}]\setminus \{1\}$ and   simplifying,  we get
\begin{eqnarray}
    &&\frac{1}{4(\frac{n}{2}-1)}\sum\limits_{i=2}^{\frac{n}{2}}\Big(A_i A_{i+\frac{n}{2}}\otimes \mathcal{B}_i \mathcal{B}_{i+\frac{n}{2}} + A_{i+\frac{n}{2}} A_i\otimes \mathcal{B}_{i+\frac{n}{2}} \mathcal{B}_i - A_i A_{i+\frac{n}{2}}\otimes \mathcal{B}_{i+\frac{n}{2}} \mathcal{B}_i - A_{i+\frac{n}{2}} A_i\otimes \mathcal{B}_i \mathcal{B}_{i+\frac{n}{2}}\Big)\ket{\psi}_{AB}=\ket{\psi}_{AB}\nonumber\\
    &&\frac{1}{4(\frac{n}{2}-1)}\sum\limits_{i=2}^{\frac{n}{2}}\big[A_i, A_{i+\frac{n}{2}}\big]\otimes \big[\mathcal{B}_i,\mathcal{B}_{i+\frac{n}{2}}\big]\ket{\psi}_{AB}=\ket{\psi}_{AB}
\end{eqnarray}
Let us consider
\begin{eqnarray} C_3 \otimes C_3 &=& \frac{1}{4(\frac{n}{2}-1)}\sum\limits_{i=2}^{\frac{n}{2}}\Big(A_i A_{i+\frac{n}{2}}\otimes \mathcal{B}_i \mathcal{B}_{i+\frac{n}{2}} + A_{i+\frac{n}{2}} A_i\otimes \mathcal{B}_{i+\frac{n}{2}} \mathcal{B}_i - A_i A_{i+\frac{n}{2}}\otimes \mathcal{B}_{i+\frac{n}{2}} \mathcal{B}_i - A_{i+\frac{n}{2}} A_i\otimes \mathcal{B}_i \mathcal{B}_{i+\frac{n}{2}}\Big)\nonumber\\
&=&\frac{1}{4(\frac{n}{2}-1)}\sum\limits_{i=2}^{\frac{n}{2}}\big[A_i, A_{i+\frac{n}{2}}\big]\otimes \big[\mathcal{B}_i,\mathcal{B}_{i+\frac{n}{2}}\big]
\end{eqnarray}
With some further simplification, we can write 
\begin{eqnarray}
    C_3 \otimes C_3 &=& (C_1 \otimes C_1)(C_2\otimes C_2)
\end{eqnarray}
  From the optimization conditions of the SOS approach, it can be further shown that $\langle[C_{\bar{i}}\otimes C_{\bar{i}}, C_{\bar{j}}\otimes C_{\bar{j}}]\rangle_{\rho_{AB}}=0, \ \forall \bar{i}\neq \bar{j}\in [d^2-1]$ along with  $\Tr[C_{\bar{i}}\otimes C_{\bar{i}} \ \rho_{AB}]=1,\forall \bar{i}\in [d^2-1]$.  Now,  any arbitrary $d$-dimensional bipartite state can be written as a function of $C_{\bar{i}}\otimes C_{\bar{i}}$ as follows.
\begin{eqnarray}\label{state 4}
    \rho_{AB} = \frac{1}{d^2} \qty[\openone_d\otimes\openone_d + C_1 \otimes C_1+C_2 \otimes C_2+C_3 \otimes C_3 + \sum_{\bar{i}=4}^{d^2-1} C_{\bar{i}} \otimes C_{\bar{i}}]
\end{eqnarray}
Again, altogether we get
\begin{eqnarray}
    C_1\otimes C_1\ket{\psi}_{AB}=C_2\otimes C_2\ket{\psi}_{AB}=C_3\otimes C_3\ket{\psi}_{AB}=\ket{\psi}_{AB}
\end{eqnarray} 
Hence, we can directly show that $\langle [C_{\bar{i}}\otimes C_{\bar{i}},C_{\bar{j}}\otimes C_{\bar{j}}]\rangle_{\rho_{AB}}=0,\forall \bar{i}\neq \bar{j}\in\{1,2,3\}$.

And the state  $\rho_{AB}$ provides the optimal quantum violation if $\Tr{({A}_i\otimes \mathcal{B}_i) \ \rho_{AB}}=1,  \ \forall i\in[n]$. Clearly, for higher values of $n$, the derivation is extremely complicated. For better understanding,  we show the derivation for  $n=4$. Following the above argument and using the optimization conditions from  Eq. (\ref{a4}), we get 
\ba
C_1 \otimes C_1 &=& {A}_1 \otimes \mathcal{B}_1, \ 
C_2 \otimes C_2={A}_3 \otimes \mathcal{B}_3, \ 
C_3 \otimes C_3 =\frac{1}{4}\big[A_2, A_{4}\big]\otimes \big[\mathcal{B}_2,\mathcal{B}_{4}\big]
\ea
Now we can further reduce $C_3 \otimes C_3$ in the following way,
\begin{eqnarray}\label{c3c3n=4}
    C_3 \otimes C_3 \ket{\psi}_{AB}&=&\frac{1}{4}\big[A_2, A_{4}\big]\otimes \big[\mathcal{B}_2,\mathcal{B}_{4}\big]\ket{\psi}_{AB}\nonumber\\
    &=&\frac{1}{4}(A_2 A_4\otimes \mathcal{B}_2 \mathcal{B}_4 + A_4 A_2\otimes \mathcal{B}_4 \mathcal{B}_2 - A_2 A_4\otimes \mathcal{B}_4 \mathcal{B}_2 - A_4 A_2\otimes \mathcal{B}_2 \mathcal{B}_4)\ket{\psi}_{AB}\nonumber\\
    &=& \frac{1}{4}(A_2 A_4-A_4 A_2)\otimes (\mathcal{B}_2 \mathcal{B}_4-\mathcal{B}_4 \mathcal{B}_2)\ket{\psi}_{AB}\nonumber\\
    &=&( A_2 A_4\otimes \mathcal{B}_2 \mathcal{B}_4)\ket{\psi}_{AB} \quad \bigg[\text{As $\langle\{A_2,A_4\}\rangle=\langle\{\mathcal{B}_2, \mathcal{B}_4 \}\rangle=0$}\bigg]
\end{eqnarray}
Again from the SOS conditions,  we know that $A_2=\frac{A_1+A_3}{\sqrt{2}}, A_4=\frac{A_3-A_1}{\sqrt{2}}, \mathcal{B}_2=\frac{\mathcal{B}_1+\mathcal{B}_3}{\sqrt{2}}$ and $\mathcal{B}_4=\frac{\mathcal{B}_3-\mathcal{B}_1}{\sqrt{2}}$. Now, using these conditions in Eq.~(\ref{c3c3n=4}) we get
\begin{eqnarray}
    C_3 \otimes C_3 &=& (A_1\otimes \mathcal{B}_1) (A_3\otimes \mathcal{B}_3)= (C_1 \otimes C_1)(C_2\otimes C_2)
\end{eqnarray}
Now, in order to proof the relation $\Tr[(A_i\otimes \mathcal{B}_i) \rho_{AB}]=1$, we need some additional conditions of $A_i$ and $\mathcal{B}_i$. From the optimization for  $n=4$, we have  $\{A_i,A_{i+x}\}=\{\mathcal{B}_i, \mathcal{B}_{i+x}\}=2\openone_d \cos{\frac{\pi x}{4}}$ only for $\rho_{AB}$ which directly implies that \ba \label{4anti}\Tr[A_i A_{i+x}]=\Tr[\mathcal{B}_i\mathcal{B}_{i+x}]=d^2\cos^2{\frac{\pi x}{4}}\ea And $\Tr[\sum_{{\bar{i}}=3}^{d^2-1} A_jC_{\bar{i}} \otimes \mathcal{B}_jC_{\bar{i}}] =0, \forall j\in[4]$. 
For proving $\Tr[A_1\otimes \mathcal{B}_1 \ \rho_{AB}]=1$,  we will use Eq. (\ref{4anti}) and calculate the following:
\begin{eqnarray}\label{A1beta1c1}
    \Tr[(A_1\otimes  \mathcal{B}_1)(C_1 \otimes C_1)]&=&d^2,\ 
    \Tr[(A_1\otimes  \mathcal{B}_1)(C_2 \otimes C_2)]= \Tr[A_1 A_3 \otimes \mathcal{B}_1 \mathcal{B}_3]
    = 0
\end{eqnarray}
Also, we have 
\begin{eqnarray}\label{A1beta1c3}
    \Tr[(A_1\otimes \mathcal{B}_1)( C_3 \otimes C_3)]&=& \frac{1}{4}\Tr[A_1A_2 A_4\otimes \mathcal{B}_1\mathcal{B}_2 \mathcal{B}_4 + A_1 A_4 A_2\otimes \mathcal{B}_1 \mathcal{B}_4 \mathcal{B}_2 - A_1 A_2 A_4\otimes \mathcal{B}_1 \mathcal{B}_4 \mathcal{B}_2 - A_1 A_4 A_2\otimes \mathcal{B}_1\mathcal{B}_2 \mathcal{B}_4]
\end{eqnarray} Since each observable is dichotomic, we have $(A_i)^2= \openone_d$ and $\Tr[A_i]=\Tr[\mathcal{B}_i]=0,\forall i\in[4]$. 
Now, using the cyclic property of the trace operator, we get $\Tr [   A_iA_jA_i]=\Tr[A_j]=0, \ \forall i\neq j\in[4]$, which gives 
\begin{eqnarray}\label{4c3_1}
    \Tr[A_1 A_2 A_4]&=& \Tr[A_1\frac{(A_3+A_1)}{\sqrt{2}}\frac{(A_3-A_1)}{\sqrt{2}}]=0, \quad 
    \Tr[A_1 A_4 A_2]= \Tr[A_1\frac{(A_3-A_1)}{\sqrt{2}}\frac{(A_3+A_1)}{\sqrt{2}}]=0
    \end{eqnarray}
Using  Eq.~(\ref{4c3_1})  in Eq.~(\ref{A1beta1c3}) we get
\begin{eqnarray}
    \Tr[A_1 C_3 \otimes \mathcal{B}_1 C_3]&=&0
\end{eqnarray}which finally implies that 
$\Tr[A_1\otimes \mathcal{B}_1 \ \rho_{AB}]=\frac{1}{d^2}\qty[d^2]=1$. 
Following  the similar procedure,  we can show that $\Tr[A_i\otimes \mathcal{B}_i \ \rho_{AB}]=1, \   \forall i\in[4]$. This shows that the state defined in Eq. (\ref{state 4}) is indeed required to obtain the optimal quantum violation $(\mathscr{C}_4)^{opt}_Q$.
%%%%%%%%%%%%%%%%%%%%%%%%%%%%%%%%%%%%%%%%%%%%%%%%%%%%%%%%%%%%%%%%%%%%%%%%%%%%%%%%%%%%%%%%%%%%%%%%%%%%%%%%%%%%%%%%%%%%%%%%%%%%%%%%%%%%%%%%%%%%%%%%%%%%%%%%%%%%%%%%%
\section{Self-testing of state and observables based on the optimal quantum value}
\subsection{For odd \texorpdfstring{$n$}{n}}\label{c5}
From the  optimization conditions,  clearly we have  $\{\mathcal{A}_i,\mathcal{A}_{n+1-i}\}=\{B_i,B_{n+1-i}\}$ for each $i\in[n]$.  Using it   in Eq.~(\ref{AiBi}),  we get
\begin{eqnarray}
    &&\frac{(\mathcal{A}_{i} \otimes B_{i} + \mathcal{A}_{n+1-i} \otimes B_{n+1-i} - \mathcal{A}_{i} \otimes B_{n+1-i} - \mathcal{A}_{n+1-i} \otimes B_{i})}{(2-\langle\{B_{i},B_{n+1-i}\}\rangle)}\ket{\psi}_{AB}= \ket{\psi}_{AB}\\
    &&\left(\frac{\mathcal{A}_i-\mathcal{A}_{n+1-i}}{\sqrt{2-\langle\{\mathcal{A}_1,\mathcal{A}_{n+1-i}\}\rangle}}\otimes \frac{B_i-B_{n+1-i}}{\sqrt{2-\langle\{B_i,B_{n+1-i}\}\rangle}}\right)\ket{\psi}_{AB}=\ket{\psi}_{AB}, \nonumber\\\label{selfti}
         &&\left(\frac{\mathcal{A}_i-\mathcal{A}_{n+1-i}}{\sqrt{2-\langle\{\mathcal{A}_i,\mathcal{A}_{n+1-i}\}\rangle}}\otimes \openone_d\right)\ket{\psi}_{AB}= \left(\openone_d\otimes \frac{B_i-B_{n+1-i}}{\sqrt{2-\langle\{B_i,B_{n+1-i}\}\rangle}}\right)\ket{\psi}_{AB}
\end{eqnarray}
 Adding  for each  $i\in\bigg[\lfloor\frac{n}{2}\rfloor\bigg]$ and   simplifying it,  we get  
 \begin{eqnarray}
     \frac{1}{\lfloor\frac{n}{2}\rfloor}\sum_{i=1}^{{\lfloor \frac{n}{2}\rfloor}}\left(\frac{\mathcal{A}_{i}-\mathcal{A}_{n+1-i}}{\sqrt{2-\langle\{\mathcal{A}_{i},\mathcal{A}_{n+1-i}\}\rangle}}\otimes \openone_d\right)\ket{\psi}_{AB}&=&\left(\openone_d\otimes \frac{1}{\lfloor\frac{n}{2}\rfloor}\sum_{i=1}^{{\lfloor \frac{n}{2}\rfloor}}\frac{B_{i}-B_{n+1-i}}{\sqrt{2-\langle\{B_{i},B_{n+1-i}\}\rangle}}\right)\ket{\psi}_{AB}, \label{ct}
 \end{eqnarray}
 which implies 
 \begin{eqnarray}
     X_A\ket{\psi}_{AB}=X_B\ket{\psi}_{AB}
 \end{eqnarray}
 where  $X_A$ and $X_B$ are given by 
\begin{eqnarray}
    X_A &=&\frac{1}{\lfloor\frac{n}{2}\rfloor}\sum_{i=1}^{{\lfloor \frac{n}{2}\rfloor}}\frac{\mathcal{A}_{i}-\mathcal{A}_{n+1-i}}{\sqrt{2-\langle\{\mathcal{A}_{i},\mathcal{A}_{n+1-i}\}\rangle}}, \quad 
    X_B = \frac{1}{\lfloor\frac{n}{2}\rfloor}\sum_{i=1}^{{\lfloor \frac{n}{2}\rfloor}}\frac{B_{i}-B_{n+1-i}}{\sqrt{2-\langle\{B_{i},B_{n+1-i}\}\rangle}}
\end{eqnarray}
Also, from the self-testing condition $(\mathcal{A}_{\frac{n+1}{2}}\otimes  B_{\frac{n+1}{2}}) \ket{\psi}_{AB}=\ket{\psi}_{AB}$, we can write 
\begin{eqnarray}
    (\mathcal{A}_{\frac{n+1}{2}}\otimes \openone_d) \ket{\psi}_{AB}&=& (\openone_d \otimes B_{\frac{n+1}{2}}) \ket{\psi}_{AB}
\end{eqnarray}
 which satisfies the following condition
  \begin{eqnarray}
     Z_A\ket{\psi}_{AB}=Z_B\ket{\psi}_{AB}
 \end{eqnarray}
 where $Z_A = \mathcal{A}_{\frac{n+1}{2}}, Z_B=B_{\frac{n+1}{2}}$. We can further show that $\{Z_A,X_A\}\ket{\psi}_{AB}=\{Z_B,X_B\}\ket{\psi}_{AB}=0$. If $\ket{\psi}_{AB}\in\mathcal{H}_A\otimes\mathcal{H}_B$ is the suitable  state and $A_i\in \mathcal{H}_A, \ B_j\in\mathcal{H}_B, \ (\forall i,j\in[n])$ are the corresponding  set of observables for obtaining the optimal quantum violation, then from the Fig.~\ref{Oddc} in main text,   it can be proved that there exist a local unitary operation  $\Phi$ and ancillary state $\ket{00}_{A'B'}$ such that
\begin{eqnarray}\label{phic}
    \Phi(\ket{\psi}_{AB}\otimes\ket{00}_{A'B'} 
 )&=&\frac{1}{4}\Bigg[(1+Z_A)(1+Z_B)\ket{\psi}_{AB}\ket{00}+X_B(1+Z_A)(1-Z_B)\ket{\psi}_{AB}\ket{01}\nonumber\\
 &&\phantom{\frac{1}{4}\Bigg[}+X_A(1-Z_A)(1+Z_B)\ket{\psi}_{AB}\ket{10}+X_AX_B(1-Z_A)(1-Z_B)\ket{\psi}_{AB}\ket{11}]
\end{eqnarray}
Using the self-testing properties in  Eq.~(\ref{circuitself}) of the main text,  we can rewrite Eq.~(\ref{phic}) as follows.
\begin{eqnarray}
    \Phi(\ket{\psi}_{AB}\otimes\ket{00}_{A'B'} 
 )&=&\ket{\chi}_{AB}\otimes\ket{\phi^+}_{A'B'}
\end{eqnarray}
where $\ket{\chi}_{AB}=\frac{1+Z_A}{\sqrt{2}}\ket{\psi}_{AB}$. This directly depicts the self-testing of a two-qubit maximally entangled state using the optimal quantum value of the Bell functional $\mathscr{C}_{n}$. Note that we can consider 
\begin{eqnarray}\label{ABxz}
   A_i&=&\sin\qty(\frac{(i-1)\pi}{n})Z_A + \cos\qty(\frac{(i-1)\pi}{n})X_A, \ B_i=\sin\qty(\frac{(2i-1)\pi}{2\ n})Z_B + \cos\qty(\frac{(2i-1)\pi}{2\ n})X_B,\ \forall i\in [n]
\end{eqnarray}
Clearly, here each  $A_i$ and $B_i$ can be written  in terms of $Z_A, Z_B, X_A$, and $ X_B$. It is sufficient to demonstrate the function of local isometry for $Z_A, Z_B, X_A$, and $ X_B$. Hence, from the self-testing circuit, we get  
\begin{eqnarray}\label{d1}
    \Phi(X_A\ket{\psi}_{AB}\otimes\ket{00}_{A'B'} 
 )&=&\frac{1}{4}\Bigg[(1+Z_A)X_A(1+Z_B)\ket{\psi}_{AB}\ket{00}+X_B(1+Z_A)X_A(1-Z_B)\ket{\psi}_{AB}\ket{01}\nonumber\\
 &&+X_A(1-Z_A)X_A(1+Z_B)\ket{\psi}_{AB}\ket{10}+X_AX_B(1-Z_A)X_A(1-Z_B)\ket{\psi}_{AB}\ket{11}]
\end{eqnarray}
Using $X_A\ket{\psi}_{AB}=X_B\ket{\psi}_{AB}$, $Z_A\ket{\psi}_{AB}=Z_B\ket{\psi}_{AB}$ and $\{Z_A,X_A\}\ket{\psi}_{AB}=\{Z_B,X_B\}\ket{\psi}_{AB}=0$  in Eq.~(\ref{d1}) we get
\begin{eqnarray}\label{D2}
    \Phi(X_A\ket{\psi}_{AB}\otimes\ket{00}_{A'B'})=\frac{1+Z_A}{\sqrt{2}}\ket{\psi}_{AB}\otimes\frac{\ket{10}+\ket{01}}{\sqrt{2}}=\ket{\chi}_{AB}\otimes (\sigma_x\otimes\openone_d)\ket{\phi^+}_{A'B'}
\end{eqnarray}
In a similar way, we can further show that
\begin{eqnarray}\label{D3} 
\Phi(X_B\ket{\psi}_{AB}\otimes\ket{00}_{A'B'})&=&\frac{1+Z_A}{\sqrt{2}}\ket{\psi}_{AB}\otimes\frac{\ket{01}+\ket{10}}{\sqrt{2}}=\ket{\chi}_{AB}\otimes (\openone_d\otimes\sigma_x)\ket{\phi^+}_{A'B'}\\\Phi(Z_A\ket{\psi}_{AB}\otimes\ket{00}_{A'B'})&=&\frac{1+Z_A}{\sqrt{2}}\ket{\psi}_{AB}\otimes\frac{\ket{00}-\ket{11}}{\sqrt{2}}=\ket{\chi}_{AB}\otimes (\sigma_z\otimes\openone_d)\ket{\phi^+}_{A'B'}\label{D4}\\ \Phi(Z_B\ket{\psi}_{AB}\otimes\ket{00}_{A'B'})&=&\frac{1+Z_A}{\sqrt{2}}\ket{\psi}_{AB}\otimes\frac{\ket{00}-\ket{11}}{\sqrt{2}}=\ket{\chi}_{AB}\otimes (\openone_d\otimes\sigma_z)\ket{\phi^+}_{A'B'}\label{D5}
\end{eqnarray}
As we can express $A_i$ and $B_i$ interms of $Z_A, Z_B, X_A,$ and $X_B$,  the similar  relations hold for $A_i$ and $B_j$, for each $i,j \in [n]$. Hence, using   Eq.~(\ref{D2}),(\ref{D3}),(\ref{D4}) and (\ref{D5}), we can show that
\begin{eqnarray}
\Phi(A_i\ket{\psi}_{AB}\otimes\ket{00}_{A'B'} 
 )&=&\ket{\chi}_{AB}\otimes (A'_i\otimes \openone_d)\ket{\phi^+}_{A'B'}\\
\Phi(B_j\ket{\psi}_{AB}\otimes\ket{00}_{A'B'} 
 )&=&\ket{\chi}_{AB}\otimes (\openone_d\otimes B'_j)\ket{\phi^+}_{A'B'}\\
 \Phi(A_i\otimes B_j\ket{\psi}_{AB}\otimes\ket{00}_{A'B'} 
 )&=&\ket{\chi}_{AB}\otimes (A'_i\otimes B'_j)\ket{\phi^+}_{A'B'}
\end{eqnarray}
The above equations self-test the set of $n$ measurement settings  $A_i$ and $B_j$ (for each $i,j\in[n])$ for Alice and Bob, respectively.
%%%%%%%%%%%%%%%%%%%%%%%%%%%%%%%%%%%%%%%%%%%%%%%%%%%%%%%%%%%%%%%%%%%%%%%%%%%%%%%%%%%%%%%%%%%%%%%%
\subsection{For even \texorpdfstring{$n$}{n}}\label{c4}
Using the optimization condition  in as Eq.~(\ref{selfn}) of the main text, for $i=1$ and $n$, we can write 
    \begin{eqnarray}\label{Le1}
        \mathcal{A}_1\otimes B_1\ket{\psi}_{AB}&=&\ket{\psi}_{AB}\\
   \label{Len}
        \mathcal{A}_n\otimes B_n\ket{\psi}_{AB}&=&\ket{\psi}_{AB}
    \end{eqnarray}
    Multiplying $\openone_d\otimes B_n B_1$ and $\openone_d\otimes B_1 B_n$ from the left side of the Eq.~(\ref{Le1}) and Eq.~(\ref{Len}) respectively, we get,
    \begin{eqnarray}\label{EA1B4}
        \mathcal{A}_1\otimes B_n\ket{\psi}_{AB}&=&\openone_d\otimes B_n B_1\ket{\psi}_{AB}\\
        \label{EA4B1}
        \mathcal{A}_n\otimes B_1\ket{\psi}_{AB}&=&\openone_d\otimes B_1 B_n\ket{\psi}_{AB}
    \end{eqnarray}
    Since the optimization condition provides $\langle\{\mathcal{A}_i,\mathcal{A}_{n+1-i}\}\rangle=\langle\{B_i,B_{n+1-i}\}\rangle$ for each $i\in[n]$,  adding Eq.~(\ref{Le1}),(\ref{Len}); and then subtracting Eq.~(\ref{EA1B4}), (\ref{EA4B1}) we get the following.
    \ba
        &&\mathcal{A}_1\otimes B_1+\mathcal{A}_n\otimes B_n-\mathcal{A}_1\otimes B_n-\mathcal{A}_n\otimes B_1\ket{\psi}_{AB}= (2\openone_d\otimes\openone_d-\openone_d\otimes\{B_1,B_n\})\ket{\psi}_{AB}\nonumber\\\na 
        &&\bra{\psi}_{AB}\mathcal{A}_1\otimes B_1+\mathcal{A}_n\otimes B_n-\mathcal{A}_1\otimes B_n-\mathcal{A}_n\otimes B_1\ket{\psi}_{AB}= (2-\langle\{B_1,B_n\}\rangle)\nonumber\\\na
        &&\bra{\psi}_{AB}\frac{\mathcal{A}_1\otimes B_1+\mathcal{A}_n\otimes B_n-\mathcal{A}_1\otimes B_n-\mathcal{A}_n\otimes B_1}{(2-\langle\{B_1,B_n\}\rangle)}\ket{\psi}_{AB}= 1\nonumber\\
        &&\frac{\mathcal{A}_1-\mathcal{A}_n}{\sqrt{2-\langle\{\mathcal{A}_1,\mathcal{A}_n\}\rangle}}\otimes \frac{B_1-B_n}{\sqrt{2-\langle\{B_1,B_n\}\rangle}}\ket{\psi}_{AB}=\ket{\psi}_{AB},   \quad \nonumber\ea 
    which implies that \ba     \left(\frac{\mathcal{A}_1-\mathcal{A}_n}{\sqrt{2-\langle\{\mathcal{A}_1,\mathcal{A}_n\}\rangle}}\otimes \openone_d\right)\ket{\psi}_{AB}&=&\left(\openone_d\otimes \frac{B_1-B_n}{\sqrt{2-\langle\{B_1,B_n\}\rangle}}\right)\ket{\psi}_{AB}
    \ea
Following the similar steps, we also get for any pair $\mathcal{A}_i,\mathcal{A}_{n+1-i}, \  i\in[n]$ we get
\ba
    \left(\frac{\mathcal{A}_i-\mathcal{A}_{n+1-i}}{\sqrt{2-\langle\{\mathcal{A}_i,\mathcal{A}_{n+1-i}\}\rangle}}\otimes \openone_d\right)\ket{\psi}_{AB}&=&\left(\openone_d\otimes \frac{B_i-B_{n+1-i}}{\sqrt{2-\langle\{B_i,B_{n+1-i}\}\rangle}}\right)\ket{\psi}_{AB}
\ea 
Adding  for each $i\in[\frac{n}{2}]$ and simplifying it we get
 \begin{eqnarray}
     \frac{1}{\frac{n}{2}}\left(\sum_{i=1}^{ \frac{n}{2}}\frac{\mathcal{A}_{i}-\mathcal{A}_{n+1-i}}{\sqrt{2-\langle\{\mathcal{A}_{i},\mathcal{A}_{n+1-i}\}\rangle}}\otimes \openone_d\right)\ket{\psi}_{AB}&=&\left(\openone_d\otimes  \frac{1}{\frac{n}{2}}\sum_{i=1}^{ \frac{n}{2}}\frac{B_{i}-B_{n+1-i}}{\sqrt{2-\langle\{B_{i},B_{n+1-i}\}\rangle}}\right)\ket{\psi}_{AB}\label{ect4}
 \end{eqnarray}
 Hence Eq.~(\ref{ect4}) satisfies  the following condition
 \begin{eqnarray}
     X_A\ket{\psi}_{AB}=X_B\ket{\psi}_{AB}
 \end{eqnarray}
 where  $X_A$ and $X_B$ are given by 
\begin{eqnarray}
    X_A &=&\frac{2}{n}\sum_{i=1}^{ \frac{n}{2}}\frac{\mathcal{A}_{i}-\mathcal{A}_{n+1-i}}{\sqrt{2-\langle\{\mathcal{A}_{i},\mathcal{A}_{n+1-i}\}\rangle}}, \ 
    X_B = \frac{2}{n}\sum_{i=1}^{ \frac{n}{2}}\frac{B_{i}-B_{n+1-i}}{\sqrt{2-\langle\{B_{i},B_{n+1-i}\}\rangle}}
\end{eqnarray}
Now adding Eq.~(\ref{Le1})-(\ref{EA4B1}) and following the similar process, we get
    \begin{eqnarray}\left(\frac{\mathcal{A}_1+\mathcal{A}_n}{\sqrt{2+\langle\{\mathcal{A}_1,\mathcal{A}_n\}\rangle}}\otimes \openone_d\right)\ket{\psi}_{AB}&=&\left(\openone_d\otimes \frac{B_1+B_n}{\sqrt{2+\langle\{B_1,B_n\}\rangle}}\right)\ket{\psi}_{AB}\label{E+A1A4B1B4}
    \end{eqnarray}
Similarly for any pair $\mathcal{A}_i,\mathcal{A}_{n+1-i}, \  i\in[n]$ we get
\ba
    \left(\frac{\mathcal{A}_i+\mathcal{A}_{n+1-i}}{\sqrt{2+\langle\{\mathcal{A}_i,\mathcal{A}_{n+1-i}\}\rangle}}\otimes \openone_d\right)\ket{\psi}_{AB}&=&\left(\openone_d\otimes \frac{B_i+B_{n+1-i}}{\sqrt{2+\langle\{B_i,B_{n+1-i}\}\rangle}}\right)\ket{\psi}_{AB}
\ea 
Adding  for each $i\in[\frac{n}{2}]$ and simplifying it we get
 \begin{eqnarray} \label{e+ct4}
     \frac{1}{\frac{n}{2}}\left(\sum_{i=1}^{ \frac{n}{2}}\frac{\mathcal{A}_{i}+\mathcal{A}_{n+1-i}}{\sqrt{2+\langle\{\mathcal{A}_{i},\mathcal{A}_{n+1-i}\}\rangle}}\otimes \openone_d\right)\ket{\psi}_{AB}&=&\left(\openone_d\otimes  \frac{1}{\frac{n}{2}}\sum_{i=1}^{ \frac{n}{2}}\frac{B_{i}+B_{n+1-i}}{\sqrt{2+\langle\{B_{i},B_{n+1-i}\}\rangle}}\right)\ket{\psi}_{AB}
 \end{eqnarray}
 Hence Eq.~(\ref{e+ct4}) satisfies the following condition
 \begin{eqnarray}
     Z_A\ket{\psi}_{AB}=Z_B\ket{\psi}_{AB}
 \end{eqnarray}
 where  $Z_A$ and $Z_B$ are given by 
\begin{eqnarray}
    Z_A &=&\frac{2}{n}\sum_{i=1}^{ \frac{n}{2}}\frac{\mathcal{A}_{i}+\mathcal{A}_{n+1-i}}{\sqrt{2+\langle\{\mathcal{A}_{i},\mathcal{A}_{n+1-i}\}\rangle}}, \ 
    Z_B = \sum_{i=1}^{ \frac{n}{2}}\frac{B_{i}+B_{n+1-i}}{\sqrt{2+\langle\{B_{i},B_{n+1-i}\}\rangle}}
\end{eqnarray}
We can further show that $\{X_A,Z_A\}\ket{\psi}_{AB}=\{X_B,Z_B\}\ket{\psi}_{AB}=0$. \textcolor{red}{}Now the state and the observables can be  self-tested by defining a local unitary operation $\Phi$ and ancillary state $\ket{00}_{A'B'}$ such that
\begin{eqnarray}\label{phi'c}
    \Phi(X_A\ket{\psi}_{AB}\otimes\ket{00}_{A'B'} 
 )&=&\frac{1}{4}\Bigg[(1+Z_A)X_A(1+Z_B)\ket{\psi}_{AB}\ket{00}+X_B(1+Z_A)X_A(1-Z_B)\ket{\psi}_{AB}\ket{01}\nonumber\\
 &&+X_A(1-Z_A)X_A(1+Z_B)\ket{\psi}_{AB}\ket{10}+X_AX_B(1-Z_A)X_A(1-Z_B)\ket{\psi}_{AB}\ket{11}]
\end{eqnarray} 
Using the self-testing properties in  Eq.~(\ref{circuitself}) of the main text,  we can rewrite Eq.~(\ref{phi'c}) as follows.
\begin{eqnarray}
    \Phi(\ket{\psi}_{AB}\otimes\ket{00}_{A'B'} 
 )&=&\ket{\chi}_{AB}\otimes\ket{\phi^+}_{A'B'}
\end{eqnarray}
where $\ket{\chi}_{AB}=\frac{1+Z_A}{\sqrt{2}}\ket{\psi}_{AB}$. Again, considering the observables, as in Eq. (\ref{ABxz}) and following a similar procedure, we get 
\begin{eqnarray}
&&\Phi(A_i\ket{\psi}_{AB}\otimes\ket{00}_{A'B'} 
 )=\ket{\chi}_{AB}\otimes (A'_i\otimes \openone_d)\ket{\phi^+}_{A'B'}\\
&&\Phi(B_j\ket{\psi}_{AB}\otimes\ket{00}_{A'B'} 
 )=\ket{\chi}_{AB}\otimes (\openone_d\otimes B'_j)\ket{\phi^+}_{A'B'}\\
 &&\Phi(A_i\otimes B_j\ket{\psi}_{AB}\otimes\ket{00}_{A'B'} 
 )=\ket{\chi}_{AB}\otimes (A'_i\otimes B'_j)\ket{\phi^+}_{A'B'}
\end{eqnarray}
This set of equations self-tests the set of $n$ measurements for  Alice and Bob.
%%%%%%%%%%%%%%%%%%%%%%%%%%%%%%%%%%%%%%%%%%%%%%%%%%%%%%%%%%%%%%%%%%%%%%%%%
\section{Detailed calculations for robust self-testing of state and observables}\label{robustselftesting}
Let us assume that the observables $\Tilde{X}_m$ and $ \Tilde{Z}_m$  are deviated from the real observables $X_m$ and  $ Z_m$  with the margins $\epsilon_m$ and $\alpha_m$ respectively. Then  mathematically, we can write 
\begin{eqnarray}\label{rob}
    ||(\Tilde{X}_m-X_m)\ket{\psi}_{AB}||\leq \alpha_m\Rightarrow  \Tilde{X}_m\approx \alpha_m \openone_d + X_m \ ; \ \ \ ||(\Tilde{Z}_m-Z_m)\ket{\psi}_{AB}||\leq \beta_m \Rightarrow  \Tilde{Z}_m\approx \beta_m \openone_d + Z_m \ ;\ \ \forall m\in\{A, B\}
\end{eqnarray}
 where $||X_m\ket{\psi}_{AB}||=||Z_m\ket{\psi}_{AB}||=1$. 
We derive the following relations using Eq.~(\ref{rob}). 
\ba\label{robzz}
    ||(\Tilde{Z}_m \Tilde{Z}_p-Z_m Z_p)\ket{\psi}_{AB}||&\approx& ||(\Tilde{Z}_m (\beta_p \openone_d + Z_p)-Z_m Z_p)\ket{\psi}_{AB}||\nonumber\\
    &\leq& \beta_p ||\Tilde{Z}_m\ket{\psi}_{AB}||+||(\Tilde{Z}_m-Z_m)Z_p\ket{\psi}_{AB}||\quad (\text{Using Tri-angular inequality})\nonumber\\
    &\leq& \beta_p ||\Tilde{Z}_m\ket{\psi}_{AB}||+\beta_m\ea 
Similarly, we can also derive  \ba \label{robxz}
    ||(\Tilde{X}_m \Tilde{Z}_p-X_m Z_p)\ket{\psi}_{AB}||&\approx& ||(\Tilde{X}_m (\beta_p \openone_d + Z_p))-X_m Z_p)\ket{\psi}_{AB}||
    \leq \beta_p ||\Tilde{X}_m\ket{\psi}_{AB}||+\alpha_m
\\ 
    ||(\Tilde{X}_m \Tilde{X}_p-X_m X_p)\ket{\psi}_{AB}||&\approx& ||(\Tilde{X}_m (\alpha_p \openone_d + X_p))-X_m X_p)\ket{\psi}_{AB}||
    \leq \alpha_p ||\Tilde{X}_m\ket{\psi}_{AB}||+\alpha_m\label{robxx}\ea 
Now using Eq.~(\ref{robxz}), we calculate the following. 
\ba \label{robxzz}
    ||(\Tilde{X}_m \Tilde{Z}_p\Tilde{Z}_k-X_m Z_p Z_k)\ket{\psi}_{AB}||&\approx& ||(\Tilde{X}_m \Tilde{Z}_p(\beta_k \openone_d + Z_k))-X_m Z_p Z_k)\ket{\psi}_{AB}||\nonumber\\
    &\leq& \beta_k ||\Tilde{X}_m\Tilde{Z}_p\ket{\psi}_{AB}||+||(\Tilde{X}_m\Tilde{Z}_p-X_mZ_p)Z_k\ket{\psi}_{AB}||\quad (\text{Using Tri-angular inequality})\nonumber\\
    &\leq& \beta_k ||\Tilde{X}_m\Tilde{Z}_p\ket{\psi}_{AB}||+\beta_p ||\Tilde{X}_m\ket{\psi}_{AB}||+\alpha_m
\end{eqnarray}
Similarly, using Eq.~(\ref{robxx}), we get 
\begin{eqnarray}\label{robxxz}
    ||(\Tilde{X}_m \Tilde{X}_p\Tilde{Z}_k-X_m X_p Z_k)\ket{\psi}_{AB}||&\approx& ||(\Tilde{X}_m \Tilde{X}_p(\beta_k \openone_d + Z_k))-X_m X_p Z_k)\ket{\psi}_{AB}||\nonumber\\
    &\leq& \beta_k ||\Tilde{X}_m \Tilde{X}_p\ket{\psi}_{AB}||+||(\Tilde{X}_m \Tilde{X}_p-X_m X_p) Z_k\ket{\psi}_{AB}||\quad (\text{Using Tri-angular inequality})\nonumber\\
    &\leq& \beta_k ||\Tilde{X}_m \Tilde{X}_p\ket{\psi}_{AB}||+\alpha_p ||\Tilde{X}_m\ket{\psi}_{AB}||+\alpha_m 
\end{eqnarray}  Using Eq.~(\ref{robxxz}), we get 
\begin{eqnarray}\label{robxxzz}
    ||(\Tilde{X}_m \Tilde{X}_p\Tilde{Z}_k\Tilde{Z}_l-X_m X_p Z_k Z_l)\ket{\psi}_{AB}||&\approx& ||(\Tilde{X}_m \Tilde{X}_p\Tilde{Z}_k(\beta_l \openone_d + Z_l))-X_m X_p Z_k Z_l)\ket{\psi}_{AB}||\nonumber\\
    &\leq& \beta_l ||\Tilde{X}_m \Tilde{X}_p\Tilde{Z}_k\ket{\psi}_{AB}||+||(\Tilde{X}_m \Tilde{X}_p\Tilde{Z}_k-X_m X_p Z_k) Z_l\ket{\psi}_{AB}||\quad (\text{Using Tri-angular inequality})\nonumber\\
    &\leq& \beta_l ||\Tilde{X}_m \Tilde{X}_p \Tilde{Z}_k\ket{\psi}_{AB}||+\beta_k ||\Tilde{X}_m \Tilde{X}_p\ket{\psi}_{AB}||+\alpha_p ||\Tilde{X}_m\ket{\psi}_{AB}||+\alpha_m
\end{eqnarray}
\footnote{Tri-angular inequality: $||M\pm N||\leq||M||+||N||$}
\subsection{Robust self-testing of the state}
For the perfect  implementation of the isometry $\Phi$, the output state  can be written as 
\ba\label{robs}
\Phi(\ket{\psi}_{AB}\otimes\ket{00}_{A'B'})&=&\frac{1}{4}\sum_{a,b\in\{0,1\}} (X_A)^a (X_B)^b  \Big(1+(-1)^a Z_A\Big) \Big(1+(-1)^b Z_B\Big) \ket{\psi}_{AB} \ket{ab}_{A'B'}\ea 
Similarly, for the imperfect implementation of the isometry, the output state  can be written as  
\ba \Tilde{\Phi}(\ket{\psi}_{AB}\otimes\ket{00}_{A'B'})&=&\frac{1}{4}\sum_{a,b\in\{0,1\}} (\Tilde{X}_A)^a (\Tilde{X}_B)^b \Big(1+(-1)^a \Tilde{Z}_A\Big) \Big(1+(-1)^b \Tilde{Z}_B\Big) \ket{\psi}_{AB} \ket{ab}_{A'B'}
\ea
This, in turn, provides that 
\begin{eqnarray}\label{robs1}
    ||\Tilde{\Phi}(\ket{\psi}_{AB}\otimes\ket{00}_{A'B'})-\Phi(\ket{\psi}_{AB}\otimes\ket{00}_{A'B'})||&=&\frac{1}{4}\sum_{a,b\in\{0,1\}} \Bigg|\Bigg|\Big[(\Tilde{X}_A)^a (\Tilde{X}_B)^b \Big(1+(-1)^a \Tilde{Z}_A\Big) \Big(1+(-1)^b \Tilde{Z}_B\Big)\nonumber\\
    &&\phantom{\frac{1}{4}\sum_{a,b\in\{0,1\}} \Bigg|\Bigg|\Big[}-(X_A)^a (X_B)^b \Big(1+(-1)^a Z_A\Big) \Big(1+(-1)^b Z_B\Big) \Big] \ket{\psi}_{AB} \ket{ab}_{A'B'}\Bigg|\Bigg|
\end{eqnarray}
Putting  $a=0, b=0$ in Eq. (\ref{robs1}) and using Eq.~(\ref{rob}),(\ref{robzz}),  we have 
\begin{eqnarray}\label{robs2}
   \Bigg|\Bigg|\Big[(1+ \Tilde{Z}_A) (1+ \Tilde{Z}_B)
    -(1+ Z_A) (1+ Z_B) \Big] \ket{\psi}_{AB} \ket{00}_{A'B'}\Bigg|\Bigg|\nonumber
    &\leq& \Big[||(\Tilde{Z}_A-Z_A)\ket{\psi}_{AB}||+||(\Tilde{Z}_B-Z_B)\ket{\psi}_{AB}||+||(\Tilde{Z}_A\Tilde{Z}_B-Z_A Z_B)\ket{\psi}_{AB}||\Big]\ket{00}_{A'B'}\nonumber\\
    &\leq&\Big[2\beta_A+\beta_B+\beta_B||\Tilde{Z}_A\ket{\psi}_{AB}||\Big]\ket{00}_{A'B'}\nonumber\\
    &\leq& f_1(\beta_A,\beta_B) \ket{00}_{A'B'}
\end{eqnarray}
Similarly, putting  $a=0, b=1$ in Eq. (\ref{robs1}) and using Eq.~(\ref{rob}), (\ref{robxz}) and (\ref{robxzz}),  we have 
\begin{eqnarray}\label{robs3}
    \Bigg|\Bigg|\Big[\Tilde{X}_B(1+ \Tilde{Z}_A) (1-\Tilde{Z}_B)
    -X_B(1+ Z_A) (1-Z_B)\Big] \ket{\psi}_{AB} \ket{01}_{A'B'}\Bigg|\Bigg|
    &\leq&\Big[\beta_B\qty(||\Tilde{X}_B\ket{\psi}_{AB}||+||\Tilde{X}_B\Tilde{Z}_A\ket{\psi}_{AB}||)+4 \alpha_B+2\beta_A ||\Tilde{X}_B\ket{\psi}_{AB}||\Big]\ket{01}_{A'B'}\nonumber\\
    &=& f_2(\alpha_B,\beta_A,\beta_B) \ket{01}_{A'B'}
\end{eqnarray}
Putting $a=1, b=0$ in Eq. (\ref{robs1}) and using Eq.~(\ref{rob}),(\ref{robxz}) and(\ref{robxzz}),  we have 
\begin{eqnarray}\label{robs4}
\Bigg|\Bigg|\Big[\Tilde{X}_A(1-\Tilde{Z}_A) (1+\Tilde{Z}_B)
    -X_A(1-Z_A) (1+Z_B)\Big] \ket{\psi}_{AB} \ket{10}_{A'B'}\Bigg|\Bigg|
    &\leq&\Big[\beta_A\qty(||\Tilde{X}_A\ket{\psi}_{AB}||+||\Tilde{X}_A\Tilde{Z}_B\ket{\psi}_{AB}||)+4 \alpha_A+2\beta_B ||\Tilde{X}_A\ket{\psi}_{AB}||\Big]\ket{10}_{A'B'}\nonumber\\
    &=& f_3(\alpha_A,\beta_A,\beta_B) \ket{10}_{A'B'}
\end{eqnarray}
Similarly, putting  $a=1, b=1$ in Eq. (\ref{robs1}) and using Eq.~(\ref{rob}), (\ref{robxz}), (\ref{robxx}), (\ref{robxzz}) and (\ref{robxxzz}), we have
\begin{eqnarray}\label{robs5}
   \Bigg|\Bigg|\Big[\Tilde{X}_A\Tilde{X}_B (1-\Tilde{Z}_A) (1-\Tilde{Z}_B)
    -X_A X_B (1-Z_A) (1-Z_B)\Big] \ket{\psi}_{AB} \ket{11}_{A'B'}\Bigg|\Bigg|\nonumber
    &=&\Big[\beta_B\qty(||\Tilde{X}_A\Tilde{X}_B\ket{\psi}_{AB}||+||\Tilde{X}_A\Tilde{X}_B\Tilde{Z}_A\ket{\psi}_{AB}||)+2\beta_A||\Tilde{X}_A\Tilde{X}_B\ket{\psi}_{AB}||\nonumber\\
    && +4\alpha_B||\Tilde{X}_A\ket{\psi}_{AB}||+4\alpha_A\Big]\ket{11}_{A'B'}\nonumber\\
    &\leq& f_4(\alpha_A,\alpha_B,\beta_A,\beta_B) \ket{11}_{A'B'}
\end{eqnarray}
 Substituting Eq.~(\ref{robs2})-(\ref{robs5})  in Eq.~(\ref{robs1}) and neglecting the ancillary part, we get
\begin{eqnarray}\label{robsF}
    ||\Tilde{\Phi}(\ket{\psi}_{AB}\otimes\ket{00}_{A'B'})-\Phi(\ket{\psi}_{AB}\otimes\ket{00}_{A'B'})||&\leq&\frac{1}{4}\Big[f_1(\beta_A,\beta_B)+f_2(\alpha_B,\beta_A,\beta_B)+f_3(\alpha_A,\beta_A,\beta_B)+f_4(\alpha_A,\alpha_B,\beta_A,\beta_B)\Big]
    = F_S(\alpha_A,\alpha_B,\beta_A,\beta_B)\nonumber
\end{eqnarray}
Clearly here, we have $\lim\limits_{\{\alpha_A,\alpha_B,\beta_A,\beta_B\}\to 0} F_S(\alpha_A,\alpha_B,\beta_A,\beta_B)=0$,    
which in turn provides \ba||\Tilde{\Phi}(\ket{\psi}_{AB}\otimes\ket{00}_{A'B'})||\approx ||\Phi(\ket{\psi}_{AB}\otimes\ket{00}_{A'B'})||\ea 
%%%%%%%%%%%%%%%%%%%%%%%%%%%%%%%%%%%%%%%%%%%%%%%%%%%%%%%%%%%%%%%%%%%%%%%%%%%%%%%%%%%%%%%%%%%%%%%%%%%%%%%%%%%%%%%%%%%%%%%%%%%%%%%%%%%%%%%%%%%%%%%%%%%%%%%%%%%%%%%%%%%%%

\subsection{Robust self-testing of observables}
In order to find the robustness of the observables, we follow a similar procedure as stated above. Thus we use Eq.~(\ref{robs1}) and calculate the robustness with observable  $X_m$ ($\forall m\in\{A, B\}$) as follows.
\begin{eqnarray}\label{robob1}
    &&||\Tilde{\Phi}(\Tilde{X}_m\ket{\psi}_{AB}\otimes\ket{00}_{A'B'})-\Phi(X_m\ket{\psi}_{AB}\otimes\ket{00}_{A'B'})||\nonumber\\
    &=&\frac{1}{4}\sum_{a,b\in\{0,1\}} \Bigg|\Bigg|\Big[(\Tilde{X}_A)^a (\Tilde{X}_B)^b \Big(1+(-1)^a \Tilde{Z}_A\Big) \Big(1+(-1)^b \Tilde{Z}_B\Big) \Tilde{X}_m\ket{\psi}_{AB} -(X_A)^a (X_B)^b  \Big(1+(-1)^a Z_A\Big) \Big(1+(-1)^b Z_B\Big) X_m\ket{\psi}_{AB}\Big]  \ket{ab}_{A'B'}\Bigg|\Bigg|\nonumber\\
    &\approx&\frac{1}{4}\sum_{a,b\in\{0,1\}} \Bigg|\Bigg|\Big[(\Tilde{X}_A)^a (\Tilde{X}_B)^b \Big(1+(-1)^a \Tilde{Z}_A\Big) \Big(1+(-1)^b \Tilde{Z}_B\Big) \Big(\alpha_m\openone_d + X_m\Big)\ket{\psi}_{AB} -(X_A)^a (X_B)^b  \Big(1+(-1)^a Z_A\Big) \Big(1+(-1)^b Z_B\Big) X_m\ket{\psi}_{AB}\Big]  \ket{ab}_{A'B'}\Bigg|\Bigg|\nonumber\\
     &\leq&\frac{\alpha_m}{4}\sum_{a,b\in\{0,1\}} \Bigg|\Bigg|(\Tilde{X}_A)^a (\Tilde{X}_B)^b \Big(1+(-1)^a \Tilde{Z}_A\Big) \Big(1+(-1)^b \Tilde{Z}_B\Big)\ket{\psi}_{AB}\ket{ab}_{A'B'}\Bigg|\Bigg|\nonumber\\
     &&+ \frac{1}{4}\sum_{a,b\in\{0,1\}} \Bigg|\Bigg|\Big[(\Tilde{X}_A)^a (\Tilde{X}_B)^b \Big(1+(-1)^a \Tilde{Z}_A\Big) \Big(1+(-1)^b \Tilde{Z}_B\Big)-(X_A)^a (X_B)^b  \Big(1+(-1)^a Z_A\Big) \Big(1+(-1)^b Z_B\Big)\Big]X_m \ket{\psi}_{AB} \ket{ab}_{A'B'}\Bigg|\Bigg|\\
     &\leq&\frac{\alpha_m}{4}\sum_{a,b\in\{0,1\}} \Bigg|\Bigg|(\Tilde{X}_A)^a (\Tilde{X}_B)^b \Big(1+(-1)^a \Tilde{Z}_A\Big) \Big(1+(-1)^b \Tilde{Z}_B\Big)\ket{\psi}_{AB}\ket{ab}_{A'B'}\Bigg|\Bigg|\na \\
     &&+ \frac{1}{4}\sum_{a,b\in\{0,1\}} \Bigg|\Bigg|\Big[(\Tilde{X}_A)^a (\Tilde{X}_B)^b \Big(1+(-1)^a \Tilde{Z}_A\Big) \Big(1+(-1)^b \Tilde{Z}_B\Big)-(X_A)^a (X_B)^b  \Big(1+(-1)^a Z_A\Big) \Big(1+(-1)^b Z_B\Big)\Big]\ket{\psi}_{AB} \ket{ab}_{A'B'}\Bigg|\Bigg|\nonumber\\
    &\leq&\frac{\alpha_m}{4}\sum_{a,b\in\{0,1\}} \Bigg|\Bigg|(\Tilde{X}_A)^a (\Tilde{X}_B)^b \Big(1+(-1)^a \Tilde{Z}_A\Big) \Big(1+(-1)^b \Tilde{Z}_B\Big)\ket{\psi}_{AB}\ket{ab}_{A'B'}\Bigg|\Bigg|+\Big|\Big|\Tilde{\Phi}(\ket{\psi}_{AB}\otimes\ket{00}_{A'B'})-\Phi(\ket{\psi}_{AB}\otimes\ket{00}_{A'B'})\Big|\Big|\nonumber\\\na 
    &=&F_{O_X}(\alpha_A,\alpha_B,\beta_A,\beta_B)
\end{eqnarray}
where  $F_{O_X}(\alpha_A,\alpha_B,\beta_A,\beta_B)=F(\alpha_m) + F_S(\alpha_A,\alpha_B,\beta_A,\beta_B),\forall m\in\{A,B\}$.  Note that here, we have \ba 
    \lim\limits _{\{\alpha_A,\alpha_B,\beta_A,\beta_B\}\to 0}  F_{O_X}(\alpha_A,\alpha_B,\beta_A,\beta_B)=0\ea  which in turn provides 
\ba ||\Tilde{\Phi}(\Tilde{X}_m\ket{\psi}_{AB}\otimes\ket{00}_{A'B'})||\approx ||\Phi(X_m\ket{\psi}_{AB}\otimes\ket{00}_{A'B'})||\ea \\
Similarly,  we again use   Eq.~(\ref{robs1}) and calculate the robustness for the other observable $Z_m$ as follows. 
 \begin{eqnarray}\label{robob2}
    &&||\Tilde{\Phi}(\Tilde{Z}_m\ket{\psi}_{AB}\otimes\ket{00}_{A'B'})-\Phi(Z_m\ket{\psi}_{AB}\otimes\ket{00}_{A'B'})||\nonumber\\
    &=&\frac{1}{4}\sum_{a,b\in\{0,1\}} \Bigg|\Bigg|\Big[(\Tilde{X}_A)^a (\Tilde{X}_B)^b \Big(1+(-1)^a \Tilde{Z}_A\Big) \Big(1+(-1)^b \Tilde{Z}_B\Big) \Tilde{Z}_m\ket{\psi}_{AB} -(X_A)^a (X_B)^b  \Big(1+(-1)^a Z_A\Big) \Big(1+(-1)^b Z_B\Big) Z_m\ket{\psi}_{AB}\Big]  \ket{ab}_{A'B'}\Bigg|\Bigg|\nonumber\\
    &\approx&\frac{1}{4}\sum_{a,b\in\{0,1\}} \Bigg|\Bigg|\Big[(\Tilde{X}_A)^a (\Tilde{X}_B)^b \Big(1+(-1)^a \Tilde{Z}_A\Big) \Big(1+(-1)^b \Tilde{Z}_B\Big) (\beta_m\openone_d + Z_m)\ket{\psi}_{AB} -(X_A)^a (X_B)^b  \Big(1+(-1)^a Z_A\Big) \Big(1+(-1)^b Z_B\Big) Z_m\ket{\psi}_{AB}\Big]  \ket{ab}_{A'B'}\Bigg|\Bigg|\nonumber\\
     &\leq&\frac{\beta_m}{4}\sum_{a,b\in\{0,1\}} \Bigg|\Bigg|(\Tilde{X}_A)^a (\Tilde{X}_B)^b \Big(1+(-1)^a \Tilde{Z}_A\Big) \Big(1+(-1)^b \Tilde{Z}_B\Big)\ket{\psi}_{AB}\ket{ab}_{A'B'}\Bigg|\Bigg|\nonumber\\
     &&+ \frac{1}{4}\sum_{a,b\in\{0,1\}} \Bigg|\Bigg|\Big[(\Tilde{X}_A)^a (\Tilde{X}_B)^b (1+(-1)^a \Tilde{Z}_A) (1+(-1)^b \Tilde{Z}_B)-(X_A)^a (X_B)^b  (1+(-1)^a Z_A) (1+(-1)^b Z_B)\Big]Z_m \ket{\psi}_{AB} \ket{ab}_{A'B'}\Bigg|\Bigg|\nonumber\\
     &\leq&\frac{\beta_m}{4}\sum_{a,b\in\{0,1\}} \Bigg|\Bigg|(\Tilde{X}_A)^a (\Tilde{X}_B)^b \Big(1+(-1)^a \Tilde{Z}_A\Big) \Big(1+(-1)^b \Tilde{Z}_B\Big)\ket{\psi}_{AB}\ket{ab}_{A'B'}\Bigg|\Bigg|\\
     &&+ \frac{1}{4}\sum_{a,b\in\{0,1\}} \Bigg|\Bigg|\Big[(\Tilde{X}_A)^a (\Tilde{X}_B)^b \Big(1+(-1)^a \Tilde{Z}_A\Big) \Big(1+(-1)^b \Tilde{Z}_B\Big)-(X_A)^a (X_B)^b  (1+(-1)^a Z_A) (1+(-1)^b Z_B)\Big]\ket{\psi}_{AB} \ket{ab}_{A'B'}\Bigg|\Bigg|\nonumber\\
    &\leq&\frac{\beta_m}{4}\sum_{a,b\in\{0,1\}} \Bigg|\Bigg|(\Tilde{X}_A)^a (\Tilde{X}_B)^b \Big(1+(-1)^a \Tilde{Z}_A\Big) \Big(1+(-1)^b \Tilde{Z}_B\Big)\ket{\psi}_{AB}\ket{ab}_{A'B'}\Bigg|\Bigg|+\Big|\Big|\Tilde{\Phi}(\ket{\psi}_{AB}\otimes\ket{00}_{A'B'})-\Phi(\ket{\psi}_{AB}\otimes\ket{00}_{A'B'})\Big|\Big| \nonumber\\
    &=&  F_{O_Z}(\alpha_A,\alpha_B,\beta_A,\beta_B) \quad (m\in\{A,B\})
\end{eqnarray}
where $F_{O_Z}(\alpha_A,\alpha_B,\beta_A,\beta_B)=F(\beta_m) + F_S(\alpha_A,\alpha_B,\beta_A,\beta_B),\forall m\in\{A,B\}$. Note that here we have \ba  \lim\limits_{\{\alpha_A,\alpha_B,\beta_A,\beta_B\}\to 0} F_{O_Z}(\alpha_A,\alpha_B,\beta_A,\beta_B)=0\ea  which in turn provides  
\begin{eqnarray}
||\Tilde{\Phi}(\Tilde{Z}_m\ket{\psi}_{AB}\otimes\ket{00}_{A'B'})||\approx ||\Phi(Z_m\ket{\psi}_{AB}\otimes\ket{00}_{A'B'})||
\end{eqnarray}
Thus, we have provided robust self-testing of the state and the set of observables of Alice and Bob.
%%%%%%%%%%%%%%%%%%%%%%%%%%%%%%%%%%%%%%%%%%%%%%%%%%%%%%%%%%%%%%%%%%%%%%%%%%%%%%%%%
\subsection{Special case: For odd \textit{n} only second party (Bob) implements imperfect observables}\label{spodd}

If we consider only second-party (Bob) implements the imperfect observables, and  the error in both is the same, i.e.,  $\alpha_B=\beta_B=\epsilon\geq 0$ then \ba  ||(\Tilde{X}_B-X_B)\ket{\psi}_{AB}||\leq \epsilon,\ ||(\Tilde{Z}_B-Z_B)\ket{\psi}_{AB}||\leq \epsilon\ea Again, if the error of each observable of the second party is $\delta$ then \ba ||(\Tilde{B}_i-B_i)\ket{\psi}_{AB}||\leq \delta\implies \Tilde{B}_i\approx B_i+\delta \ \openone_d, \ \forall i\in[n], \ \delta \geq 0\ea \tcr{}According to \cite{Bamps2015},  in the presence of noise, the observable is  $\Tilde{Z}_B=\Tilde{B}_{\frac{n+1}{2}}$, which implies that 
\ba ||(\Tilde{Z}_B-Z_B)\ket{\psi}_{AB}||=||(\Tilde{B}_{\frac{n+1}{2}}-B_{\frac{n+1}{2}})\ket{\psi}_{AB}||\leq \delta \Rightarrow  \delta\approx\epsilon\ea  Hence, the operator $\mathcal{L}_{n,i}$ of Eq. (\ref{lni}) from the main text becomes $\Tilde{\mathcal{L}}_{n,i}$, and thus we get 

\begin{eqnarray}\label{noisygamman}
    \Tr[\Tilde{\Gamma}_n\  \rho_{AB}]&=& \frac{1}{2}\sum_{i=1}^{n} \nu_{n,i} \bra{\psi}_{AB}\Tilde{\mathcal{L}}_{n,i}^{\dag} \Tilde{\mathcal{L}}_{n,i}\ket{\psi}_{AB}= \frac{1}{2}\sum_{i=1}^{n} \nu_{n,i}  \ \xi_{n,i}^2=\xi
\end{eqnarray}
where $\Tilde{\mathcal{L}}_{n,i}$ defined as 
\ba \Tilde{\mathcal{L}}_{n,i}&=&\mathcal{A}_i \otimes \openone_d -\openone_d \otimes \Tilde{B}_i, \forall i\in[n]\quad \qty(\text{where $\mathcal{A}_i=\frac{A_i+A_{i+1}}{\nu_{n,i}}$})
\ea 
which further implies that 
\begin{eqnarray}\na 
    ||\Tilde{\mathcal{L}}_{n,i}-\mathcal{L}_{n,i}\ket{\psi}_{AB}||&\leq&||\openone_d\otimes(\Tilde{B}_i-B_i)\ket{\psi}_{AB}||\leq \epsilon \\
    \Tilde{\mathcal{L}}_{n,i} &\approx& \mathcal{L}_{n,i}+\openone_d\otimes\epsilon \openone_d
\end{eqnarray}
Hence, we get  
\begin{eqnarray}\label{noisyLij} \Tr[\Tilde{\mathcal{L}}_{n,i}^{\dag} \Tilde{\mathcal{L}}_{n,i}\rho_{AB}]&=&\langle (\mathcal{L}_{n,i}^\dagger+\openone_d \otimes \epsilon \openone_d)(\mathcal{L}_{n,i}+\openone_d \otimes \epsilon \openone_d)\rangle_{\rho_{AB}}\nonumber\\
    &=& \langle \mathcal{L}_{n,i}^\dagger \mathcal{L}_{n,i} \rangle_{\rho_{AB}}+\epsilon \langle \mathcal{L}_{n,i}^\dagger \rangle_{\rho_{AB}}+\epsilon \langle \mathcal{L}_{n,i} \rangle_{\rho_{AB}}+\epsilon^2
\end{eqnarray}
Since in the ideal scenario $\langle \mathcal{L}_{n,i}^\dagger \mathcal{L}_{n,i} \rangle_{\rho_{AB}}=\langle \mathcal{L}_{n,i}^\dagger\rangle_{\rho_{AB}}=\langle \mathcal{L}_{n,i} \rangle_{\rho_{AB}}=0$, we have

\ba \Tr[\Tilde{\mathcal{L}}_{n,i}^{\dag} \Tilde{\mathcal{L}}_{n,i}\rho_{AB}]&=&\epsilon^2 
\end{eqnarray}
The optimization condition for deriving $(\mathscr{
C}_{n})_{Q}^{opt}$ gives $\nu_{n,i}=2\cos{\frac{\pi}{2n}}$. Using  $\nu_{n,i}$ and Eq.~(\ref{noisyLij}) in Eq.~(\ref{noisygamman}),  we get 
\begin{eqnarray}\label{Efun}
    \frac{1}{2}\sum_{i=1}^{n} \nu_{n,i} \bra{\psi}_{AB}\Tilde{\mathcal{L}}_{n,i}^{\dag} \Tilde{\mathcal{L}}_{n,i}\ket{\psi}_{AB}=\xi\Rightarrow n \ \epsilon^2 \ \cos{\frac{\pi}{2n}}=\xi\Rightarrow \epsilon =\sqrt{\frac{\xi }{n \cos \left(\frac{\pi }{2 n}\right)}}
\end{eqnarray}
In this case, the robust self-testing of the state from Eq.~(\ref{robs1}) provides 
\begin{eqnarray}\label{stateep}
    ||\Tilde{\Phi}(\ket{\psi}_{AB}&\otimes&\ket{00}_{A'B'})-\Phi(\ket{\psi}_{AB}\otimes\ket{00}_{A'B'})||\nonumber\\
    &=&\frac{1}{4}\sum_{a,b\in\{0,1\}} \Bigg|\Bigg|\Big[({X}_A)^a (\Tilde{X}_B)^b \Big(1+(-1)^a {Z}_A\Big) \Big(1+(-1)^b \Tilde{Z}_B\Big) -(X_A)^a (X_B)^b \Big(1+(-1)^a Z_A\Big) \Big(1+(-1)^b Z_B\Big) \Big] \ket{\psi}_{AB} \ket{ab}_{A'B'}\Bigg|\Bigg|\nonumber\\
    &\leq&\frac{1}{4}\sum_{b\in\{0,1\}} \Bigg|\Bigg|\Big[\qty( (\Tilde{X}_B)^b  \Big(1+(-1)^b \Tilde{Z}_B\Big) - (X_B)^b  \Big(1+(-1)^b Z_B\Big))\sum_{a\in\{0,1\}}\Big(1+(-1)^a {Z}_A\Big)\Big] \ket{\psi}_{AB} \ket{ab}_{A'B'}\Bigg|\Bigg|\nonumber\\
    &\leq&\frac{1}{4}\Bigg[\sum_{b\in\{0,1\}} \Bigg|\Bigg|\qty( (\Tilde{X}_B)^b  \Big(1+(-1)^b \Tilde{Z}_B\Big) - (X_B)^b  \Big(1+(-1)^b Z_B\Big))\sum_{a\in\{0,1\}} \ket{\psi}_{AB} \ket{ab}_{A'B'}\Bigg|\Bigg|\nonumber\\
    &&\phantom{\frac{1}{4}\Bigg[}+\sum_{b\in\{0,1\}} \Bigg|\Bigg|\qty( (\Tilde{X}_B)^b  \Big(1+(-1)^b \Tilde{Z}_B\Big) - (X_B)^b  \Big(1+(-1)^b Z_B\Big))\sum_{a\in\{0,1\}}(-1)^a {Z}_A \ket{\psi}_{AB} \ket{ab}_{A'B'}\Bigg|\Bigg|\Bigg]\nonumber\\
    &\leq&\frac{1}{4}\Bigg[\sum_{b\in\{0,1\}} 2\Bigg|\Bigg|\qty( (\Tilde{X}_B)^b  \Big(1+(-1)^b \Tilde{Z}_B\Big) - (X_B)^b  \Big(1+(-1)^b Z_B\Big))\sum_{a\in\{0,1\}} \ket{\psi}_{AB} \ket{ab}_{A'B'}\Bigg|\Bigg|\nonumber\\
     &\leq&\sum_{b\in\{0,1\}} \Bigg|\Bigg|\Big[ (\Tilde{X}_B)^b  \Big(1+(-1)^b \Tilde{Z}_B\Big) - (X_B)^b  \Big(1+(-1)^b Z_B\Big)\Big] \ket{\psi}_{AB} \Bigg|\Bigg|\ \ \ \ \qty( \text{Neglecting the ancillary part  $\ket{ab}_{A'B'}$} )\nonumber\\
     &\leq&\sum_{b\in\{0,1\}} \Bigg|\Bigg|\Big[\qty(( \Tilde{X}_B)^b-(X_B)^b) + (-1)^b \qty((\Tilde{X}_B)^b \Tilde{Z}_B-(X_B)^b Z_B)\Big] \ket{\psi}_{AB} \Bigg|\Bigg|\ \ \ \nonumber\\
     &\leq& ||(\Tilde{X}_B-X_B)\ket{\psi}_{AB}||+||(\Tilde{Z}_B- Z_B)\ket{\psi}_{AB}||+
     ||(\Tilde{X}_B \Tilde{Z}_B-X_B Z_B)\ket{\psi}_{AB}||
\end{eqnarray}
Again we get 
\begin{eqnarray}\label{xbzbsp}
    ||(\Tilde{X}_B \Tilde{Z}_B-X_B Z_B)\ket{\psi}_{AB}||&\leq&||(\Tilde{X}_B (Z_B+\epsilon \openone_d)_B-X_B Z_B)\ket{\psi}_{AB}||\nonumber\\
    &\leq& ||(\Tilde{X}_B-X_B)\ket{\psi}_{AB}||+\epsilon||\Tilde{X}_B\ket{\psi}_{AB}||\nonumber\\
    &\leq& \epsilon+\epsilon||(\epsilon\openone_d+X_B)\ket{\psi}_{AB}||\nonumber\\
    &\leq&2\epsilon +\epsilon^2
\end{eqnarray}
Substituting Eq.~(\ref{xbzbsp})  in Eq.~(\ref{stateep}) and using $ ||(\Tilde{X}_B-X_B)\ket{\psi}_{AB}||\leq \epsilon,||(\Tilde{Z}_B-Z_B)\ket{\psi}_{AB}||\leq \epsilon$, we get 
\begin{eqnarray}\label{staterobep}
    ||\Tilde{\Phi}(\ket{\psi}_{AB}\otimes\ket{00}_{A'B'})-\Phi(\ket{\psi}_{AB}\otimes\ket{00}_{A'B'})||\leq 4\epsilon +\epsilon^2
\end{eqnarray}
Similarly, the  robust self-testing of the observable from Eq.~(\ref{robob1}) provides 

\begin{eqnarray}
&&||\Tilde{\Phi}(\Tilde{X}_B\ket{\psi}_{AB}\otimes\ket{00}_{A'B'})-\Phi(X_B\ket{\psi}_{AB}\otimes\ket{00}_{A'B'})||\nonumber\\
    &&\leq\frac{\epsilon}{4}\sum_{a,b\in\{0,1\}} \Bigg|\Bigg|(X_A)^a (\Tilde{X}_B)^b \Big(1+(-1)^a Z_A\Big) \Big(1+(-1)^b \Tilde{Z}_B\Big)\ket{\psi}_{AB}\ket{ab}_{A'B'}\Bigg|\Bigg|+\Big|\Big|\Tilde{\Phi}(\ket{\psi}_{AB}\otimes\ket{00}_{A'B'})-\Phi(\ket{\psi}_{AB}\otimes\ket{00}_{A'B'})\Big|\Big|\nonumber\\
    &&\leq\frac{\epsilon}{4}\sum_{b\in\{0,1\}} \Bigg|\Bigg| (\Tilde{X}_B)^b  \Big(1+(-1)^b \Tilde{Z}_B\Big)\sum_{a\in\{0,1\}} (X_A)^a \Big(1+(-1)^a Z_A\Big)\ket{\psi}_{AB}\ket{ab}_{A'B'}\Bigg|\Bigg|+4\epsilon+\epsilon^2\nonumber\\
    &&\leq\frac{\epsilon}{4}\sum_{b\in\{0,1\}} \Bigg|\Bigg| (\Tilde{X}_B)^b  \Big(1+(-1)^b \Tilde{Z}_B\Big) (1+Z_A+X_A-X_A Z_A)\ket{\psi}_{AB}\ket{ab}_{A'B'}\Bigg|\Bigg|+4\epsilon+\epsilon^2\nonumber\\
    &&\leq\epsilon\sum_{b\in\{0,1\}} \Bigg|\Bigg| (\Tilde{X}_B)^b  \Big(1+(-1)^b \Tilde{Z}_B\Big) \ket{\psi}_{AB}\Bigg|\Bigg|+4\epsilon+\epsilon^2\nonumber\ \ \ \ \ \qty( \text{Neglecting the ancillary part  $\ket{ab}_{A'B'}$} )\\
    &&\leq \epsilon \Bigg|\Bigg| \qty(1+\Tilde{Z}_B +\Tilde{X}_B-\Tilde{X}_B\Tilde{Z}_B) \ket{\psi}_{AB}\Bigg|\Bigg|+4\epsilon+\epsilon^2\nonumber\\
     &&\leq \epsilon \qty(1+||\Tilde{X}_B \ket{\psi}_{AB}||+||\Tilde{Z}_B \ket{\psi}_{AB}||+||\Tilde{X}_B\Tilde{Z}_B \ket{\psi}_{AB}||)+4\epsilon+\epsilon^2\nonumber\\
     &&\leq \epsilon \qty(1+||(X_B+\epsilon \openone_d) \ket{\psi}_{AB}||+||(Z_B+\epsilon \openone_d)\ket{\psi}_{AB}||+||(X_B+\epsilon \openone_d)(Z_B+\epsilon \openone_d) \ket{\psi}_{AB}||)+4\epsilon+\epsilon^2\nonumber\\
      &&\leq \epsilon^3+5\epsilon^2+8\epsilon
\end{eqnarray}
Similarly, 
\begin{eqnarray}
      &&||\Tilde{\Phi}(\Tilde{Z}_B\ket{\psi}_{AB}\otimes\ket{00}_{A'B'})-\Phi(Z_B\ket{\psi}_{AB}\otimes\ket{00}_{A'B'})||\leq \epsilon^3+5\epsilon^2+8\epsilon
\end{eqnarray}
Re-collecting all the results
\begin{eqnarray}
    &&||\Tilde{\Phi}(\ket{\psi}_{AB}\otimes\ket{00}_{A'B'})-\Phi(\ket{\psi}_{AB}\otimes\ket{00}_{A'B'})||\leq 4\epsilon +\epsilon^2\\
    &&||\Tilde{\Phi}(\Tilde{O}\ket{\psi}_{AB}\otimes\ket{00}_{A'B'})-\Phi(O\ket{\psi}_{AB}\otimes\ket{00}_{A'B'})||\leq \epsilon^3+5\epsilon^2+8\epsilon\quad \forall O\in\{X_B,Z_B\}
\end{eqnarray}
and the relation between $\epsilon$ and the observed deviation $\xi$ i.e.,  $\epsilon =\sqrt{\frac{\xi }{n \cos \left(\frac{\pi }{2 n}\right)}}$, we determine the robust self-testing bounds for both state and observables.  We define the relative observed violations $r$ as a function of the number of inputs $n$ and observed deviation $\xi$ as follows.
\begin{eqnarray}
    &&r=\frac{\Tilde{(\mathscr{C}_{n})}_{Q}-(\mathscr{C}_n)_C}{(\mathscr{C}_{n})^{opt}_{Q}-(\mathscr{C}_n)_C}= 1-\frac{\xi}{2 n \cos{\frac{\pi}{2n}}-2n+2}
\end{eqnarray}
Alternatively, $\xi$ can be written in terms of $r$ and $n$ as
\begin{eqnarray}
        &&\xi =(1-r) \left(2 n \cos \left(\frac{\pi }{2 n}\right)-2n+2)\right)
\end{eqnarray}
where $\Tilde{(\mathscr{C}_{n})}_{Q}$ and $ (\mathscr{C}_{n})^{opt}_{Q}$ are defined in Eqs.~(\ref{Cn noise}) and (\ref{cnopt}) respectively. Here $(\mathscr{C}_n)_c =2n-2$ is the classical bound of the chained Bell inequality. 
Hence, the trace distance between the observed and ideal state in terms of the relative operator $r$ is
\begin{eqnarray}
  && ||\Tilde{\Phi}(\ket{\psi}_{AB}\otimes\ket{00}_{A'B'})-\Phi(\ket{\psi}_{AB}\otimes\ket{00}_{A'B'})||\leq f_s(r) \\ &&f_s(r)=\frac{2 (n-1) (r-1) \sec \left(\frac{\pi }{2 n}\right)}{n}+4 \sqrt{2} \sqrt{\frac{(r-1) \left((n-1) \sec \left(\frac{\pi }{2 n}\right)-n\right)}{n}}-2 (r-1)
\end{eqnarray}
 Now, according to Fuchs-Van de Graaf, the approximate relation between the trace distance $f_s(r)$ and robust fidelity $F_s(r)$ is 
 \begin{eqnarray}
     2\qty(1-\sqrt{F_s(r)})\leq f_s(r)\leq 2\sqrt{1-F_s(r)}
 \end{eqnarray}
 This gives  the lower bound of fidelity in terms of trace distance as 
 \begin{eqnarray}
     F_s(r)\geq \qty(1-\frac{1}{2}f_s(r))^2
 \end{eqnarray}
Similarly, for the observables, we can define the robust  fidelity $F_o(r)$ in terms of trace distance $f_o(r)$, i.e., 
  \begin{eqnarray}
     F_o(r)\geq \qty(1-\frac{1}{2}f_o(r))^2
 \end{eqnarray}
 where \ba f_o(r)=2 \sqrt{2} \left(\dfrac{(r-1) \left((n-1) \sec \left(\frac{\pi }{2 n}\right)-n\right)}{n}\right)^{3/2}+\dfrac{10 (r-1) \left((n-1) \sec \left(\frac{\pi }{2 n}\right)-n\right)}{n}+8 \sqrt{2} \sqrt{\dfrac{(r-1) \left((n-1) \sec \left(\frac{\pi }{2 n}\right)-n\right)}{n}}\ea

\begin{figure}
    \centering
    \includegraphics[width=12cm, height=4.5cm]{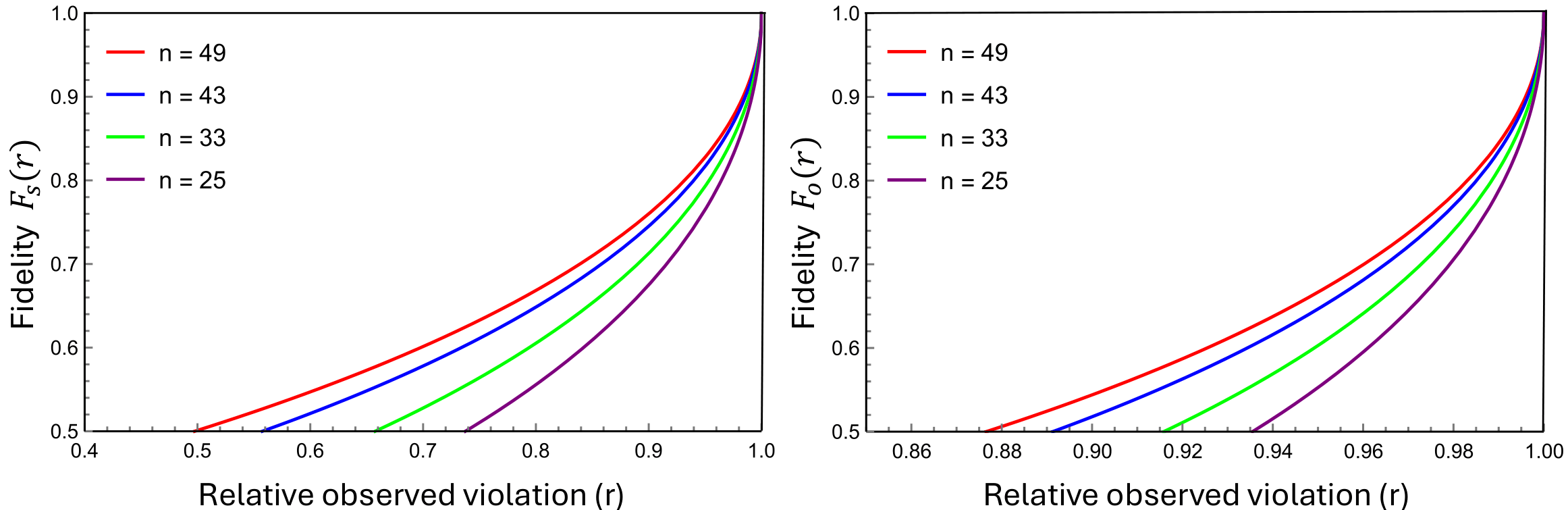}
    \caption{Trade-off between relative observed violation ($r$) and the fidelity for state $F_s(r)$ and observer $F_o(r), \ \forall O\in\{X_B,Z_B\}$ with respect to the measurement settings ($n$). Here, we are taking only that region where fidelity converges to one. As fidelity can't be greater than one.}\label{highn}
\end{figure} 
%%%%%%%%%%%%%%%%%%%%%%%%%%%%

%%%%%%%%%%%%%%%%%%%%%%%%%%%%%%%%%%%%%%%%%%%%%%%%%%%%%%%%
\section{Detailed calculations for randomness}\label{randoma}
We demonstrate two bits of maximum randomness based on the chained Bell inequality. For an odd $n$, one choice of measurement settings for Alice and Bob is $A_i$ and $B_{i+\frac{n-1}{2}}$, which ensures that there are no additional measurement settings to maximize randomness. Now, we evaluate $\bra{\psi_{AB}}A_i\otimes B_{i+\frac{n-1}{2}}\ket{\psi}_{AB}, \forall i\in[n]$. By putting $j=i+\frac{n-1}{2}$ means $x=\frac{n-1}{2}$ in Eq.~(\ref{rand0}), we get
\begin{eqnarray}\label{maxr}
    \bra{\psi}_{AB}A_i\otimes B_{i+\frac{n-1}{2}}\ket{\psi}_{AB}
    &=&\frac{1}{2}\bra{\psi}_{AB}\qty(\frac{\{A_i,A_{i+\frac{n-1}{2}}\}+\{A_i,A_{i+\frac{n+1}{2}}\}}{2\cos{\frac{\pi}{2n}}})\otimes\openone_d\ket{\psi}_{AB}=\frac{2\cos{\frac{\pi(n-1)}{2n}}+2\cos{\frac{\pi(n+1)}{2n}}}{4\cos{\frac{\pi}{2n}}}=0
\end{eqnarray}
This means from Eq.~(\ref{probaibj}) that  $P\left(a,b|A_i,B_{i+\frac{n-1}{2}}\right)=\frac{1}{4}$, which implies $\mathcal{R}_{max}=2$.

Now, one may ask whether $\bra{\psi}_{AB}A_i\otimes B_{i+\frac{n-1}{2}}\ket{\psi}_{AB}$ correlations are self-tested. We answer this question with an example where we consider the chained Bell test for $n=3$ and we take $i=2$. Hence,

\begin{eqnarray}\label{randa2b3}
    \bra{\psi}_{AB}A_2\otimes B_3\ket{\psi}_{AB}&=&\bra{\psi}_{AB}\frac{A_1+A_3}{\sqrt{3}}\otimes B_3\ket{\psi}_{AB}\quad (\text{optimization condition for n=3})\nonumber\\
    &=& \frac{\bra{\psi}_{AB}A_1\otimes B_3\ket{\psi}_{AB}+\bra{\psi}_{AB}A_3\otimes B_3\ket{\psi}_{AB}}{\sqrt{3}}
\end{eqnarray}
These two expectation values are present in the actual Bell functional for $n=3$, so they are already self-tested. We may then argue that though the correlation required for achieving maximum randomness are not included in the Bell functional, they can still be self-tested via the optimal quantum violation of chain Bell inequality.  Again, from Eq.~(\ref{rand0}), we know that $\bra{\psi}_{AB}A_1\otimes B_3\ket{\psi}_{AB}=\frac{-\cos{\frac{\pi}{3}}-1}{2 \cos{\frac{\pi}{6}}}$ and $\bra{\psi}_{AB}A_3\otimes B_3\ket{\psi}_{AB}=\frac{\cos{\frac{\pi}{3}}+1}{2 \cos{\frac{\pi}{6}}}$. Hence, putting these in Eq.~(\ref{randa2b3}) we get
\begin{eqnarray}
    \bra{\psi}_{AB}A_2\otimes B_3\ket{\psi}_{AB}&=&\bra{\psi}_{AB}\frac{A_1+A_3}{\sqrt{3}}\otimes B_3\ket{\psi}_{AB}=0
\end{eqnarray}
which further implies $P(a,b|A_2,B_3)=\frac{1}{4}$. Hence, we get the maximum randomness $\mathcal{R}_{max}=2$.
%%%%%%%%%%%%%%%%%%%%%%%%%%%%%%%%%%%%%%%%%%%%%%%%%%%%%%%%%
\subsection{Derivation of  Eq.~(\ref{trcos}) i.e., \texorpdfstring{$A_i\otimes B_i\ket{\psi}_{AB} = A_{i+1}\otimes B_i\ket{\psi}_{AB}$}{Equation for Ai Bi Psi}}
Putting $j=i$ (i.e., $x=0$) in Eq.~(\ref{rand0}) we get
\begin{eqnarray}\label{ra2}
    \bra{\psi}_{AB}A_i\otimes B_i\ket{\psi}_{AB}
    &=&\frac{1}{2}\bra{\psi}_{AB}\qty(\frac{2\openone_d+\{A_i,A_{i+1}\}}{2\cos{\frac{\pi}{2n}}})\otimes\openone_d\ket{\psi}_{AB}=\frac{1+\cos{\frac{\pi}{n}}}{2\cos{\frac{\pi}{2n}}}
\end{eqnarray}
Following a similar way, Putting $i=i+1$ and $j=i$ in Eq.~(\ref{rand0'}) we get 
\begin{eqnarray}\label{ra3}
    \bra{\psi}_{AB}A_{i+1}\otimes B_i\ket{\psi}_{AB}
    &=&\frac{1}{2}\bra{\psi}_{AB}\qty(\frac{2\openone_d+\{A_i,A_{i+1}\}}{2\cos{\frac{\pi}{2n}}})\otimes\openone_d\ket{\psi}_{AB}=\frac{1+\cos{\frac{\pi}{n}}}{2\cos{\frac{\pi}{2n}}}
\end{eqnarray}
Hence from Eq.~(\ref{ra2}) and (\ref{ra3}),  we get
\begin{eqnarray}
    \bra{\psi}_{AB}A_i\otimes B_i\ket{\psi}_{AB}&=&\bra{\psi}_{AB}A_{i+1}\otimes B_i\ket{\psi}_{AB}\nonumber\\
    A_i\otimes B_i\ket{\psi}_{AB}&=&A_{i+1}\otimes B_i\ket{\psi}_{AB}\label{r72}
\end{eqnarray}
\subsection{Robustness of certified randomness }\label{Rob RC}
In the experimental scenario, achieving the optimal quantum violation is nearly impossible due to noisy systems. In this scenario, we introduce equal noise parameters for each of the observables of Bob, while assuming the state is perfect. Thus, Bell's functional takes the form
 \ba  \Tilde{\mathscr{C}_{n}}=\sum\limits_{i=1}^{n}(A_i+A_{i+1}
 )\Tilde{B}_i\leq {2n-2}
  \ea
 
We can write the noisy observables in the following form
\begin{eqnarray}
    ||(\Tilde{B}_i-B_i)\ket{\psi}_{AB}||\leq \epsilon&&\implies  \Tilde{B}_i \approx B_i+ \epsilon \openone_d\nonumber\\
    &&\implies \Tilde{B}_i= \frac{ B_i+ \epsilon \openone_d}{\sqrt{1+\epsilon^2}} \quad (\text{where $\Tr[B_i \ \rho_{AB}]=0$})
\end{eqnarray}
where $\Tilde{B}_i$ is a dichotomic observable, hence we include the normalization part which turns $\Tilde{B}_i= \frac{ B_i+ \epsilon \openone_d}{\sqrt{1+\epsilon^2}}$ as $\Tr[B_i \ \rho_{AB}]=0$. For a given measurement settings $A_i$ and $B_j$, the probability $P\left(a,b|A_i,\Tilde{B}_j\right)$ is defined as follows.

\begin{eqnarray}
P\left(a,b|A_i,\Tilde{B}_j\right)&=&\Tr[ \bigg(\frac{\openone_d+a A_i}{2}\otimes \frac{\openone_d+b\Tilde{B}_j}{2}\bigg)\ \rho_{AB}]\quad (\text{where $a,b\in \pm 1$})\nonumber\\
&=&\frac{\sqrt{1+\epsilon^2}+b\epsilon+a b \langle A_i\otimes B_j\rangle_{\rho_{AB}}}{4\sqrt{1+\epsilon^2}} \quad (\text{As $\langle A_i\rangle_{\rho_{AB}}=\langle B_j\rangle_{\rho_{AB}}=0$})
\end{eqnarray}
where from Eq.~(\ref{Efun}) $\epsilon =\sqrt{\frac{\xi }{n \cos \left(\frac{\pi }{2 n}\right)}}$. Now, from Eq.~(\ref{probaibj}) the maximum probability we can obtain

\begin{eqnarray}
&&P\left(a,b|A_i,\Tilde{B}_j\right)=\frac{\sqrt{1+\epsilon^2}+b\epsilon+a b \langle A_i\otimes B_j\rangle_{\rho_{AB}}}{4\sqrt{1+\epsilon^2}}, \quad \forall j= i \ \text{or} \ i+1\nonumber, \\
&&P_{max}\left(a,b|A_i,\Tilde{B}_j\right)=\frac{\sqrt{1+\epsilon^2}+\epsilon+ \cos{\frac{\pi}{2 n}}}{4\sqrt{1+\epsilon^2}},  \quad \forall j= i \ \text{or} \ i+1
\end{eqnarray}
Hence, the least  achievable randomness is given  by 
\begin{eqnarray}
\Tilde{\mathcal{R}}_{\text{min}} = \log_2 \left(\frac{4\sqrt{1+\epsilon^2}}{\sqrt{1+\epsilon^2}+\epsilon+\cos{\frac{\pi}{2n}}}\right)
\end{eqnarray}

Now, we calculate $\bra{\psi_{AB}}A_i\otimes \Tilde{B}_{i+\frac{n-1}{2}}\ket{\psi}_{AB}, \forall i\in[n]$ as follows. Using  Eq.~(\ref{maxr}) $\bra{\psi}_{AB}A_i\otimes B_{i+\frac{n-1}{2}}\ket{\psi}_{AB}=0$ and the relations $\bra{\psi}_{AB}A_i\otimes \openone_d\ket{\psi}_{AB}=\bra{\psi}_{AB}\openone_d\otimes B_{i+\frac{n-1}{2}}\ket{\psi}_{AB}=0$ for the maximally entangled state, we get 
\begin{eqnarray}
&&P\left(a,b|A_i,\Tilde{B}_{i+\frac{n-1}{2}}\right)=\frac{\sqrt{1+\epsilon^2}+b\epsilon+a b \langle A_i\otimes \Tilde{B}_{i+\frac{n-1}{2}}\rangle_{\rho_{AB}}}{4\sqrt{1+\epsilon^2}}\nonumber\\
&&P_{max}\left(a,b|A_i,\Tilde{B}_{i+\frac{n-1}{2}}\right)=\frac{\sqrt{1+\epsilon^2}+\epsilon}{4\sqrt{1+\epsilon^2}}
\end{eqnarray}
Hence, in the presence of noise, the maximum randomness $\mathcal{R}_{\text{max}}$ changes to 
\begin{eqnarray}
\Tilde{\mathcal{R}}_{\text{max}} = \log_2 \left(\frac{4\sqrt{1+\epsilon^2}}{\sqrt{1+\epsilon^2}+\epsilon}\right)
\end{eqnarray}

\end{widetext}

\bibliography{references} 

@article{Rai2021,
  title = {Device-independent bounds from Cabello's nonlocality argument},
  author = {Rai, Ashutosh and Pivoluska, Matej and Plesch, Martin and Sasmal, Souradeep and Banik, Manik and Ghosh, Sibasish},
  journal = {Phys. Rev. A},
  volume = {103},
  issue = {6},
  pages = {062219},
  numpages = {7},
  year = {2021},
  month = {Jun},
  publisher = {American Physical Society},
  doi = {10.1103/PhysRevA.103.062219},
  url = {https://link.aps.org/doi/10.1103/PhysRevA.103.062219}
}

@article{Rai2022,
  title = {Self-testing quantum states via nonmaximal violation in Hardy's test of nonlocality},
  author = {Rai, Ashutosh and Pivoluska, Matej and Sasmal, Souradeep and Banik, Manik and Ghosh, Sibasish and Plesch, Martin},
  journal = {Phys. Rev. A},
  volume = {105},
  issue = {5},
  pages = {052227},
  numpages = {8},
  year = {2022},
  month = {May},
  publisher = {American Physical Society},
  doi = {10.1103/PhysRevA.105.052227},
  url = {https://link.aps.org/doi/10.1103/PhysRevA.105.052227}
}

@article{Wooltron2022,
  title = {Tight Analytic Bound on the Trade-Off between Device-Independent Randomness and Nonlocality},
  author = {Wooltorton, Lewis and Brown, Peter and Colbeck, Roger},
  journal = {Phys. Rev. Lett.},
  volume = {129},
  issue = {15},
  pages = {150403},
  numpages = {6},
  year = {2022},
  month = {Oct},
  publisher = {American Physical Society},
  doi = {10.1103/PhysRevLett.129.150403},
  url = {https://link.aps.org/doi/10.1103/PhysRevLett.129.150403}
}

@article{Singh2025prepare,
doi = {10.1088/1367-2630/ae1f35},
url = {https://doi.org/10.1088/1367-2630/ae1f35},
year = {2025},
month = {nov},
publisher = {IOP Publishing},
volume = {27},
number = {11},
pages = {114520},
author = {Singh, Ritesh K and Sasmal, Souradeep and Nautiyal, S and Pan, A K},
title = {Self-testing in a constrained prepare-measure scenario sans assuming quantum dimension},
journal = {New Journal of Physics}
}

@article{Tavakoli2018,
  title = {Self-testing quantum states and measurements in the prepare-and-measure scenario},
  author = {Tavakoli, Armin and Kaniewski, J\ifmmode \mbox{\k{e}}\else \k{e}\fi{}drzej and V\'ertesi, Tam\'as and Rosset, Denis and Brunner, Nicolas},
  journal = {Phys. Rev. A},
  volume = {98},
  issue = {6},
  pages = {062307},
  numpages = {13},
  year = {2018},
  month = {Dec},
  publisher = {American Physical Society},
  doi = {10.1103/PhysRevA.98.062307},
  url = {https://link.aps.org/doi/10.1103/PhysRevA.98.062307}
}

@Article{Zhang2019,
author={Zhang, Wen-Hao
and Chen, Geng
and Yin, Peng
and Peng, Xing-Xiang
and Hu, Xiao-Min
and Hou, Zhi-Bo
and Zhou, Zhi-Yuan
and Yu, Shang
and Ye, Xiang-Jun
and Zhou, Zong-Quan
and Xu, Xiao-Ye
and Tang, Jian-Shun
and Xu, Jin-Shi
and Han, Yong-Jian
and Liu, Bi-Heng
and Li, Chuan-Feng
and Guo, Guang-Can},
title={Experimental demonstration of robust self-testing for bipartite entangled states},
journal={npj Quantum Information},
year={2019},
month={Jan},
day={11},
volume={5},
number={1},
pages={4},
issn={2056-6387},
doi={10.1038/s41534-018-0120-0},
url={https://doi.org/10.1038/s41534-018-0120-0}
}

@Article{Hu2023,
author={Hu, Xiao-Min
and Xie, Yi
and Arora, Atul Singh
and Ai, Ming-Zhong
and Bharti, Kishor
and Zhang, Jie
and Wu, Wei
and Chen, Ping-Xing
and Cui, Jin-Ming
and Liu, Bi-Heng
and Huang, Yun-Feng
and Li, Chuan-Feng
and Guo, Guang-Can
and Roland, J{\'e}r{\'e}mie
and Cabello, Ad{\'a}n
and Kwek, Leong-Chuan},
title={Self-testing of a single quantum system from theory to experiment},
journal={npj Quantum Information},
year={2023},
month={Oct},
day={19},
volume={9},
number={1},
pages={103},
issn={2056-6387},
doi={10.1038/s41534-023-00769-7},
url={https://doi.org/10.1038/s41534-023-00769-7}
}

@article{Zhang2018,
  title = {Experimentally Robust Self-testing for Bipartite and Tripartite Entangled States},
  author = {Zhang, Wen-Hao and Chen, Geng and Peng, Xing-Xiang and Ye, Xiang-Jun and Yin, Peng and Xiao, Ya and Hou, Zhi-Bo and Cheng, Ze-Di and Wu, Yu-Chun and Xu, Jin-Shi and Li, Chuan-Feng and Guo, Guang-Can},
  journal = {Phys. Rev. Lett.},
  volume = {121},
  issue = {24},
  pages = {240402},
  numpages = {6},
  year = {2018},
  month = {Dec},
  publisher = {American Physical Society},
  doi = {10.1103/PhysRevLett.121.240402},
  url = {https://link.aps.org/doi/10.1103/PhysRevLett.121.240402}
}

@article{Wagner2020,
  doi = {10.22331/q-2020-03-19-243},
  url = {https://doi.org/10.22331/q-2020-03-19-243},
  title = {Device-independent characterization of quantum instruments},
  author = {Wagner, Sebastian and Bancal, Jean-Daniel and Sangouard, Nicolas and Sekatski, Pavel},
  journal = {{Quantum}},
  issn = {2521-327X},
  publisher = {{Verein zur F{\"{o}}rderung des Open Access Publizierens in den Quantenwissenschaften}},
  volume = {4},
  pages = {243},
  month = mar,
  year = {2020}
}

@article{SekatskiPrl2023,
  title = {Toward the Device-Independent Certification of a Quantum Memory},
  author = {Sekatski, Pavel and Bancal, Jean-Daniel and Ioannou, Marie and Afzelius, Mikael and Brunner, Nicolas},
  journal = {Phys. Rev. Lett.},
  volume = {131},
  issue = {17},
  pages = {170802},
  numpages = {6},
  year = {2023},
  month = {Oct},
  publisher = {American Physical Society},
  doi = {10.1103/PhysRevLett.131.170802},
  url = {https://link.aps.org/doi/10.1103/PhysRevLett.131.170802}
}

@article{SekatskiPrl2018,
  title = {Certifying the Building Blocks of Quantum Computers from Bell's Theorem},
  author = {Sekatski, Pavel and Bancal, Jean-Daniel and Wagner, Sebastian and Sangouard, Nicolas},
  journal = {Phys. Rev. Lett.},
  volume = {121},
  issue = {18},
  pages = {180505},
  numpages = {5},
  year = {2018},
  month = {Nov},
  publisher = {American Physical Society},
  doi = {10.1103/PhysRevLett.121.180505},
  url = {https://link.aps.org/doi/10.1103/PhysRevLett.121.180505}
}

@article{Paul2024,
  title = {Self-testing of multiple unsharpness parameters through sequential violations of a noncontextual inequality},
  author = {Paul, Rajdeep and Sasmal, Souradeep and Pan, A. K.},
  journal = {Phys. Rev. A},
  volume = {110},
  issue = {1},
  pages = {012444},
  numpages = {21},
  year = {2024},
  month = {Jul},
  publisher = {American Physical Society},
  doi = {10.1103/PhysRevA.110.012444},
  url = {https://link.aps.org/doi/10.1103/PhysRevA.110.012444}
}

@article{Singh2025,
  title = {Self-testing of the $m$-partite Greenberger-Horne-Zeilinger state and observables using the Svetlichny inequality},
  author = {Singh, Ritesh K. and Sasmal, Souradeep and Pan, A. K.},
  journal = {Phys. Rev. A},
  volume = {112},
  issue = {3},
  pages = {032404},
  numpages = {30},
  year = {2025},
  month = {Sep},
  publisher = {American Physical Society},
  doi = {10.1103/x6ct-vmnx},
  url = {https://link.aps.org/doi/10.1103/x6ct-vmnx}
}

@article{Smania2020,
author = {Massimiliano Smania and Piotr Mironowicz and Mohamed Nawareg and Marcin Paw{\l}owski and Ad\'{a}n Cabello and Mohamed Bourennane},
journal = {Optica},
number = {2},
pages = {123--128},
publisher = {Optica Publishing Group},
title = {Experimental certification of an informationally complete quantum measurement in a device-independent protocol},
volume = {7},
month = {Feb},
year = {2020},
url = {https://opg.optica.org/optica/abstract.cfm?URI=optica-7-2-123},
doi = {10.1364/OPTICA.377959},
}

@article{Gomez2016,
  title = {Device-Independent Certification of a Nonprojective Qubit Measurement},
  author = {G\'omez, Esteban S. and G\'omez, Santiago and Gonz\'alez, Pablo and Ca\~nas, Gustavo and Barra, Johanna F. and Delgado, Aldo and Xavier, Guilherme B. and Cabello, Ad\'an and Kleinmann, Matthias and V\'ertesi, Tam\'as and Lima, Gustavo},
  journal = {Phys. Rev. Lett.},
  volume = {117},
  issue = {26},
  pages = {260401},
  numpages = {5},
  year = {2016},
  month = {Dec},
  publisher = {American Physical Society},
  doi = {10.1103/PhysRevLett.117.260401},
  url = {https://link.aps.org/doi/10.1103/PhysRevLett.117.260401}
}

@article{McKague2012,
doi = {10.1088/1751-8113/45/45/455304},
url = {https://dx.doi.org/10.1088/1751-8113/45/45/455304},
year = {2012},
month = {oct},
publisher = {IOP Publishing},
volume = {45},
number = {45},
pages = {455304},
author = {M McKague and T H Yang and V Scarani},
title = {Robust self-testing of the singlet},
journal = {Journal of Physics A: Mathematical and Theoretical}
}

@article{supic2020input,
doi = {10.1088/1367-2630/ab90d1},
url = {https://dx.doi.org/10.1088/1367-2630/ab90d1},
year = {2020},
month = {jul},
publisher = {IOP Publishing},
volume = {22},
number = {7},
pages = {073006},
author = {Ivan Šupić and Matty J Hoban and Laia Domingo Colomer and Antonio Acín},
title = {Self-testing and certification using trusted quantum inputs},
journal = {New Journal of Physics}
}

@article{Braunstein1990,
title = {Wringing out better Bell inequalities},
journal = {Annals of Physics},
volume = {202},
number = {1},
pages = {22-56},
year = {1990},
issn = {0003-4916},
doi = {https://doi.org/10.1016/0003-4916(90)90339-P},
url = {https://www.sciencedirect.com/science/article/pii/000349169090339P},
author = {Samuel L Braunstein and Carlton M Caves}
}

@article{Wehner2006,
  title = {Tsirelson bounds for generalized Clauser-Horne-Shimony-Holt inequalities},
  author = {Wehner, Stephanie},
  journal = {Phys. Rev. A},
  volume = {73},
  issue = {2},
  pages = {022110},
  numpages = {5},
  year = {2006},
  month = {Feb},
  publisher = {American Physical Society},
  doi = {10.1103/PhysRevA.73.022110},
  url = {https://link.aps.org/doi/10.1103/PhysRevA.73.022110}
}

@article{Gomez2018,
  title = {Experimental nonlocality-based randomness generation with nonprojective measurements},
  author = {G\'omez, S. and Mattar, A. and G\'omez, E. S. and Cavalcanti, D. and Far\'{\i}as, O. Jim\'enez and Ac\'{\i}n, A. and Lima, G.},
  journal = {Phys. Rev. A},
  volume = {97},
  issue = {4},
  pages = {040102},
  numpages = {5},
  year = {2018},
  month = {Apr},
  publisher = {American Physical Society},
  doi = {10.1103/PhysRevA.97.040102},
  url = {https://link.aps.org/doi/10.1103/PhysRevA.97.040102}
}

@article{Jordanlemma,
  author = {Jordan, Camille},
  title = {Essay on Geometry in $n$ Dimensions},
  journal = {Bulletin of the French Mathematical Society},
  pages = {103--174},
  publisher = {French Mathematical Society},
  volume = {3},
  year = {1875},
  doi = {10.24033/bsmf.90},
  language = {en},
  url = {https://www.numdam.org/articles/10.24033/bsmf.90/}
}

@Article{Supic2023,
author={{\v{S}}upi{\'{c}}, Ivan
and Bowles, Joseph
and Renou, Marc-Olivier
and Ac{\'i}n, Antonio
and Hoban, Matty J.},
title={Quantum networks self-test all entangled states},
journal={Nature Physics},
year={2023},
month={May},
day={01},
volume={19},
number={5},
pages={670-675},
issn={1745-2481},
doi={10.1038/s41567-023-01945-4},
url={https://doi.org/10.1038/s41567-023-01945-4}
}

@article{Dhara2013,
  title = {Maximal quantum randomness in Bell tests},
  author = {Dhara, Chirag and Prettico, Giuseppe and Ac\'{\i}n, Antonio},
  journal = {Phys. Rev. A},
  volume = {88},
  issue = {5},
  pages = {052116},
  numpages = {5},
  year = {2013},
  month = {Nov},
  publisher = {American Physical Society},
  doi = {10.1103/PhysRevA.88.052116},
  url = {https://link.aps.org/doi/10.1103/PhysRevA.88.052116}
}

@article{Pironio2009,
doi = {10.1088/1367-2630/11/4/045021},
url = {https://dx.doi.org/10.1088/1367-2630/11/4/045021},
year = {2009},
month = {apr},
publisher = {},
volume = {11},
number = {4},
pages = {045021},
author = {Stefano Pironio and Antonio Acín and Nicolas Brunner and Nicolas Gisin and Serge Massar and Valerio Scarani},
title = {Device-independent quantum key distribution secure against collective attacks},
journal = {New Journal of Physics}
}

@article{Acin2021,
  title = {Bell Nonlocality Is Not Sufficient for the Security of Standard Device-Independent Quantum Key Distribution Protocols},
  author = {Farkas, M\'at\'e and Balanz\'o-Juand\'o, Maria and \L{}ukanowski, Karol and Ko\l{}ody\ifmmode \acute{n}\else \'{n}\fi{}ski, Jan and Ac\'{\i}n, Antonio},
  journal = {Phys. Rev. Lett.},
  volume = {127},
  issue = {5},
  pages = {050503},
  numpages = {6},
  year = {2021},
  month = {Jul},
  publisher = {American Physical Society},
  doi = {10.1103/PhysRevLett.127.050503},
  url = {https://link.aps.org/doi/10.1103/PhysRevLett.127.050503}
}

@article{Agresti2021,
  title = {Experimental Robust Self-Testing of the State Generated by a Quantum Network},
  author = {Agresti, Iris and Polacchi, Beatrice and Poderini, Davide and Polino, Emanuele and Suprano, Alessia and \ifmmode \check{S}\else \v{S}\fi{}upi\ifmmode \acute{c}\else \'{c}\fi{}, Ivan and Bowles, Joseph and Carvacho, Gonzalo and Cavalcanti, Daniel and Sciarrino, Fabio},
  journal = {PRX Quantum},
  volume = {2},
  issue = {2},
  pages = {020346},
  numpages = {19},
  year = {2021},
  month = {Jun},
  publisher = {American Physical Society},
  doi = {10.1103/PRXQuantum.2.020346},
  url = {https://link.aps.org/doi/10.1103/PRXQuantum.2.020346}
}

@article{Acin2012,
  title = {Randomness versus Nonlocality and Entanglement},
  author = {Ac\'{\i}n, Antonio and Massar, Serge and Pironio, Stefano},
  journal = {Phys. Rev. Lett.},
  volume = {108},
  issue = {10},
  pages = {100402},
  numpages = {5},
  year = {2012},
  month = {Mar},
  publisher = {American Physical Society},
  doi = {10.1103/PhysRevLett.108.100402},
  url = {https://link.aps.org/doi/10.1103/PhysRevLett.108.100402}
}

@Article{Coladangelo2017,
author={Coladangelo, Andrea
and Goh, Koon Tong
and Scarani, Valerio},
title={All pure bipartite entangled states can be self-tested},
journal={Nature Communications},
year={2017},
month={May},
day={26},
volume={8},
number={1},
pages={15485},
issn={2041-1723},
doi={10.1038/ncomms15485},
url={https://doi.org/10.1038/ncomms15485}
}

@article{Roy2023,
doi = {10.1088/1367-2630/acb4b5},
url = {https://dx.doi.org/10.1088/1367-2630/acb4b5},
year = {2023},
month = {feb},
publisher = {IOP Publishing},
volume = {25},
number = {1},
pages = {013040},
author = {Prabuddha Roy and A K Pan},
title = {Device-independent self-testing of unsharp measurements},
journal = {New Journal of Physics},
}

@article{Brunner2014,
   author = {Nicolas Brunner and Daniel Cavalcanti and Stefano Pironio and Valerio Scarani and Stephanie Wehner},
   doi = {10.1103/RevModPhys.86.419},
   issn = {15390756},
   issue = {2},
   journal = {Reviews of Modern Physics},
   month = {4},
   pages = {419-478},
   publisher = {American Physical Society},
   title = {Bell nonlocality},
   volume = {86},
   year = {2014},
}

@article{Abhyoudai2023,
  title = {Robust certification of unsharp instruments through sequential quantum advantages in a prepare-measure communication game},
  author = {S. S., Abhyoudai and Mukherjee, Sumit and Pan, A. K.},
  journal = {Phys. Rev. A},
  volume = {107},
  issue = {1},
  pages = {012411},
  numpages = {18},
  year = {2023},
  month = {Jan},
  publisher = {American Physical Society},
  doi = {10.1103/PhysRevA.107.012411},
  url = {https://link.aps.org/doi/10.1103/PhysRevA.107.012411}
}

@article{Kaniewski2016,
  title = {Analytic and Nearly Optimal Self-Testing Bounds for the Clauser-Horne-Shimony-Holt and Mermin Inequalities},
  author = {Kaniewski, J\ifmmode \mbox{\k{e}}\else \k{e}\fi{}drzej},
  journal = {Phys. Rev. Lett.},
  volume = {117},
  issue = {7},
  pages = {070402},
  numpages = {6},
  year = {2016},
  month = {Aug},
  publisher = {American Physical Society},
  doi = {10.1103/PhysRevLett.117.070402},
  url = {https://link.aps.org/doi/10.1103/PhysRevLett.117.070402}
}

@article{Bamps2015,
  title = {Sum-of-squares decompositions for a family of Clauser-Horne-Shimony-Holt-like inequalities and their application to self-testing},
  author = {Bamps, C\'edric and Pironio, Stefano},
  journal = {Phys. Rev. A},
  volume = {91},
  issue = {5},
  pages = {052111},
  numpages = {13},
  year = {2015},
  month = {May},
  publisher = {American Physical Society},
  doi = {10.1103/PhysRevA.91.052111},
  url = {https://link.aps.org/doi/10.1103/PhysRevA.91.052111}
}

@article{Paulunitary,
  title = {Semi-device-independent self-testing of unitary operations},
  author = {Paul, Rajdeep and Roy, Prabuddha and Pan, A. K.},
  journal = {Phys. Rev. A},
  volume = {112},
  issue = {6},
  pages = {062216},
  numpages = {6},
  year = {2025},
  month = {Dec},
  publisher = {American Physical Society},
  doi = {10.1103/94pf-njhr},
  url = {https://link.aps.org/doi/10.1103/94pf-njhr}
}

@article{Clauser1969,
  title = {Proposed Experiment to Test Local Hidden-Variable Theories},
  author = {Clauser, John F. and Horne, Michael A. and Shimony, Abner and Holt, Richard A.},
  journal = {Phys. Rev. Lett.},
  volume = {23},
  issue = {15},
  pages = {880--884},
  numpages = {0},
  year = {1969},
  month = {Oct},
  publisher = {American Physical Society},
  doi = {10.1103/PhysRevLett.23.880},
  url = {https://link.aps.org/doi/10.1103/PhysRevLett.23.880}
}

@Article{Hu2018,
author={Hu, Meng-Jun
and Zhou, Zhi-Yuan
and Hu, Xiao-Min
and Li, Chuan-Feng
and Guo, Guang-Can
and Zhang, Yong-Sheng},
title={Observation of non-locality sharing among three observers with one entangled pair via optimal weak measurement},
journal={npj Quantum Information},
year={2018},
month={Dec},
day={03},
volume={4},
number={1},
pages={63},
issn={2056-6387},
doi={10.1038/s41534-018-0115-x},
url={https://doi.org/10.1038/s41534-018-0115-x}
}

@Article{Sarkar2021,
author={Sarkar, Shubhayan
and Saha, Debashis
and Kaniewski, J{\k{e}}drzej
and Augusiak, Remigiusz},
title={Self-testing quantum systems of arbitrary local dimension with minimal number of measurements},
journal={npj Quantum Information},
year={2021},
month={Oct},
day={14},
volume={7},
number={1},
pages={151},
issn={2056-6387},
doi={10.1038/s41534-021-00490-3},
url={https://doi.org/10.1038/s41534-021-00490-3}
}

@Article{Panwar2023,
author={Panwar, Ekta
and Pandya, Palash
and Wie{\'{s}}niak, Marcin},
title={An elegant scheme of self-testing for multipartite Bell inequalities},
journal={npj Quantum Information},
year={2023},
month={Jul},
day={17},
volume={9},
number={1},
pages={71},
issn={2056-6387},
doi={10.1038/s41534-023-00735-3},
url={https://doi.org/10.1038/s41534-023-00735-3}
}

@article{Acin2020,
  title = {Scalable Bell Inequalities for Qubit Graph States and Robust Self-Testing},
  author = {Baccari, F. and Augusiak, R. and \ifmmode \check{S}\else \v{S}\fi{}upi\ifmmode \acute{c}\else \'{c}\fi{}, I. and Tura, J. and Ac\'{\i}n, A.},
  journal = {Phys. Rev. Lett.},
  volume = {124},
  issue = {2},
  pages = {020402},
  numpages = {6},
  year = {2020},
  month = {Jan},
  publisher = {American Physical Society},
  doi = {10.1103/PhysRevLett.124.020402},
  url = {https://link.aps.org/doi/10.1103/PhysRevLett.124.020402}
}

@article{Xiao2023,
  title = {Device-independent randomness based on a tight upper bound of the maximal quantum value of chained inequality},
  author = {Xiao, Youwang and Li, XinHui and Wang, Jing and Li, Ming and Fei, Shao-Ming},
  journal = {Phys. Rev. A},
  volume = {107},
  issue = {5},
  pages = {052415},
  numpages = {7},
  year = {2023},
  month = {May},
  publisher = {American Physical Society},
  doi = {10.1103/PhysRevA.107.052415},
  url = {https://link.aps.org/doi/10.1103/PhysRevA.107.052415}
}

@article{Bell1964,
  title = {On the Einstein Podolsky Rosen paradox},
  author = {Bell, J. S.},
  journal = {Physics Physique Fizika},
  volume = {1},
  issue = {3},
  pages = {195--200},
  numpages = {6},
  year = {1964},
  month = {Nov},
  publisher = {American Physical Society},
  doi = {10.1103/PhysicsPhysiqueFizika.1.195},
  url = {https://link.aps.org/doi/10.1103/PhysicsPhysiqueFizika.1.195}
}

@article{Acin2007,
  title = {Device-Independent Security of Quantum Cryptography against Collective Attacks},
  author = {Ac\'{\i}n, Antonio and Brunner, Nicolas and Gisin, Nicolas and Massar, Serge and Pironio, Stefano and Scarani, Valerio},
  journal = {Phys. Rev. Lett.},
  volume = {98},
  issue = {23},
  pages = {230501},
  numpages = {4},
  year = {2007},
  month = {Jun},
  publisher = {American Physical Society},
  doi = {10.1103/PhysRevLett.98.230501},
  url = {https://link.aps.org/doi/10.1103/PhysRevLett.98.230501}
}

@article{Sarkar2022,
  title = {Self-testing of multipartite Greenberger-Horne-Zeilinger states of arbitrary local dimension with arbitrary number of measurements per party},
  author = {Sarkar, Shubhayan and Augusiak, Remigiusz},
  journal = {Phys. Rev. A},
  volume = {105},
  issue = {3},
  pages = {032416},
  numpages = {19},
  year = {2022},
  month = {Mar},
  publisher = {American Physical Society},
  doi = {10.1103/PhysRevA.105.032416},
  url = {https://link.aps.org/doi/10.1103/PhysRevA.105.032416}
}

@article{Colbeck2011,
doi = {10.1088/1751-8113/44/9/095305},
url = {https://dx.doi.org/10.1088/1751-8113/44/9/095305},
year = {2011},
month = {feb},
publisher = {},
volume = {44},
number = {9},
pages = {095305},
author = {Colbeck, Roger and Kent, Adrian},
title = {Private randomness expansion with untrusted devices},
journal = {Journal of Physics A: Mathematical and Theoretical}
}

@article{Pan2021,
  title = {Oblivious communication game, self-testing of projective and nonprojective measurements, and certification of randomness},
  author = {Pan, A. K.},
  journal = {Phys. Rev. A},
  volume = {104},
  issue = {2},
  pages = {022212},
  numpages = {14},
  year = {2021},
  month = {Aug},
  publisher = {American Physical Society},
  doi = {10.1103/PhysRevA.104.022212},
  url = {https://link.aps.org/doi/10.1103/PhysRevA.104.022212}
}

@Article{Pironio2010,
author={Pironio, S.
and Ac{\'i}n, A.
and Massar, S.
and de la Giroday, A. Boyer
and Matsukevich, D. N.
and Maunz, P.
and Olmschenk, S.
and Hayes, D.
and Luo, L.
and Manning, T. A.
and Monroe, C.},
title={Random numbers certified by Bell's theorem},
journal={Nature},
year={2010},
month={Apr},
day={01},
volume={464},
number={7291},
pages={1021-1024},
issn={1476-4687},
doi={10.1038/nature09008},
url={https://doi.org/10.1038/nature09008}
}

@article{Wooltorton2024,
  title = {Device-Independent Quantum Key Distribution with Arbitrarily Small Nonlocality},
  author = {Wooltorton, Lewis and Brown, Peter and Colbeck, Roger},
  journal = {Phys. Rev. Lett.},
  volume = {132},
  issue = {21},
  pages = {210802},
  numpages = {6},
  year = {2024},
  month = {May},
  publisher = {American Physical Society},
  doi = {10.1103/PhysRevLett.132.210802},
  url = {https://link.aps.org/doi/10.1103/PhysRevLett.132.210802}
}

@article{Munshi2022,
  title = {Characterizing nonlocal correlations through various $n$-locality inequalities in a quantum network},
  author = {Munshi, Sneha and Pan, A. K.},
  journal = {Phys. Rev. A},
  volume = {105},
  issue = {3},
  pages = {032216},
  numpages = {11},
  year = {2022},
  month = {Mar},
  publisher = {American Physical Society},
  doi = {10.1103/PhysRevA.105.032216},
  url = {https://link.aps.org/doi/10.1103/PhysRevA.105.032216}
}

@Article{Renou2021,
author={Renou, Marc-Olivier
and Trillo, David
and Weilenmann, Mirjam
and Le, Thinh P.
and Tavakoli, Armin
and Gisin, Nicolas
and Ac{\'i}n, Antonio
and Navascu{\'e}s, Miguel},
title={Quantum theory based on real numbers can be experimentally falsified},
journal={Nature},
year={2021},
month={Dec},
day={01},
volume={600},
number={7890},
pages={625-629},
issn={1476-4687},
doi={10.1038/s41586-021-04160-4},
url={https://doi.org/10.1038/s41586-021-04160-4}
}

@article{Mohan2019,
doi = {10.1088/1367-2630/ab3773},
url = {https://dx.doi.org/10.1088/1367-2630/ab3773},
year = {2019},
month = {aug},
publisher = {IOP Publishing},
volume = {21},
number = {8},
pages = {083034},
author = {Karthik Mohan and Armin Tavakoli and Nicolas Brunner},
title = {Sequential random access codes and self-testing of quantum measurement instruments},
journal = {New Journal of Physics}
}

@article{Miklin2020,
  title = {Semi-device-independent self-testing of unsharp measurements},
  author = {Miklin, Nikolai and Borka\l{}a, Jakub J. and Paw\l{}owski, Marcin},
  journal = {Phys. Rev. Res.},
  volume = {2},
  issue = {3},
  pages = {033014},
  numpages = {15},
  year = {2020},
  month = {Jul},
  publisher = {American Physical Society},
  doi = {10.1103/PhysRevResearch.2.033014},
  url = {https://link.aps.org/doi/10.1103/PhysRevResearch.2.033014}
}

@article{Supic2016,
doi = {10.1088/1367-2630/18/3/035013},
url = {https://dx.doi.org/10.1088/1367-2630/18/3/035013},
year = {2016},
month = {apr},
publisher = {IOP Publishing},
volume = {18},
number = {3},
pages = {035013},
author = {Šupić, I and Augusiak, R and Salavrakos, A and Acín, A},
title = {Self-testing protocols based on the chained Bell inequalities},
journal = {New Journal of Physics}
}

@article{colbeck2011arxiv,
  title={Quantum And Relativistic Protocols For Secure Multi-Party Computation},
  author={Roger Colbeck},
  journal={arXiv: Quantum Physics},
  year={2009},
  url={https://api.semanticscholar.org/CorpusID:118140072}
}

@inproceedings{mayers1998,
author = {Mayers, Dominic and Yao, Andrew},
title = {Quantum Cryptography with Imperfect Apparatus},
year = {1998},
isbn = {0818691727},
publisher = {IEEE Computer Society},
address = {USA},
booktitle = {Proceedings of the 39th Annual Symposium on Foundations of Computer Science},
pages = {503},
series = {FOCS '98},
url = {https://dl.acm.org/doi/10.5555/795664.796390}
}

@article{Wu2016,
  title = {Device-independent parallel self-testing of two singlets},
  author = {Wu, Xingyao and Bancal, Jean-Daniel and McKague, Matthew and Scarani, Valerio},
  journal = {Phys. Rev. A},
  volume = {93},
  issue = {6},
  pages = {062121},
  numpages = {5},
  year = {2016},
  month = {Jun},
  publisher = {American Physical Society},
  doi = {10.1103/PhysRevA.93.062121},
  url = {https://link.aps.org/doi/10.1103/PhysRevA.93.062121}
}

@article{Bowles2018,
  title = {Device-Independent Entanglement Certification of All Entangled States},
  author = {Bowles, Joseph and \ifmmode \check{S}\else \v{S}\fi{}upi\ifmmode \acute{c}\else \'{c}\fi{}, Ivan and Cavalcanti, Daniel and Ac\'{\i}n, Antonio},
  journal = {Phys. Rev. Lett.},
  volume = {121},
  issue = {18},
  pages = {180503},
  numpages = {6},
  year = {2018},
  month = {Oct},
  publisher = {American Physical Society},
  doi = {10.1103/PhysRevLett.121.180503},
  url = {https://link.aps.org/doi/10.1103/PhysRevLett.121.180503}
}

@article{Bowles2018/2,
  title = {Self-testing of Pauli observables for device-independent entanglement certification},
  author = {Bowles, Joseph and \ifmmode \check{S}\else \v{S}\fi{}upi\ifmmode \acute{c}\else \'{c}\fi{}, Ivan and Cavalcanti, Daniel and Ac\'{\i}n, Antonio},
  journal = {Phys. Rev. A},
  volume = {98},
  issue = {4},
  pages = {042336},
  numpages = {24},
  year = {2018},
  month = {Oct},
  publisher = {American Physical Society},
  doi = {10.1103/PhysRevA.98.042336},
  url = {https://link.aps.org/doi/10.1103/PhysRevA.98.042336}
}

@article{mayers2004s,
  title={Self testing quantum apparatus},
  author={Dominic Mayers and Andrew Chi-Chih Yao},
  journal={Quantum Inf. Comput.},
  year={2003},
  volume={4},
  pages={273-286},
  url={https://api.semanticscholar.org/CorpusID:14069874}
}

@article{McKague2016,
doi = {10.1088/1367-2630/18/4/045013},
url = {https://dx.doi.org/10.1088/1367-2630/18/4/045013},
year = {2016},
month = {apr},
publisher = {IOP Publishing},
volume = {18},
number = {4},
pages = {045013},
author = {Matthew McKague},
title = {Self-testing in parallel},
journal = {New Journal of Physics}
}

@incollection{McKague2014,
       booktitle = {Theory of Quantum Computation, Communication, and Cryptography: 6th Conference, TQC 2011, Revised Selected Papers (Lecture Notes in Computer Science, Volume 6745)},
           title = {Self-testing graph states},
           pages = {104--120},
          author = {Matthew McKague},
       publisher = {Springer},
          editor = {D Bacon and M Roetteler and M Martin-Delgado},
            year = {2014},
         address = {Germany},
             doi = {10.1007/978-3-642-54429-3\_7},
             url = {https://eprints.qut.edu.au/107389/}
}

@book{Wilde2013,
  title={Quantum Information Theory},
  author={Wilde, M.},
  isbn={9781107034259},
  lccn={2012047378},
  series={Quantum Information Theory},
  url={https://books.google.co.in/books?id=T36v2Sp7DnIC},
  year={2013},
  publisher={Cambridge University Press}
}

@article{Wu2014,
  title = {Robust self-testing of the three-qubit $W$ state},
  author = {Wu, Xingyao and Cai, Yu and Yang, Tzyh Haur and Le, Huy Nguyen and Bancal, Jean-Daniel and Scarani, Valerio},
  journal = {Phys. Rev. A},
  volume = {90},
  issue = {4},
  pages = {042339},
  numpages = {10},
  year = {2014},
  month = {Oct},
  publisher = {American Physical Society},
  doi = {10.1103/PhysRevA.90.042339},
  url = {https://link.aps.org/doi/10.1103/PhysRevA.90.042339}
}

@article{Supic2020rev,
  doi = {10.22331/q-2020-09-30-337},
  url = {https://doi.org/10.22331/q-2020-09-30-337},
  title = {Self-testing of quantum systems: a review},
  author = {{\v{S}}upi{\'{c}}, Ivan and Bowles, Joseph},
  journal = {{Quantum}},
  issn = {2521-327X},
  publisher = {{Verein zur F{\"{o}}rderung des Open Access Publizierens in den Quantenwissenschaften}},
  volume = {4},
  pages = {337},
  month = sep,
  year = {2020}
}

@article{Munshi2023PRA,
  title = {Self-testing of an unbounded number of mutually commuting local observables},
  author = {Munshi, Sneha and Pan, A. K.},
  journal = {Phys. Rev. A},
  volume = {108},
  issue = {6},
  pages = {062607},
  numpages = {11},
  year = {2023},
  month = {Dec},
  publisher = {American Physical Society},
  doi = {10.1103/PhysRevA.108.062607},
  url = {https://link.aps.org/doi/10.1103/PhysRevA.108.062607}
}
\end{document}